%% file: ms.tex
\documentclass[a4paper,fleqn,usenatbib,useAMS]{mnras}

\usepackage{graphicx}
\usepackage{fixltx2e}
\usepackage{natbib}
\usepackage{color}
\usepackage{booktabs}
\usepackage{multirow}
\usepackage{bm}

\usepackage[T1]{fontenc}
\usepackage{ae,aecompl}

\def\Teff{$T_{\rm eff}$}
\def\logg{$\log g$}
\def\Vsini{$V_r$\,sin\,$i$}
\def\Vm{$V_{\rm m}$}
\def\xi{$X_{i}$}
\def\kmps{km\,s$^{\rm-1}$}
\newcommand{\pion}[2]{{#1}\,{\sc {#2}}}

\bibpunct{(}{)}{,}{a}{}{;}

\title[The metal rich abundance pattern]
{The metal rich abundance pattern -- spectroscopic properties and abundances for 107 main-sequence stars}

\author[O.M. Ivanyuk et al.]
{O.M. Ivanyuk$^{1}$,
J.S. Jenkins$^{2}$,
Ya.V. Pavlenko$^{1,3}$,
H.R.A. Jones$^{3}$,
D.J. Pinfield$^{3}$ \\
\mbox{}
$^{1}$Main Astronomical Observatory, National Academy of Sciences of Ukraine, Holosiyiv Wood, Kyiv-127, 03680, Ukraine \\
$^{2}$Departamento de Astronom\'ia, Universidad de Chile, Camino el Observatorio 1515, Las Condes, Santiago, Chile \\
$^{3}$Centre for Astrophysics Research, University of Hertfordshire, College Lane, Hatfield, Hertfordshire, AL10 9AB, UK}

\date{}
\pubyear{2017}
\begin{document}
\label{firstpage}
\pagerange{\pageref{firstpage}--\pageref{lastpage}}
\maketitle

\begin{abstract}
We report results from the high resolution spectral analysis of the 107 metal rich (mostly [Fe/H]$\ge$7.67\,dex) target stars from the Calan-Hertfordshire Extrasolar Planet Search program observed with HARPS. Using our procedure of finding the best fit to the absorption line profiles in the observed spectra, we measure the abundances of Na, Mg, Al, Si, Ca, Ti, Cr, Mn, Fe, Ni, Cu, and Zn, and we then compare them with known results from different authors. Most of our abundances agree with these works at the level of $\pm$0.05\,dex or better for the stars we have in common. However, we do find systematic differences that make direct inferences difficult. Our analysis suggests that the selection of line lists and atomic line data along with the adopted continuum level influence these differences the most. At the same time, we confirm the positive trends of abundances versus metallicity for Na, Mn, Ni, and to a lesser degree, Al. A slight negative trend is observed for Ca, whereas Si and Cr tend to follow iron. Our analysis allows us to determine the positively skewed normal distribution of projected rotational velocities with a maximum peaking at 3\,\kmps. Finally, we obtained a Gaussian distribution of microturbulent velocities that has a maximum at 1.2\,\kmps and a full width at half maximum $\Delta v_{1/2}$=0.35\,\kmps, indicating that metal rich dwarfs and subgiants in our sample have a very restricted range in microturbulent velocity.
\end{abstract}


\section{Introduction}
\label{_introduction}
Understanding the chemical make-up of stars like the Sun is fundamental to our understanding of star formation and stellar evolution. \citet{edva93} undertook a classical study of the chemical evolution of the galactic disc, deriving abundances for a sample of 189 nearby field F and G dwarfs. They found a number of interesting relationships in the data, for example, that the strongest age dependent abundance comes from Ba, which they attribute to the efficient s-element synthesis in low-mass AGB stars that enrich the interstellar medium long after star formation. They confirmed that metal-poor stars ([Fe/H]$<$-0.4\,dex) are relatively overabundant in $\alpha$-elements. Since the [$\alpha$/Fe] for these stars shows a gradient that decreases with increasing galactocentric distance of the orbits, they show that star formation was probably more vigorous and started first in the inner parts of the galactic disc.

Since this work, various other samples have been studied, a number of which were performed by the exoplanet community \citep[e.g.][]{vale05,bond08,neve09}. These works have concentrated on studying the abundance distributions of exoplanet host stars compared with non-exoplanet hosts and a number of interesting trends have been found. Along with the well established dependence of giant planet detection probability on host star metallicity \citep{gonz97,sant08} these abundance analyses indicate that various other atomic abundances are likely enhanced in exoplanet hosts compared to non-exoplanet hosts (e.g. Si \& Ni), at least for the gas giant population. These over-abundances are explained in the framework of the core accretion scenario of planet formation \citep{livi03,ida08} where the more disc material present, the higher the probability of planet formation.

In general, the relative yields of different elements in the atmospheres of stars change with time, due to the processes of nucleosynthesis in the Galaxy. These processes are still poorly known, yet performing high quality analyses of homogeneous stellar samples can shed light on the fine underlying processes that are occurring, following the formation of planetary systems.

In recent years, series of large-scale abundance analyses have been published \citep{vale05,luck06,jenk08,adib12,bens14,brew16}, mainly driven by the availability of high resolution spectral data from observational campaigns that are dedicated to searching for planets and the proliferation of software that allows us to automatically process the spectra of hundreds of stars in an automatic, or semi-automatic, manner.

The analysis we provide here is used to determine the chemical abundances in the atmospheres of metal rich stars, along with their relative behaviour versus iron. It is worth noting that our sample is one of the generally homogeneous samples \citep[e.g.][]{adib12} in comparison to similar studies, which cover up to a thousand stars, but obtained using different instruments with different resolutions and pipeline reduction techniques \citep[e.g.][]{bens14}. On the other hand, some homogeneous studies cover rather smaller number of stars \citep[e.g.][]{felt98,brun02,gill06} or with lower resolution \citep[e.g.][]{luck06}. In this work we combined both large enough sample and high resolution to provide us with a good statistical sample to investigate.

The layout of the manuscript is as follows, in Section \ref{_observations} we discuss the observational data, in Section \ref{_abundance_method} we describe the method used to derive the abundances, \Teff\, and \logg, our selection criteria of the examined lines, the influence of the microturbulence and rotation. Section \ref{_results} contains the description of our results, and in Section \ref{_comparison} we compare these results to the previously published works. Finally, in Section \ref{_discussion} we discuss the implication of our results in the field, and summarise our findings in Section \ref{_summary}.

\section{Observations}
\label{_observations}

The Calan-Hertfordshire Extrasolar Planet Search (CHEPS) program \citep{jenk09} is monitoring a sample of the metal rich stars in the southern hemisphere, to improve the statistics for planets orbiting such stars, along with searching for short period planets that have a high probability to transit their host star. All stars in the sample were initially selected from the {\it Hipparcos} catalogue to have $V$-band magnitudes in the range 7.5 to 9.5. This range ensures that the sample is not overlapping with existing planet search programs, since most solar-like stars with magnitudes below 7.5 are already being examined as part of other planet search programs, and the upper limit is set to ensure that the stars are bright enough to allow the best follow-up to search for secondary eclipses and transmission spectroscopy, essentially bridging the gap between the long-term precision radial velocity programs, and the fainter samples that photometric transit surveys are generally biased towards.

More specifically, the CHEPS sample primarily contains objects that were drawn from a larger southern sample observed using the ESO-FEROS spectrograph \citep[see][]{jenk08,jenk11}. The secondary selection of these targets was focused on the inactive (log\,$R'_{\rm{HK}}\le$-4.5\,dex) and metal rich ([Fe/H]$\ge$0.1\,dex) subset of this sample \citep{jenk08} to ensure the most radial velocity stable targets, and to make use of the known increase in the fraction of planet-host stars with increasing metallicity, mentioned above. We note that, recent work has shown that the fraction of metal rich stars hosting low-mass planets may not follow the metallicity trend observed in the gas giant population \citep{buch12,jenk13a}.

Our subset of targets have been followed-up using the HARPS at La Silla in Chile. In our analysis, we use high S/N ($>$100) and high-resolution spectra (R$\sim$120,000) of predominantly single stars from the CHEPS sample that have been well characterised photometrically based on {\it Hipparcos} data \citep{jenk08,jenk09,jenk11}.

\begin{figure*}
\centering
\includegraphics[width=55mm]{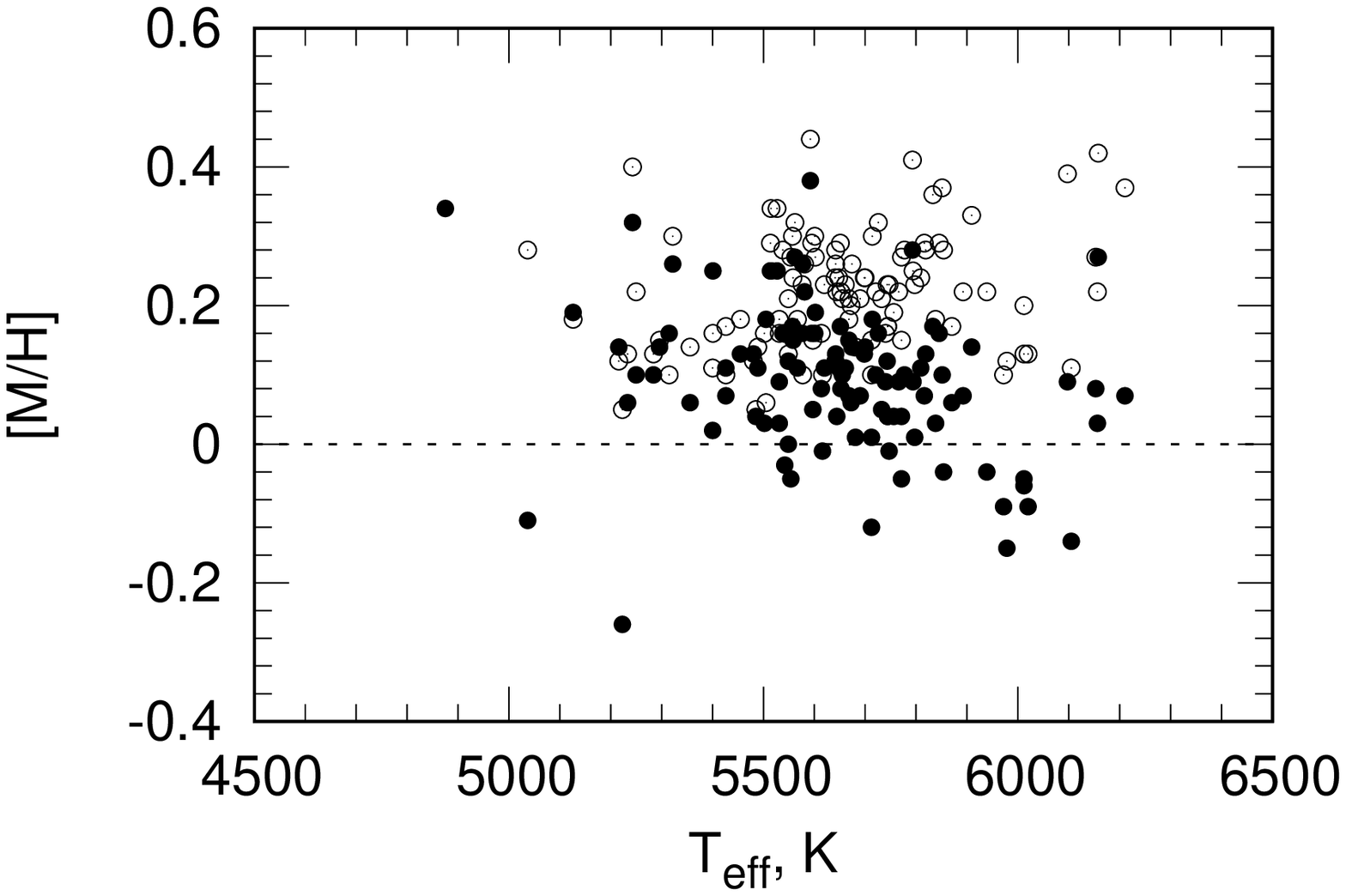} 
\includegraphics[width=55mm]{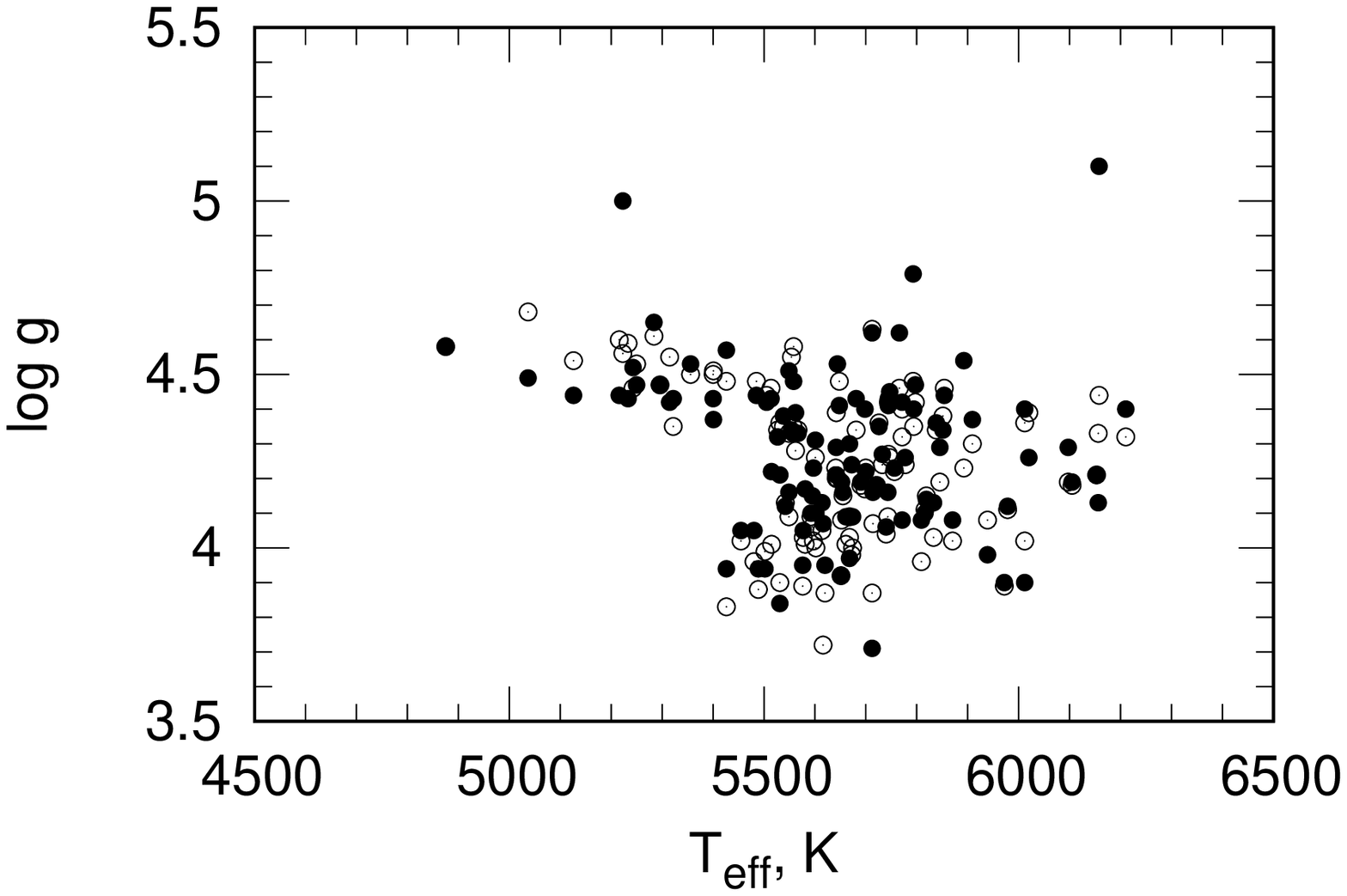} 
\includegraphics[width=55mm]{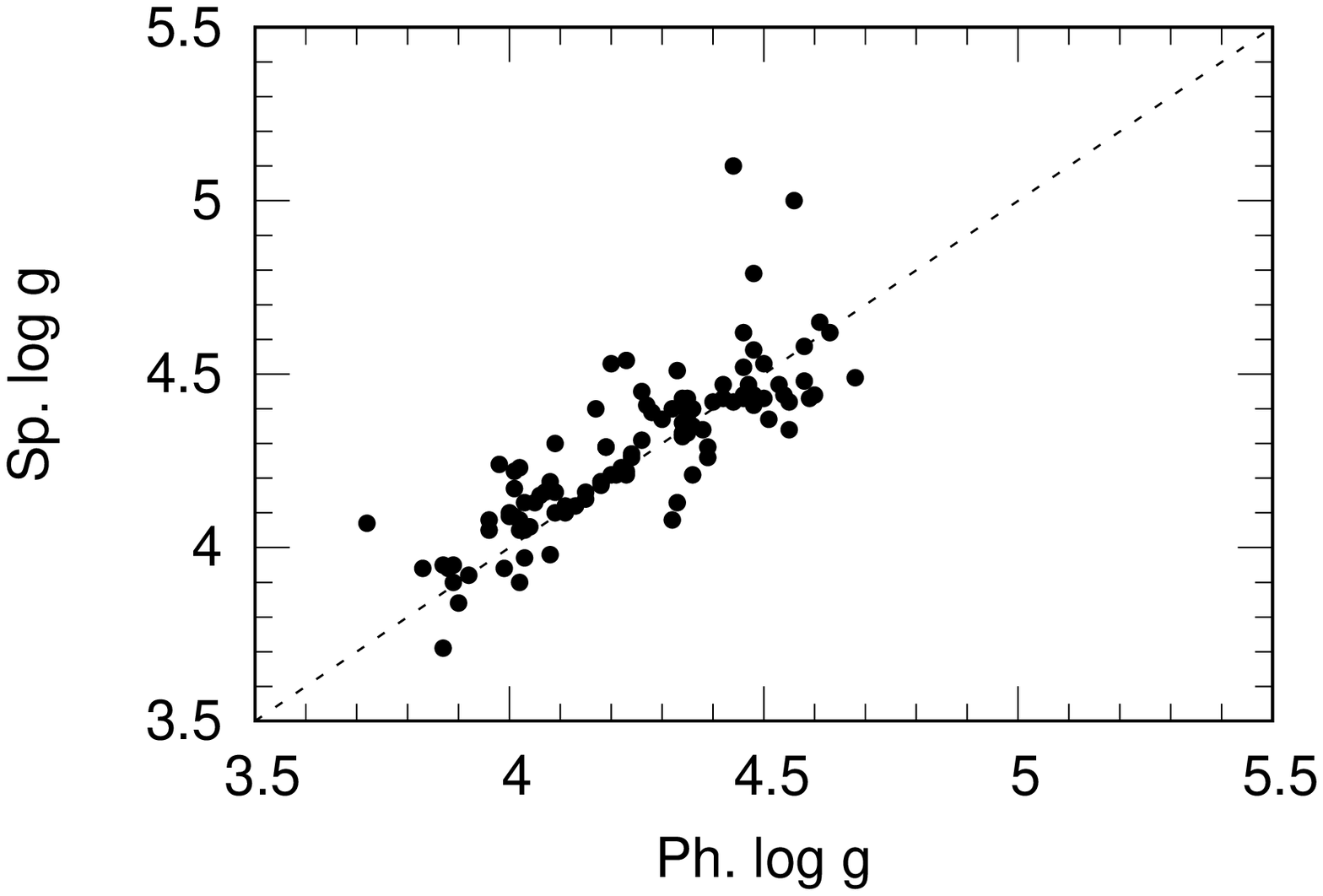} 
\includegraphics[width=55mm]{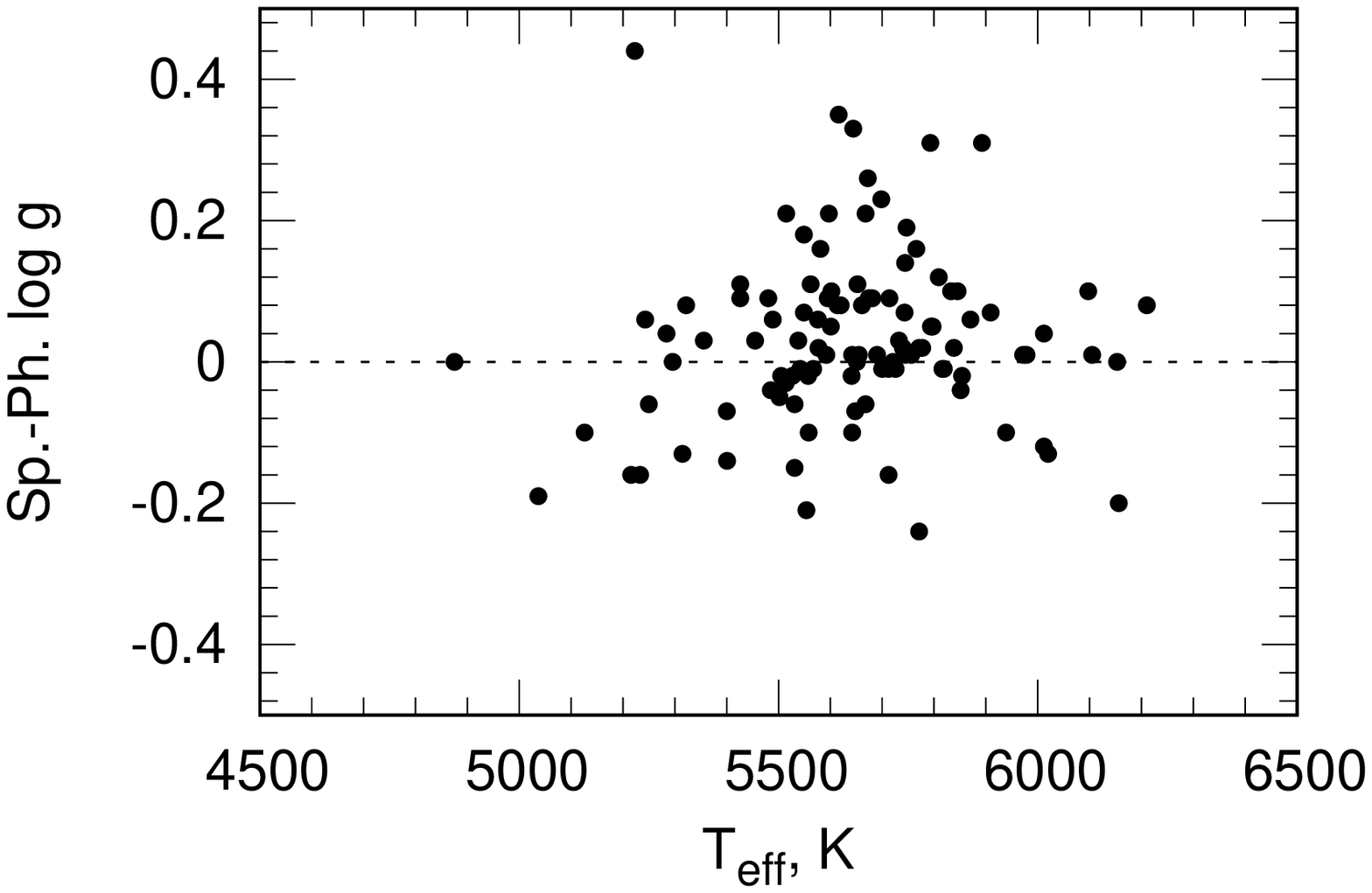} 
\includegraphics[width=55mm]{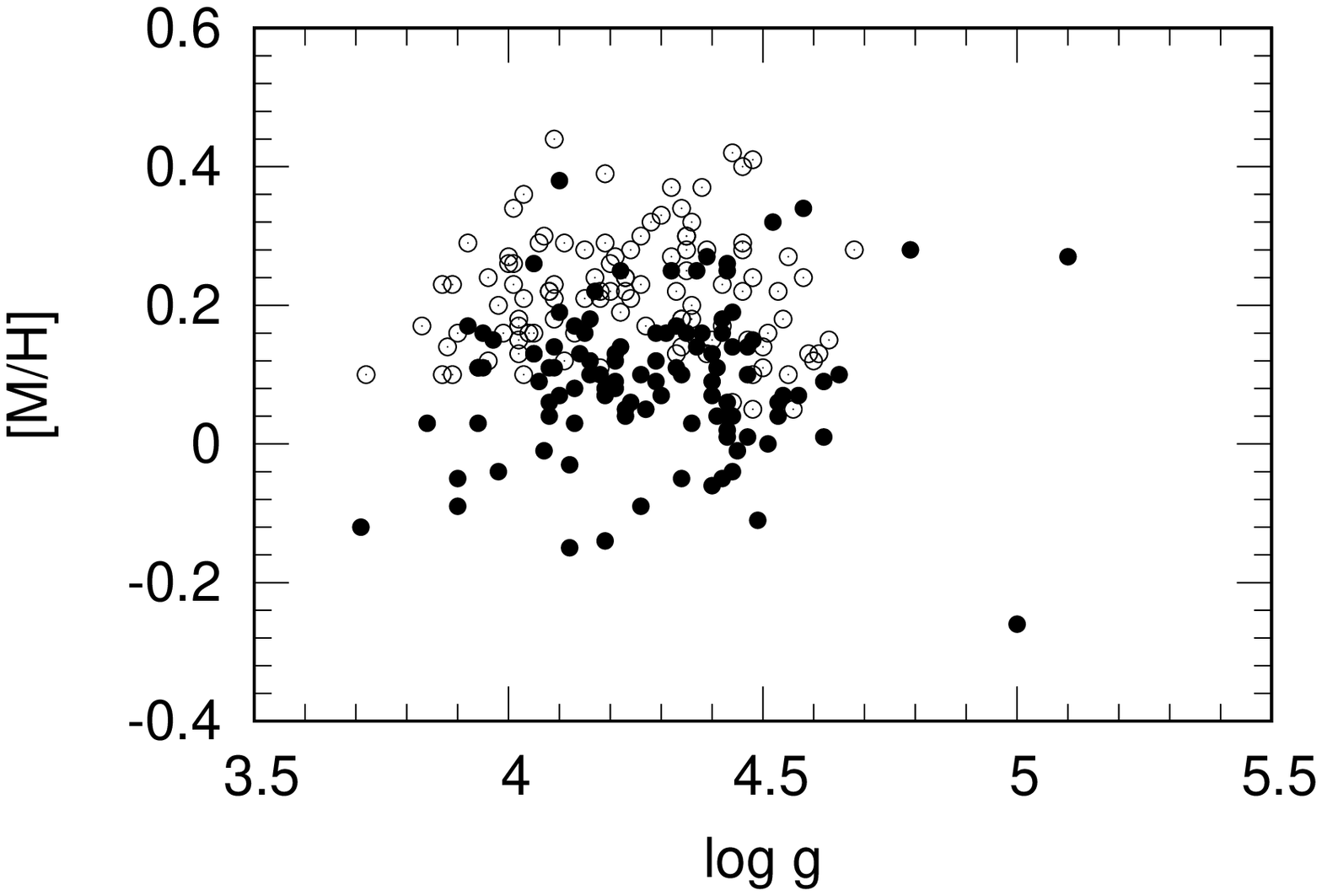} 
\includegraphics[width=55mm]{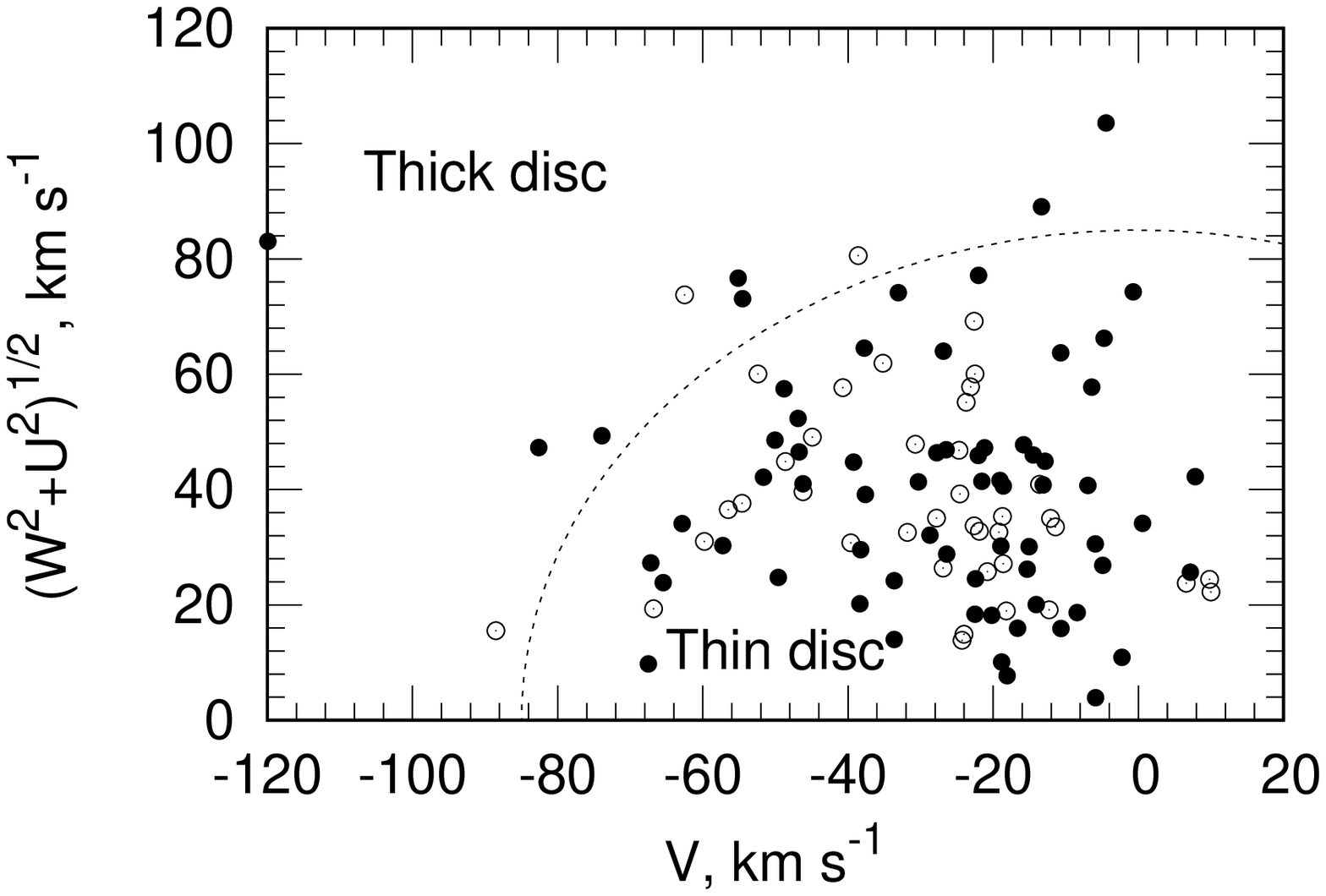} 
\caption{The general properties of the sample. The HARPS data, based on our spectroscopic \logg\, values, is shown with filled circles. Mean error bars of our \Teff, \logg, and [Fe/H] determinations are $\pm$50\,K, $\pm$0.2\,dex, and $\pm$0.1\,dex, respectively. The FEROS data, based on the photometric determinations of the \logg\, parameter \protect\citep{jenk08}, is shown with open circles. {\it Bottom right:} the Toomre diagram for 67 stars in the sample is shown, using the data taken from \protect\cite{murg13} (filled circles) and for the remaining 40 stars from \protect\cite{jenk11} (empty circles). The boundary between the thin and thick disc of the Galaxy is defined by following \protect\cite{fuhr04}.}
\label{_figure_general_properties}
\end{figure*}

The general properties of the sample are shown in Fig. \ref{_figure_general_properties}. \cite{jenk08} carried out the photometric determinations of the surface gravity parameter for our stars (shown with open circles), and we compare these values to our spectroscopically measured \logg\, (filled circles). We work within a narrow \Teff\, range between 5200 to 6200\,K, with the majority of our stars being metal rich and on the main sequence (\logg$>$4.1), although a few are slightly metal deficient ([Fe/H]$>$-0.2). Most of the stars in our sample belong to the thin disc galactic population; see the Toomre diagram at the bottom right of Fig. \ref{_figure_general_properties}.

\section{Synthetic spectra analysis}
\label{_abundance_method}

\subsection{Abundances and surface gravities}

We used a modified numerical scheme developed by \citet{pavl17} that allows to determine the atomic abundances, rotational velocities, microturbulences and surface gravities in the stellar atmospheres from high resolution and high S/N spectra, all in the framework of iterative approach, i.e. at each new step, the model atmospheres and synthetic spectra were recomputed for the metallicities and gravities derived before in our procedure, and for the fixed effective temperatures that were set photometrically.

The method is based on the minimisation of differences in the profiles of computed and observed lines. We used the two independent procedures for the final model parameters. Firstly, we required there to be no dependence of the \pion{Fe}{i} abundance on the line strengths to obtain \Vm. Secondly, we required the agreement between the abundances of \pion{Fe}{i} and \pion{Fe}{ii} obtained for the previously found \Vm\, to determine \logg.

For our starting point we used the photometrically determined metallicities and \logg\, \citep[see][]{jenk08}, then we ran a few iterations to determine [\pion{Fe}{i}/H], \Vm, [\pion{Fe}{ii}/H] and \logg. The final step was the determination of abundances and \Vsini for the model atmosphere parameters found in the procedure before.

All synthetic spectra were computed by the \textsc{wita6} program \citep{pavl97} using 1D local thermodynamic equilibrium (LTE) model atmospheres computed with \textsc{sam12} code \citep{pavl03} combined with the minimization routine \textsc{abel8} \citep{pavl12,pavl17}. The lines to be fitted in the stellar spectra were compiled using the solar spectrum by \cite{kuru84} and atomic line data from \textsc{vald-2} and \textsc{vald-3} \citep{kupk99,ryab15}.

In our analysis we made a few basic assumptions: the level of activity is low in these atmospheres, in other words, the continuum in the optical range forms in the photosphere of the star. Both the micro- and macroturbulent velocity fields are similar to the solar case. For simplicity we adopt that $V_{\rm macro}$ and \Vm\, do not change with the depth in these atmospheres. The convection zone is similar to the solar convection zone, even in the case of super metal rich stars. The minimum of the atmospheric temperature is located above the line formation region. Spots on the surface of these stars are not sufficiently numerous to contribute to the formation of absorption lines or continuum, in terms of the emitted fluxes. Naturally, if any of these assumptions is not valid for the physical state of the studied stars, we will obtain a spread of abundances and other determined parameters, even in the case of ideally determined \Teff\, and \logg.

\subsection{Effective temperatures}

The effective temperatures were estimated using the photometric methods \citep[see][]{jenk08}, exploiting the existing large photometric databases, along with the latest relationships between stellar broad-band colours and their photospheric effective temperatures. Namely, we used the Johnson $V$-band photometry that was taken from the {\it Hipparcos} database \citep{perr97} as our optical anchor point, and combined this with near-infrared photometry from the 2MASS database \citep{stru06}, in particular the $K_s$-band magnitude that gives a sufficiently large wavelength baseline to sample the shape of the spectral energy distribution. We then used the relationships provided in \citet{casa10} to calculate accurate effective temperatures and place the stars on an HR-diagram to calculate the photometric \logg\, values using the Y2 evolutionary models \citep{dema04}.

We choose to use these photometric \Teff\, for the computations, rather than update them spectroscopically. We believe this constrains uncertainties which may come from the quality of fits to the spectra and atomic line data, and these may also affect the results of our determination of abundances, microturbulent velocities, and surface gravities.

\subsection{Line lists}
\label{_line_lists}

We provide our analysis for the pre-selected list of spectral lines extracted from \textsc{vald-2} and \textsc{vald-3} \citep{pisk95,kupk99,ryab15}. To create the list of 'reliable' lines for each element, we computed the solar synthetic spectra and convolved them to get the effective resolving power of R$\sim$70,000, which is the effective resolution due to macroturbulent velocity field in the solar atmosphere.

The theoretical spectra were fitted to the selected observed lines using the minimization routine \textsc{abel8}. At this stage, we excluded severely blended lines, too weak lines with residual fluxes lower than 20 per cent and strong lines with $r_{nu}>$0.7 to minimize the distortion of the results due to noise and possible NLTE effects in the cores of the strong lines \citep{mash11}.

We computed our synthetic spectra using the damping constants provided by \textsc{vald} \citep{kupk99}. For the absorption lines without damping constants we computed them using the Unsold formulas \citep{unso55}.

The fitting part of each line profile was adopted manually to reduce the uncertainties introduced by wing blending. In some cases, very close blends of the lines of the same element were fitted together, resulting in a single spectral range to fit multiple lines. The final list of fitting ranges for the neutral and ionized atoms available on CDS.

\subsection{Microturbulent velocities}
\label{_method_microturbulence}

\begin{figure}
\centering\includegraphics[width=53mm, angle=-90]{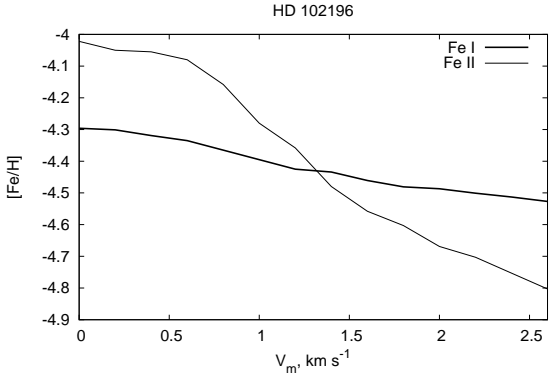} 
\caption{The dependence of \pion{Fe}{i} and \pion{Fe}{ii} abundances on microturbulent velocity obtained from the fits of our synthetic spectra for the model atmosphere \Teff/\logg=6012/3.90 to the observed HD 102196 spectrum.}
\label{_figure_method_microturbulence}
\end{figure}

The microturbulent velocity is an important parameter in 1D line-profile fitting techniques as there is an evident trend with abundances that causes uncertainties in their measurements. The dependence of \pion{Fe}{i} and \pion{Fe}{ii} abundances on microturbulent velocity for HD 102196 is shown in Fig. \ref{_figure_method_microturbulence}. Here, \pion{Fe}{i} and \pion{Fe}{ii} abundances differ by up to $\pm$0.1\,dex at \Vm\,$=$1.0\,\kmps but the same at 1.4\,\kmps.

We carried out a set of synthetic spectra fits to \pion{Fe}{i} and \pion{Fe}{ii} lines independently, using the grid of adopted microturbulent velocities in range from 0.0 to 2.6\,\kmps, with a step of 0.2\,\kmps. In fact, we followed the procedure of \citet{pavl12}, but here we investigated the dependence of $D_a=\partial a/\partial r_{\nu_0}$, where $a$ and $r_{\nu_0}$ are the abundance and the central intensity of the corresponding iron absorption line computed for the grid of \Vm, the $D_a$ is known to show the dependence on \Vm. Therefore, we determine our \Vm\, at $D_a$=0.

Our procedure is based on fitting to the line profiles, excluding the shallow parts of the wings, which could be more affected with blending by the weak lines of other elements. Furthermore, the wings of lines are more affected by the pressure broadening than the microturbulent velocities. Accounting both these factors should improve the determination of \Vm\, (see Fig. \ref{_figure_microturbulence_comparison}).

\subsection{Rotational velocities}

The profiles of absorption lines in the stellar spectra affected by the rotation of a star, depending on the magnitude of rotation. The rotational profile, in our case, determined by the formula of \cite{gray76}, is of a different shape compared to the instrumental broadening and macroturbulence. In our analysis, each line of the synthetic spectra was convolved with a profile of a different \Vsini, and was fitted to the observed spectra until the best result was found. Despite we used only \pion{Fe}{i} lines for the determination of \Vsini, the procedure was also applied for every other line of all our elements to get a better fit to the observed line profiles.

\section{Results}
\label{_results}

\subsection{Rotational velocities}

\begin{figure*}
\centering
\includegraphics[width=55mm]{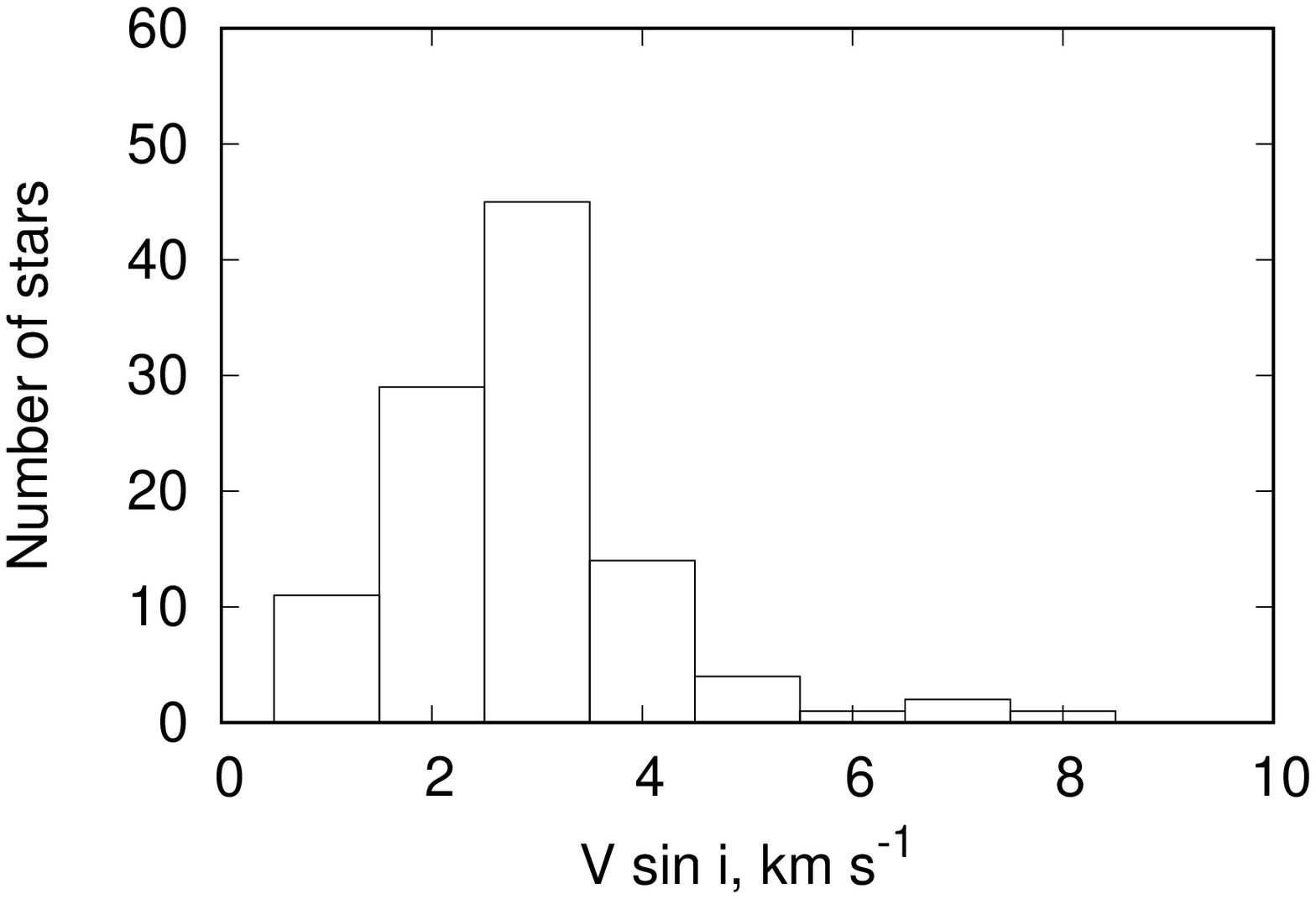} 
\includegraphics[width=55mm]{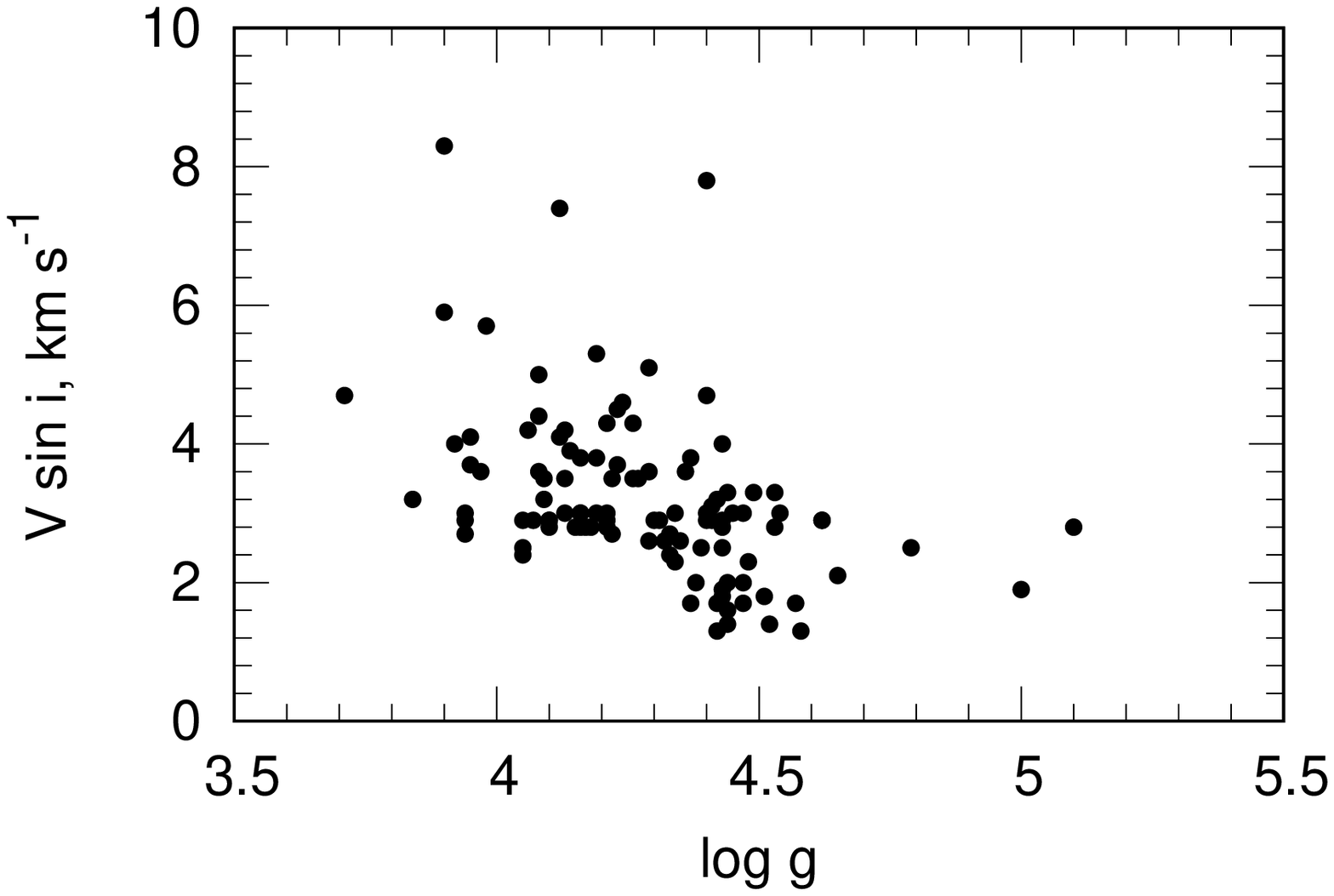} 
\caption{{\it Left:} the histogram shows the distribution of rotational velocities \Vsini\, for the stars in our sample. {\it Right:} the dependence of \Vsini\, versus \logg.}
\label{_figure_rotation}
\end{figure*}

We found that the majority of our stars are slow rotators, with \Vsini$<$4\,\kmps (Fig. \ref{_figure_rotation}). Likely, a few stars with larger rotational velocity present the cases where sin\,$i \sim$\,1, rather than large $V_r$ itself, since these stars are of lower activities, with only a few fast rotating active stars left over.

On the other hand, \Vsini\, correlates with \logg, older stars rotate slower, in agreement with the theory, and vice versa, which is in line with their ages given by the preliminary CHEPS selection. We can see an uprising trend of \Vsini\, for the stars of lower \logg, which represent a younger population.

For the elements other than \pion{Fe}{i}, \Vsini\, was used rather as the adjusting parameter. Nevertheless, it agreed for the properly fitted lines of different elements within an uncertainty of less than 1\,\kmps. We can claim a clear measurement for the values of \Vsini$>$2\,\kmps. Uniform results for the iron lines give credibility to our results.

\subsection{Solar abundances}

We carried out a differential analysis with respect to the Sun as a star, since our line lists were selected based on a comparison of the synthetic spectra to the observed solar spectrum by \cite{kuru84}. This allowed us to minimize possible blending effects, at least in the solar case.

We performed a quantitative analysis of the solar abundances to verify our procedure. Computations were done for the initial model atmosphere with \Teff/\logg/[Fe/H]=5777/4.44/-4.40, computed using \textsc{sam12} \citep{pavl03}.

For the iron ions, we obtained N(\pion{Fe}{i})=-4.42$\pm$0.03\,dex and N(\pion{Fe}{ii})=-4.46$\pm$0.02\,dex, which again highlights the known problem of the measured difference between \pion{Fe}{i} and \pion{Fe}{ii} abundances obtained in the framework of the classical approach \citep[see][]{holw90,shch15,shch16}. The abundances for the other elements (see the first column of Table \ref{_table_solar}) are in agreement with those in the literature \citep[e.g.][]{ande89,aspl09}.

We also determined the microturbulence and projected rotational velocity for the Sun as a star, measuring \Vm=1.0$\pm$0.2\,\kmps and \Vsini=1.69$\pm$0.09\,\kmps, which also agree with known results. The solar abundances determined for the different atomic line data and line lists are discussed in Section \ref{_comparison}.

\subsection{Stellar abundances}
\label{_abundances}

The complete table of abundances available on CDS. In our analysis, we used the parameter $X_f$ to describe the average slope of the distribution of abundances for a given element relative to iron. We computed $X_f$ using a standard least squares approach to approximate the dependence of [X/Fe] versus [Fe/H] by a linear function. By definition, $X_f$ characterises, to first-order, the relative changes of the yield of elements with respect to the iron.

\begin{figure*}
\centering
\includegraphics[width=55mm]{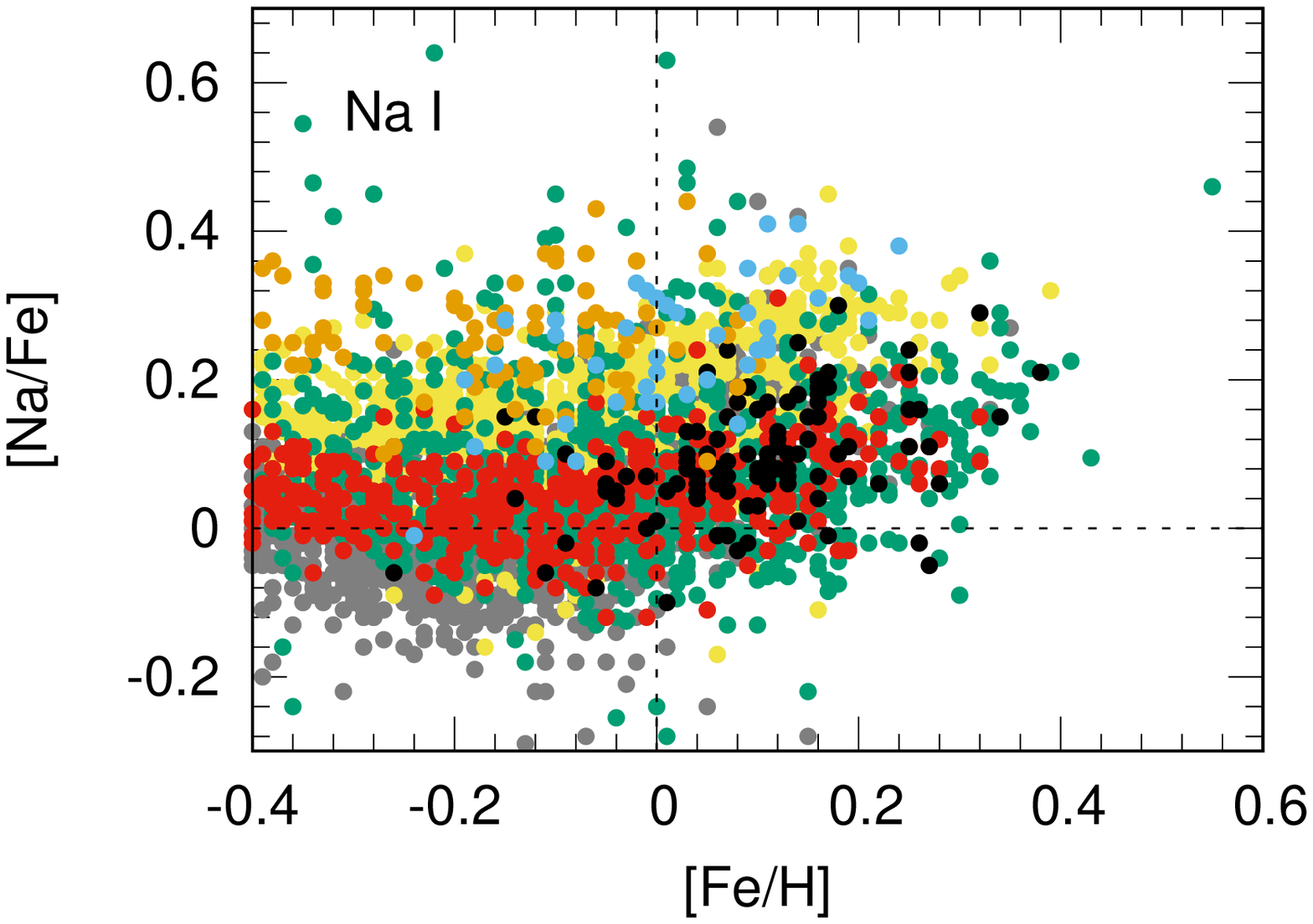} 
\includegraphics[width=55mm]{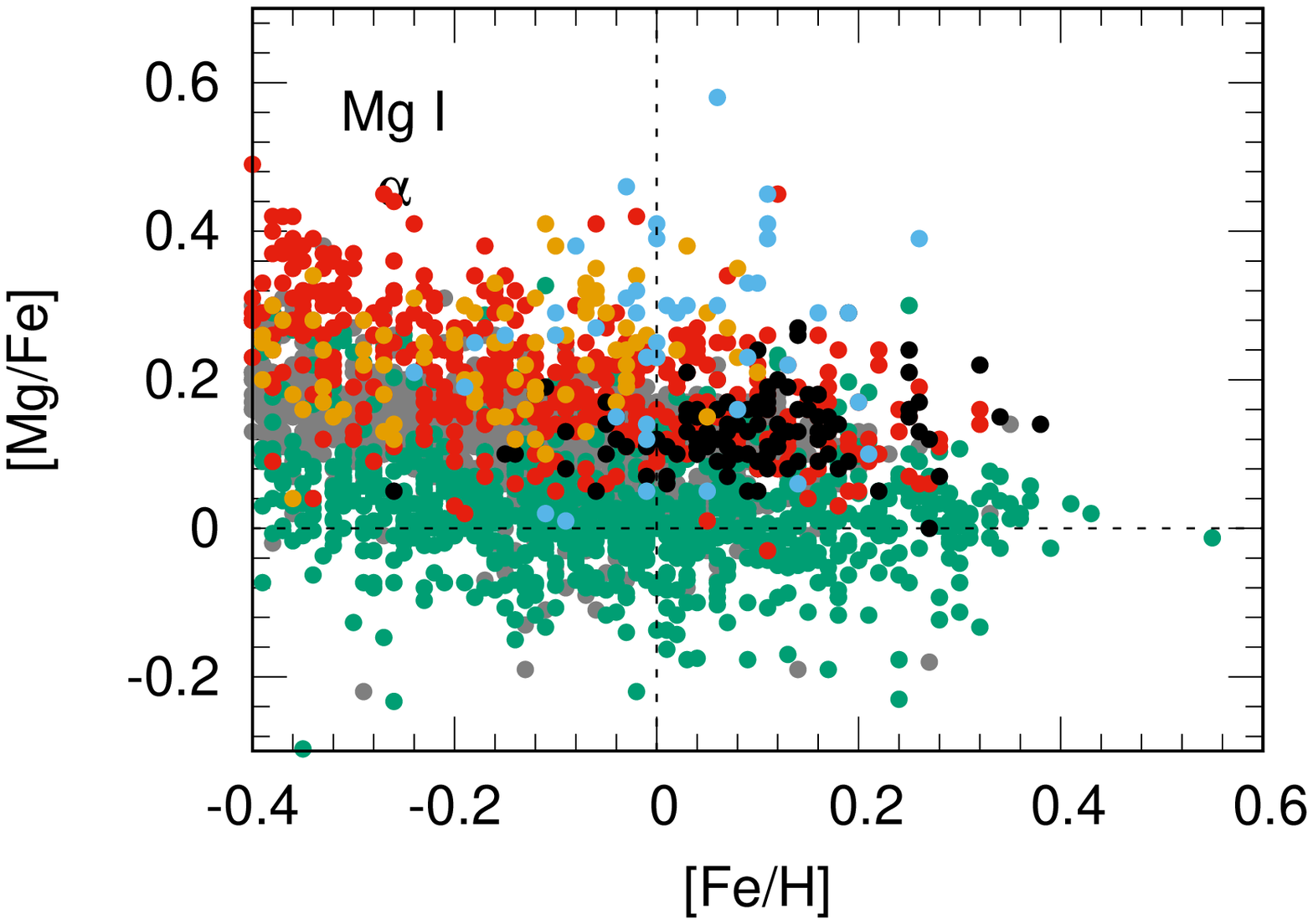} 
\includegraphics[width=55mm]{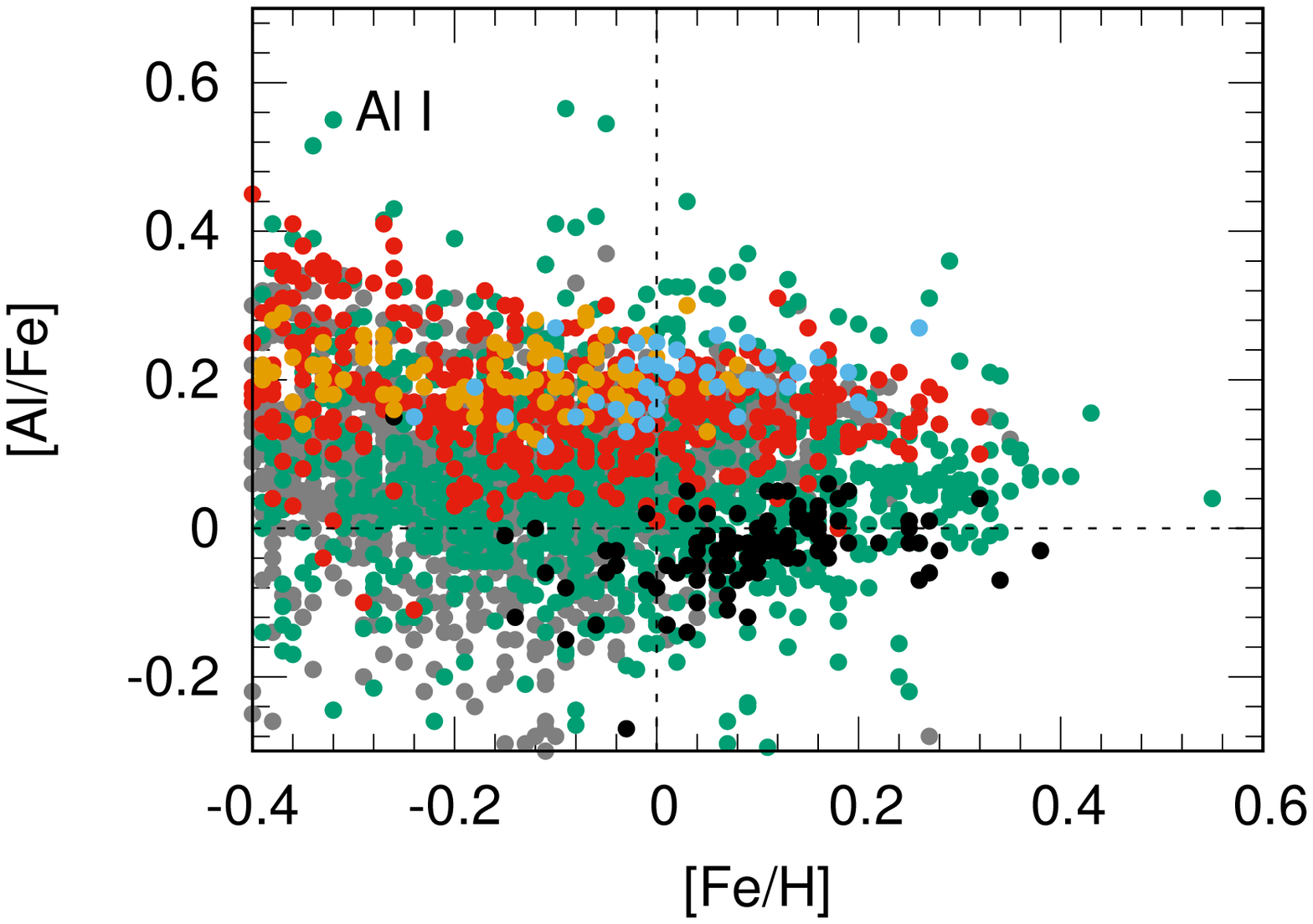} 
\includegraphics[width=55mm]{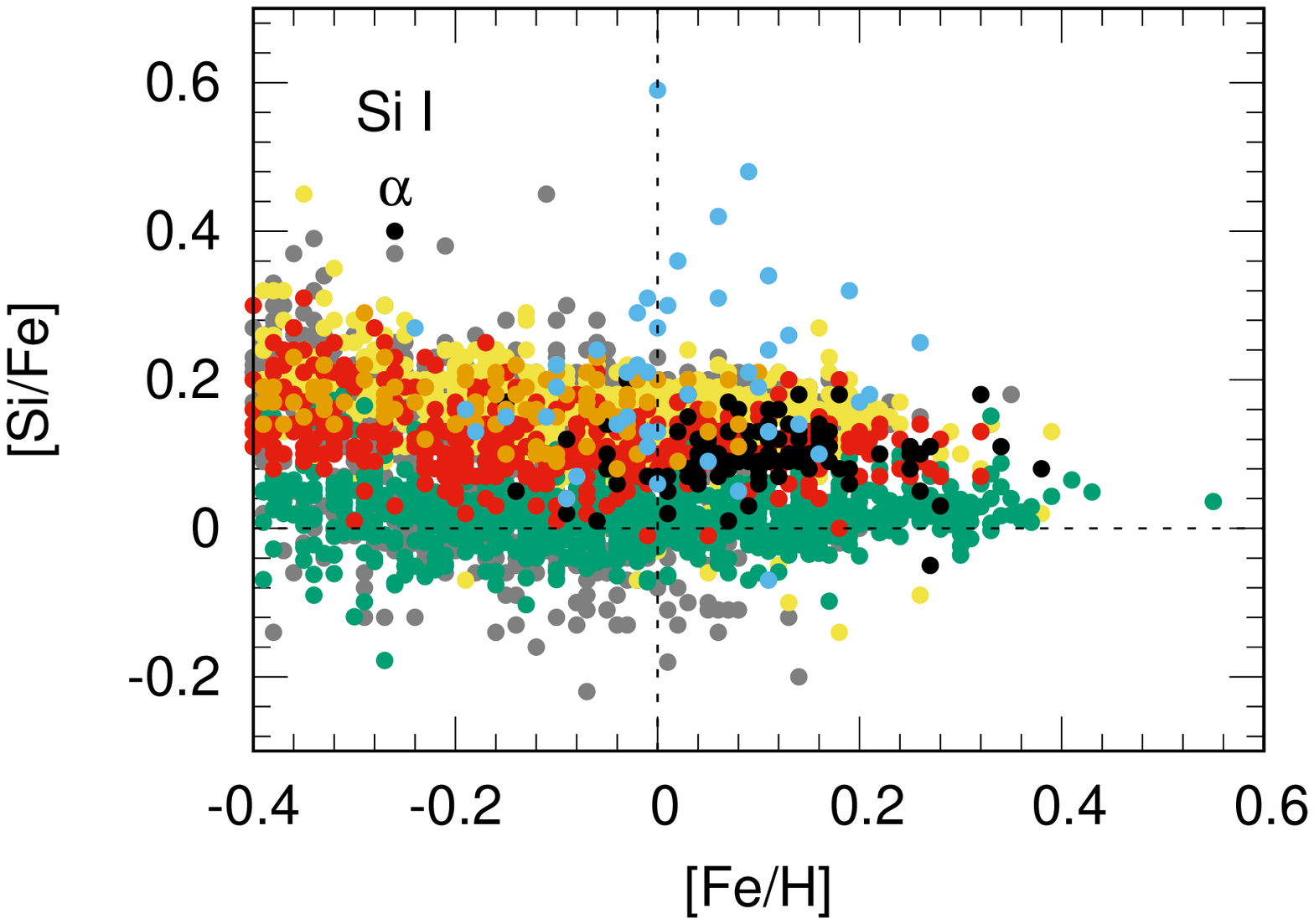} 
\includegraphics[width=55mm]{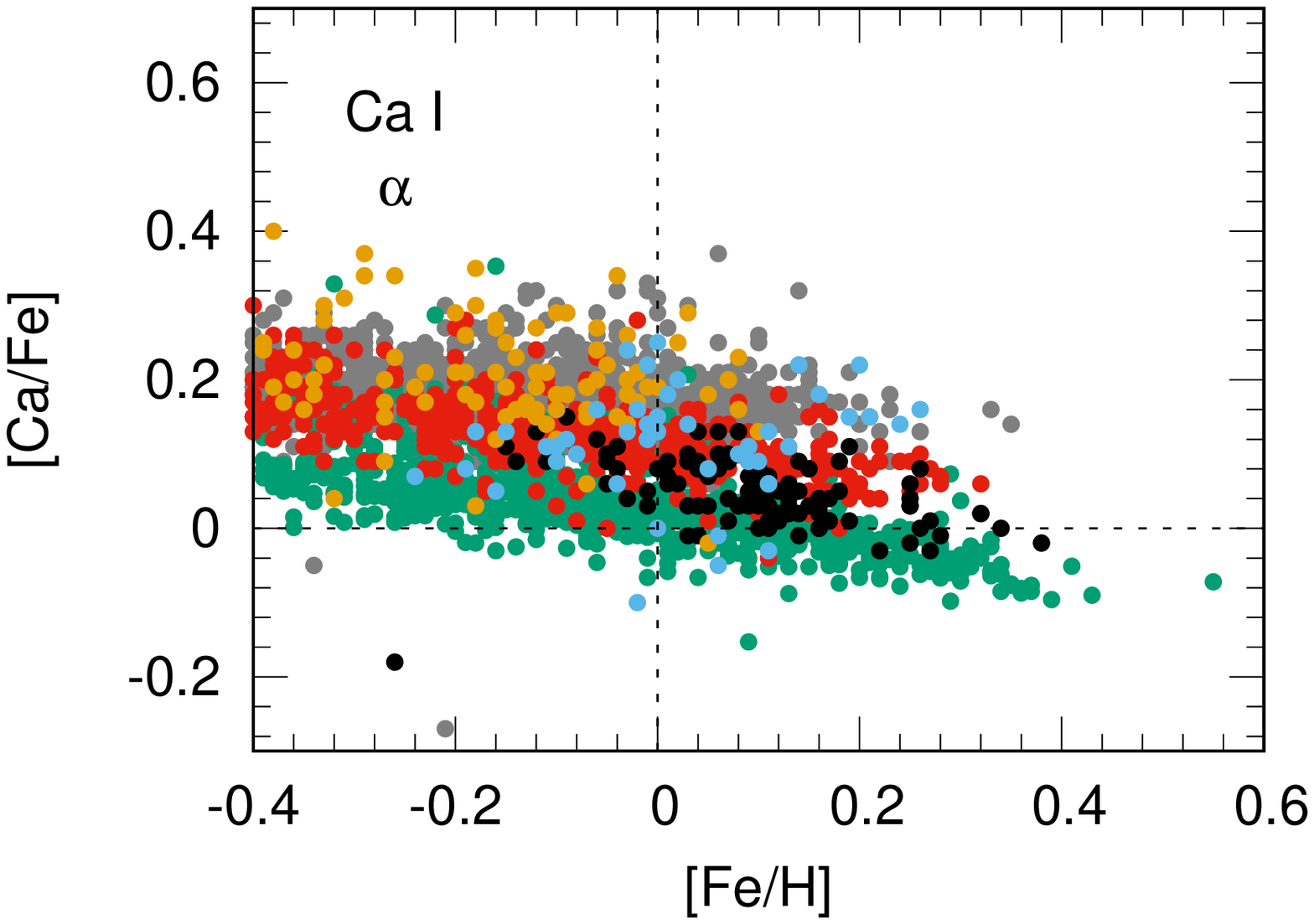} 
\includegraphics[width=55mm]{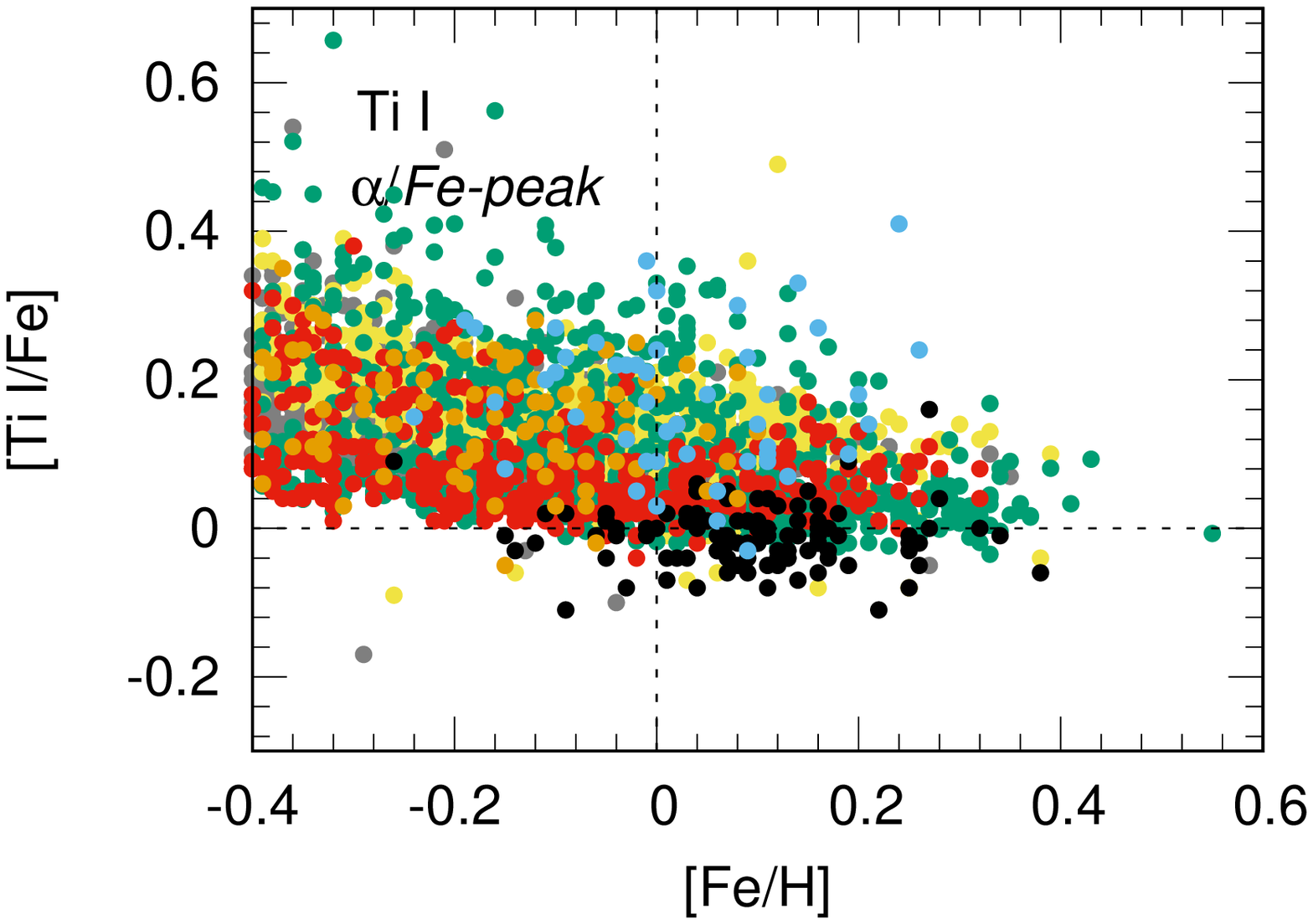} 
\includegraphics[width=55mm]{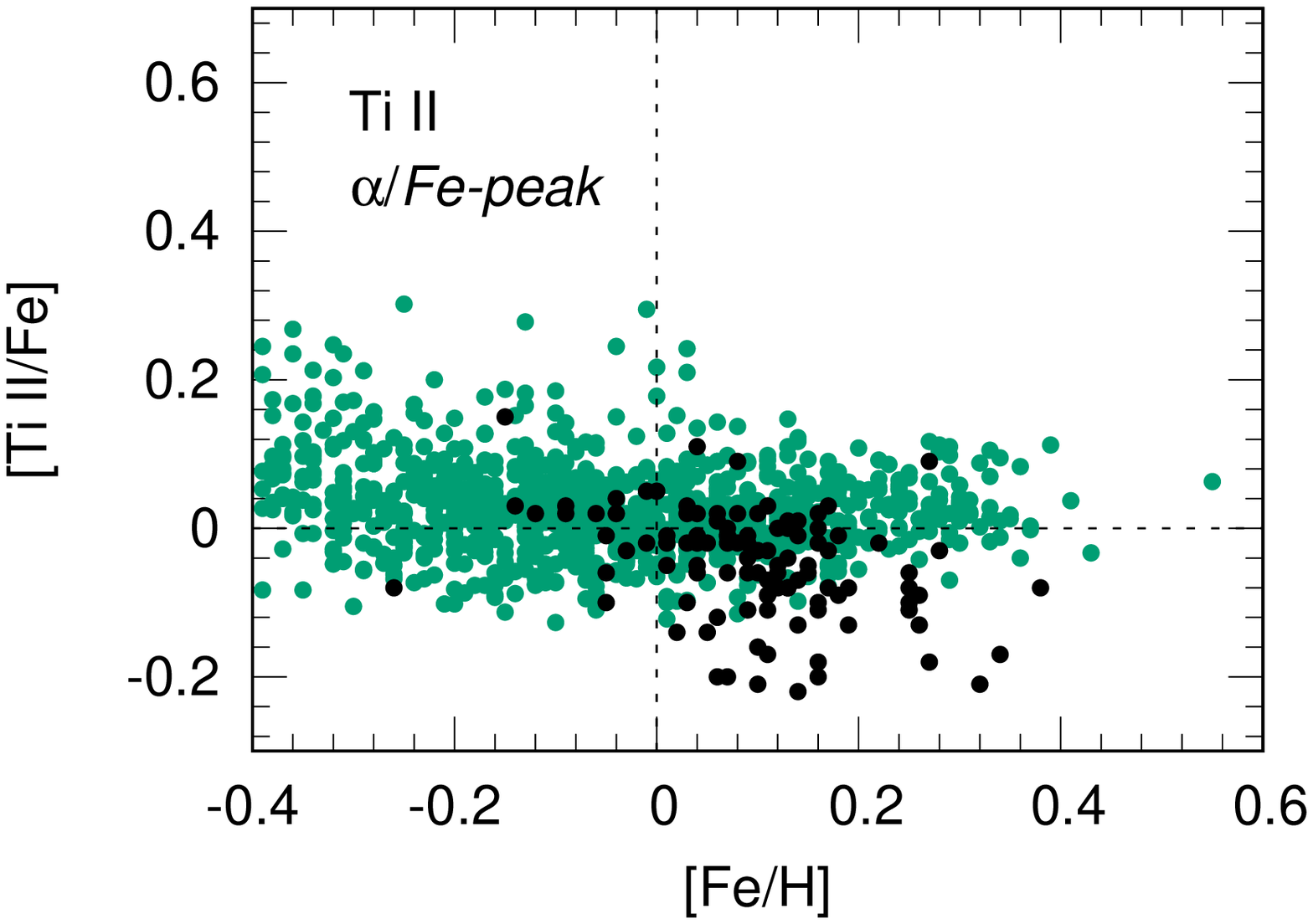} 
\includegraphics[width=55mm]{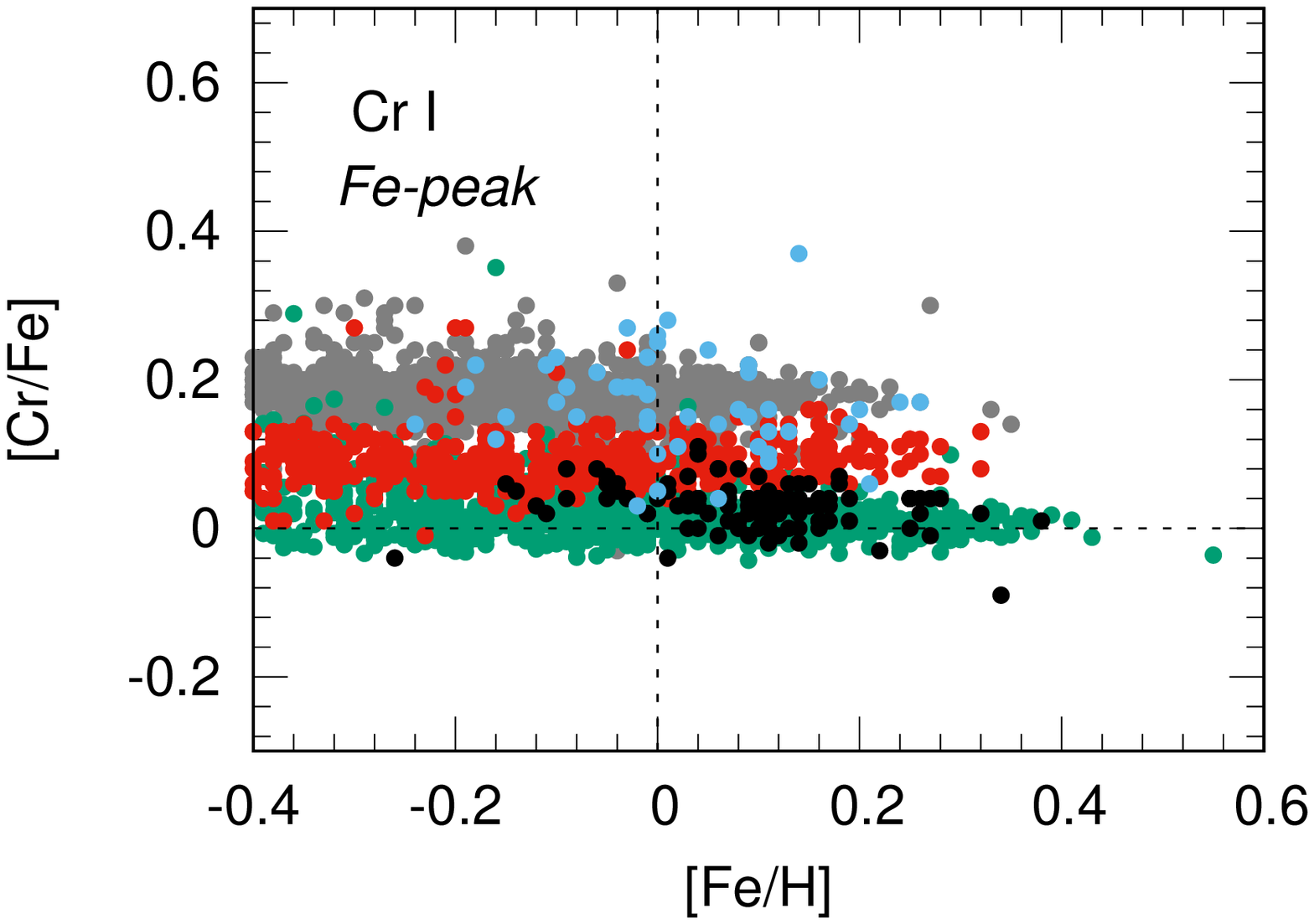} 
\includegraphics[width=55mm]{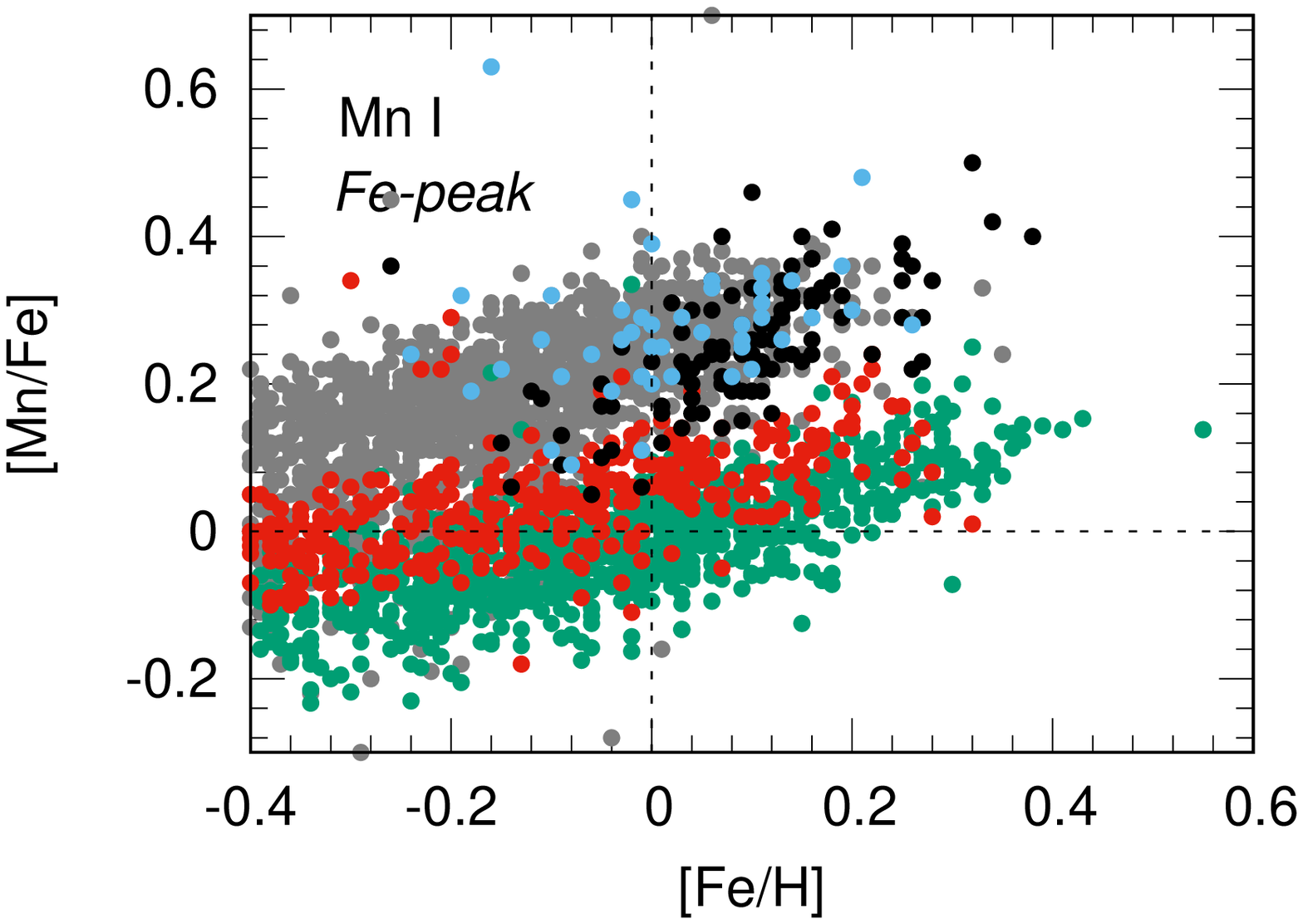} 
\includegraphics[width=55mm]{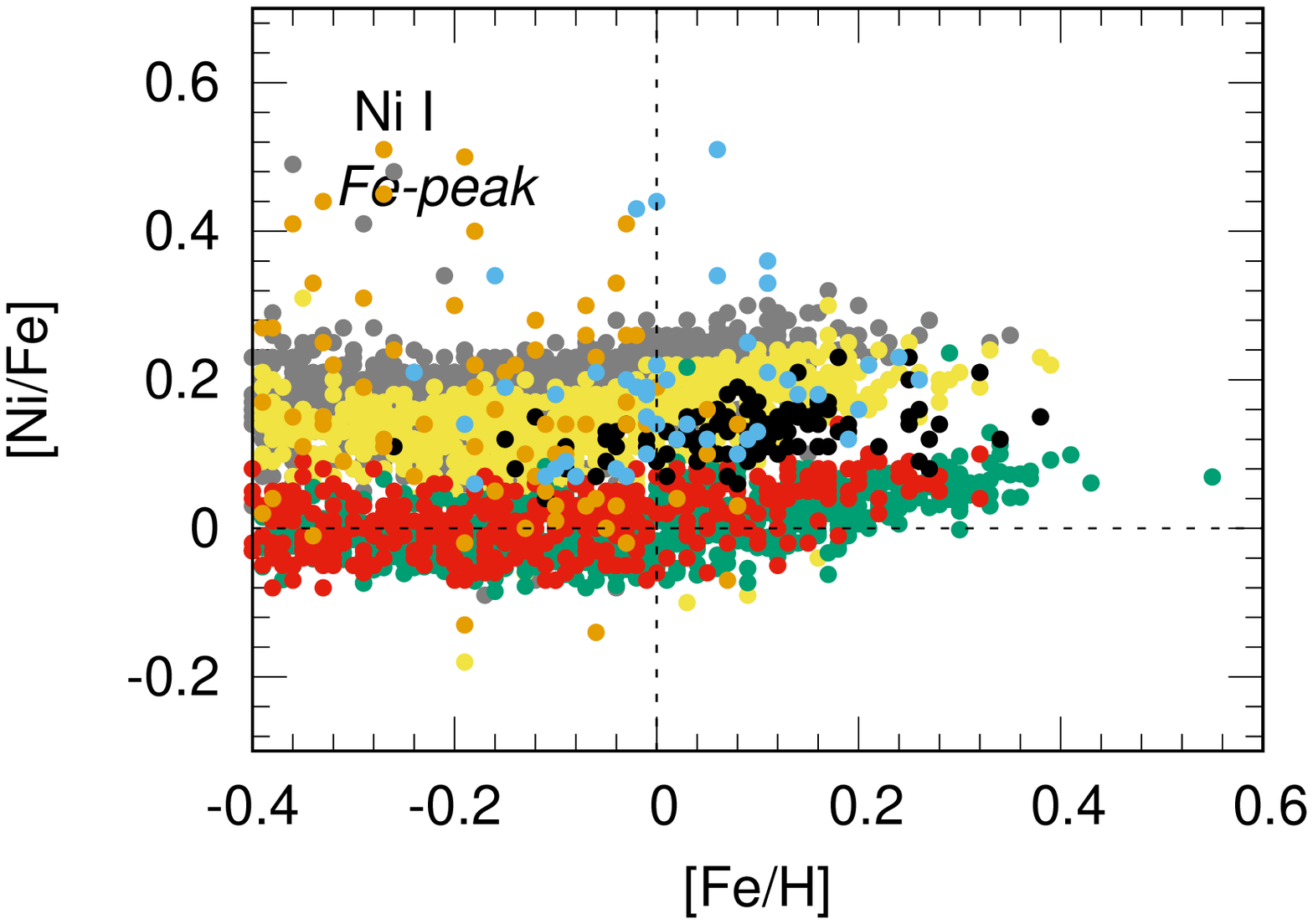} 
\includegraphics[width=55mm]{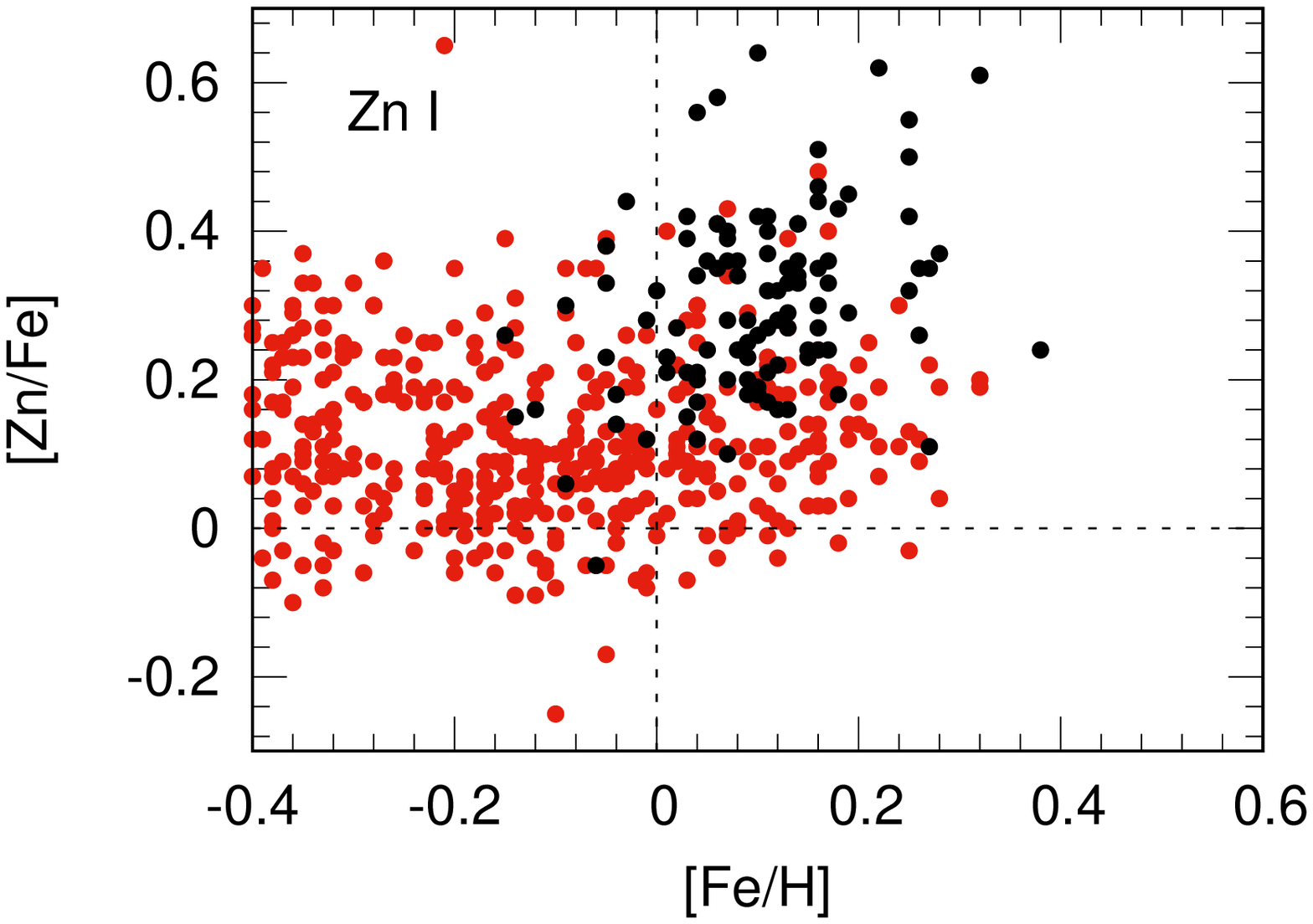} 
\caption{The comparison of [X/Fe] versus [Fe/H] in the samples by \protect\cite{edva93} -- shown with orange, \protect\cite{felt98} -- with blue, \protect\cite{vale05} -- with yellow, \protect\cite{adib12} -- with green, \protect\cite{bens14} \protect\citep[][in case of Mn]{batt15} -- with red, \protect\cite{brew16} -- with grey, and this work -- with the black circles. The reference solar abundances for each work are converted to the scale of \protect\cite{ande89}.}
\label{_figure_samples}
\end{figure*}

In order to make a useful comparison with other works we needed to translate onto a common scale. We found the easiest way to achieve this was to adopt the scale of \cite{ande89} rather than in terms of the solar abundances determined by our procedure. In some papers, abundances are provided in relation to derived solar scales in order to compensate for the differences in procedures \citep[e.g.][]{bens14,brew16}. Other authors adopt \cite{ande89} or its later derivatives \citep[e.g.][]{vale05,adib12}, though they did not consider their derived solar abundances. In order to compare the different samples studied by different authors we converted abundances for the various comparison samples to the scale of \cite{ande89} (Fig. \ref{_figure_samples}). This also allowed us to assess the general accuracy level of stellar abundance determinations from the combined spread of abundances.

In Fig. \ref{_figure_overiron} we show the dependence of abundances relative to iron with their corresponding error bars. These uncertainties depend on the number of lines for each element, quality of fit, the scatter of abundances determined from the fits to different lines, the local continuum level, and the atomic line data taken from \textsc{vald}. The uncertainties for [Fe/H] are not plotted to make the plots easier to read. The average [Fe/H] uncertainty is $\pm$0.02\,dex.

Below we discuss the specific results for the selected elements:

{\it Fe, Z = 26}. Differences up to 0.1\,dex between \pion{Fe}{i} and \pion{Fe}{ii} abundances are due to the iterative nature of our computations. Previously determined abundances can change after adoption of the refined model atmosphere. In other words, it reflects the limit of accuracy implied by the model atmosphere, atomic line data, and line list.

{\it Na, Z = 11}. The distribution of the Na abundance versus iron is shown in Fig. \ref{_figure_samples}. The computed $X_f$ for the metal rich stars in our sample and those of \cite{adib12} and \cite{bens14} show a well defined positive slope (Table \ref{_table_abundance_slopes}). The abundance distribution of Na is shifted toward larger abundances by 0.1--0.2\,dex compared to Fe. The general trend is similar to those obtained in all comparison works with higher abundances toward higher metallicities (Fig. \ref{_figure_samples}).

{\it Mg, Z = 12}. Only up to 7 lines of Mg I were used in the analysis. We found similar result to that of sodium, with a notable over-abundance for our sample, and a mean abundance of [Mg/Fe]=0.13$\pm$0.05\,dex. The shift is larger than the formal accuracy of our abundance determination procedure, but may depend on the adopted continuum level and line list. We note that $X_f>0$ for the stars in our sample. However, it can be seen in Fig. \ref{_figure_samples} that this could be due to border effects, as this trend is rather marginal. \cite{bens14} show the same order of over-abundance, but also a certain downward trend with metallicity, and for the sample of \cite{adib12} we see that their Mg distribution is in agreement with [Fe/H]. At the same time, \cite{brew16} show the results similar to our.

{\it Al, Z = 13}. In our work aluminium show a definite positive slope. To some degree our findings similar to those of \cite{adib12} and \cite{brew16}. But different to \cite{edva93,felt98} and \cite{bens14} who show no definite dependence on metallicity, and higher mean [Al/Fe] for their samples.

{\it Si, Z = 14}. The results for silicon agree well with most of the studies in comparison, showing the average [Si/Fe] over-abundance to be on the order of 0.1\,dex and no noticeable trend with metallicity. On the other hand, \cite{adib12} show no excess of silicon. It is interesting to note that for the same lines in the spectra, our results for \textsc{vald-2} show 0.1\,dex lower abundances than for \textsc{vald-3}, highlighting the importance of reliable atomic line data.

{\it Ca, Z = 20}. Calcium is another element for which we see a high level of agreement between the different authors \citep{edva93,felt98,adib12,bens14,brew16}. In all these works there is a clear negative trend with metallicity (Table \ref{_table_abundance_slopes}). On the whole, [Ca/Fe]$>$0, but in the super metal rich domain, we note calcium deficiency. However, various mean abundances in different works (see Fig. \ref{_figure_samples}) reflect that this deficiency is within the accuracy limit of the modern methods of abundance determination.

{\it Ti, Z = 22}. In our sample \pion{Ti}{i} follows the iron abundance and shows no dependence with metallicity. On the other hand, all comparison works show a larger spread of stars toward higher [Ti/Fe] abundances and a weak negative trend. At the same time, \cite{adib12} show no metallicity dependence for \pion{Ti}{ii}. For our stars, the \pion{Ti}{ii} lines show a large scatter, which makes it difficult to come to any firm conclusions in regards to the metallicity trends. Similar to \pion{Si}{i} and \pion{Fe}{ii}, the lines of ionized titanium give rise to a 0.1\,dex higher abundance when the line parameters are taken from \textsc{vald-3}.

{\it Cr, Z = 24}. Chromium is one of the few elements with reliable results that show no significant scatter between the different lines and stars in our sample. We can observe a very weak negative trend with iron. The same results are shown by \cite{adib12}. \cite{bens14} and \cite{brew16} show higher abundances for their stars and a weak positive trend, which could be an indication of the sample bias effect. Once again, the same chromium lines give different abundances using the different versions of \textsc{vald}, but unlike \pion{Si}{i}, \pion{Ti}{ii}, and \pion{Fe}{ii}, we see a 0.05\,dex lower mean abundance for the \textsc{vald-3} data.

{\it Mn, Z = 25}. Manganese abundances show a clear positive gradient with metallicity, much like Zn, though with a larger scatter despite the relatively large number of lines used in the analysis. Manganese exhibits a higher abundance than iron, on average, by 0.3\,dex. A positive trend was also found in all comparison works. As for most of the other elements, the abundance measured by \cite{bens14} is slightly higher than that of \cite{adib12}. And \cite{felt98,brew16} show the results similar to our (Fig. \ref{_figure_samples}).

{\it Ni, Z = 28}. [Ni/Fe] also shows an upward trend with iron of the same order in all the samples in comparison. All the authors show different average over-abundances for Ni: up to 0.05\,dex for \cite{adib12} and \cite{bens14}, 0.1\,dex for \cite{felt98} and our work, and 0.15\,dex for \cite{vale05} and \cite{brew16}.

{\it Cu, Z = 29}. The distribution of Cu with [Fe/H] exhibits a large scatter (Fig. \ref{_figure_overiron}), mainly affected by the quality of the line list and number of lines. Formally speaking, within the range of metallicities [Fe/H]=-0.1\,--\,+0.2, there are two groups of stars, with mean [Cu/Fe] abundances of $\sim$+0.1 and -0.05\,dex, respectively, however with uncertainties at the level of $\pm$0.4\,dex. The lower average abundances and defined positive trend were found by \cite{silv15}. The same trend could be observed for the $\sim$0.0\,dex group of stars in our sample.

{\it Zn, Z = 30}. For Zn we found the same significant differences in the abundance distribution with iron that were found for manganese. Unfortunately, we used only one line of Zn in our analysis, so the visible slope of [Zn/Fe] versus [Fe/H] should be considered as a preliminary result at this point. More work should be done to confirm this result using a wider set of Zn atomic lines. A large scatter and similar abundance are also shown by \cite{bens14} and \cite{niss15}.

Our analysis shows that the different atomic line data can have a strong impact on the average abundances measured for similar samples of stars (e.g. see silicon in this work). On the other hand, observed trends change rather marginally. Comparisons to our work and those in the literature highlights the similar tendencies for the Na, Si, Ca, Cr, Mn, and Ni; for the elements with a small spread like Si, Ca, Cr, Ni we see various average abundances between the different samples, and similar trends with iron in all comparison works. And for the Mg, Al, and Ti we see weak trends with iron in all comparison works due to large spread of abundances.

\subsection{The dependencies on \Teff\, and \logg}
\label{_dependencies}

In the framework of our work, we investigated the dependence of our abundance results on the effective temperature and surface gravity. The results of these tests are shown in Figs. \ref{_figure_temperature_dependence} and \ref{_figure_gravity_dependence} in Appendix. We see the evidence for a weak trend of abundances versus \Teff\, in Fig. \ref{_figure_temperature_dependence}. These rather indistinct trends may be explained by uncertainties in the choice of the photometric \Teff\, of the metal rich stars of earlier spectral classes, as well as limitations of the current model atmospheres and atomic line data. In each case we obtained the trends of abundances versus \Teff\, for the elements with rather low ionization potentials, i.e. Mg, Al and Si, with the Pearson product-moment correlation $\rho$(\Teff)$\sim$0.4--0.5 (Table \ref{_table_abundance_slopes}). The same we also observe for Mn and Zn. However, because our data doesn't show a sufficient level of homoscedasticity, it is hard to make definite conclusions in these regards.

We note the non-linear behaviour of the [X/H] versus \Teff\, dependence for practically all our elements in the stars with \Teff\, close to 6000\,K. This apparent 'phantom' gap is poorly constrained due to low number of stars in this temperature range.

What is more important, we do not see any evident trends of the abundances versus the adopted surface gravities (Table \ref{_table_abundance_slopes}). In some sense, this provides us with the evidence that our adopted procedure performs in the way we had previously envisaged it would.

However, sensitivity of the spectroscopic \Teff\, and \logg\, to the quality of fits and line data (these are discussed in Section \ref{_comparison}) emphasizes the importance of at least one independent variable.

\section{Comparison to other authors}
\label{_comparison}

There have been a number of extensive works to determine abundances of main sequence dwarfs. They follow similar approaches to the analysis of the observed high-resolution stellar spectra: LTE, 1D model atmospheres. We tabulate these in Table \ref{_table_abundance_slopes_procedures}.

To compare our results to other authors we note that our sample primarily consists of metal rich stars; the samples of some other authors consist of a larger total number of targets, and they maintain a broader metallicity range. Most of our objects are single stars, where our spectroscopic data have been observed solely using HARPS, using a photometric pre-selection, and thus our sample is homogeneous.

The most of comparison works were done using the equivalent width analysis, we used the synthetic profile fitting. In comparison to \cite{vale05} and \cite{brew16}, who utilized synthetic spectra fitting for the broad spectral ranges and fixed microturbulent velocity, we adopted line-by-line fitting and microturbulence as a free parameter in addition to the other differences.

As noted in subsection \ref{_abundances}, some authors used their own solar abundances for the [X/Fe] distribution analysis, whereas we use the \cite{ande89} abundances in our study. In our work, we separate the relative and absolute solar scales. The first case is defined by the reference solar abundances to which the absolute values are translated. And the second is dependent on the adopted line lists and gf-values. The latter we identify as the differences in the atomic line data or line lists, but not in the adopted solar scales.

\begin{table*}
\centering
\caption{The procedures used for the abundance analysis in some recent works. Note, that the papers by \protect\cite{neve09,sous11,adib12} belong to the same series of works. Here, R labels the resolution of the observed spectra, EW -- equivalent width measurements, SS -- synthetic spectra fitting, and \xi stands for the abundances.}
\label{_table_abundance_slopes_procedures}
\begin{tabular}{cccc}
\hline\hline
      Paper      &          Targets         &               Procedure              &           Output          \\\hline
  \cite{vale05}  &        1040 HIRES,       &   SS, \textsc{atlas9},\textsc{sme}   &   \xi,\Teff,\logg,\Vsini  \\
                 &  UCLES, Hamilton, R 70K  &                                      &                           \\
  \cite{jenk08}  &   353  FEROS, R  46K     &   SS, \textsc{sam12},\textsc{wita6}  &   \xi,\Vsini              \\
  \cite{neve09}  &   451  HARPS, R 120K     &   EW, \textsc{moog},\textsc{ares}    &   \xi,\Teff,\logg,\Vm     \\
  \cite{sous11}  &   582  HARPS, R 120K     &   EW, \textsc{moog},\textsc{ares}    &   \xi,\Teff,\logg,\Vm     \\
  \cite{adib12}  &  1111  HARPS, R 120K     &   EW, \textsc{moog},\textsc{ares}    &   \xi,\Teff,\logg,\Vm     \\
  \cite{bens14}  &    60  FEROS, R  48K     &   EW, \textsc{marcs},\textsc{iraf}   &   \xi,\Teff,\logg,\Vm     \\
                 &     5  HARPS, R 120K     &                                      &                           \\
                 &    27   UVES, R 110K     &                                      &                           \\
                 &    31   UVES, R  80K     &                                      &                           \\
                 &    52  SOFIN, R  80K     &                                      &                           \\
                 &     6   FIES, R  67K     &                                      &                           \\
                 &   374   MIKE, R  65K     &                                      &                           \\
                 &    49   MIKE, R  42K     &                                      &                           \\
                 &    79   MIKE, R  55K     &                                      &                           \\
  \cite{silv15}  &   309 ELODIE, R  42K     &   EW, \textsc{moog},\textsc{ares}    &   \xi,\Teff,\logg,\Vm     \\
  \cite{brew16}  &  1626  HIRES, R  70K     &   SS, \textsc{atlas9},\textsc{sme}   &   \xi,\Teff,\logg,\Vsini  \\
    This work    &   107  HARPS, R 120K     &   SS, \textsc{sam12},\textsc{abel8}  &   \xi,\logg,\Vsini,\Vm    \\\hline
\end{tabular}                                                     
\end{table*}

\subsection{Microturbulent velocities}

\begin{figure*}
\centering
\includegraphics[width=55mm]{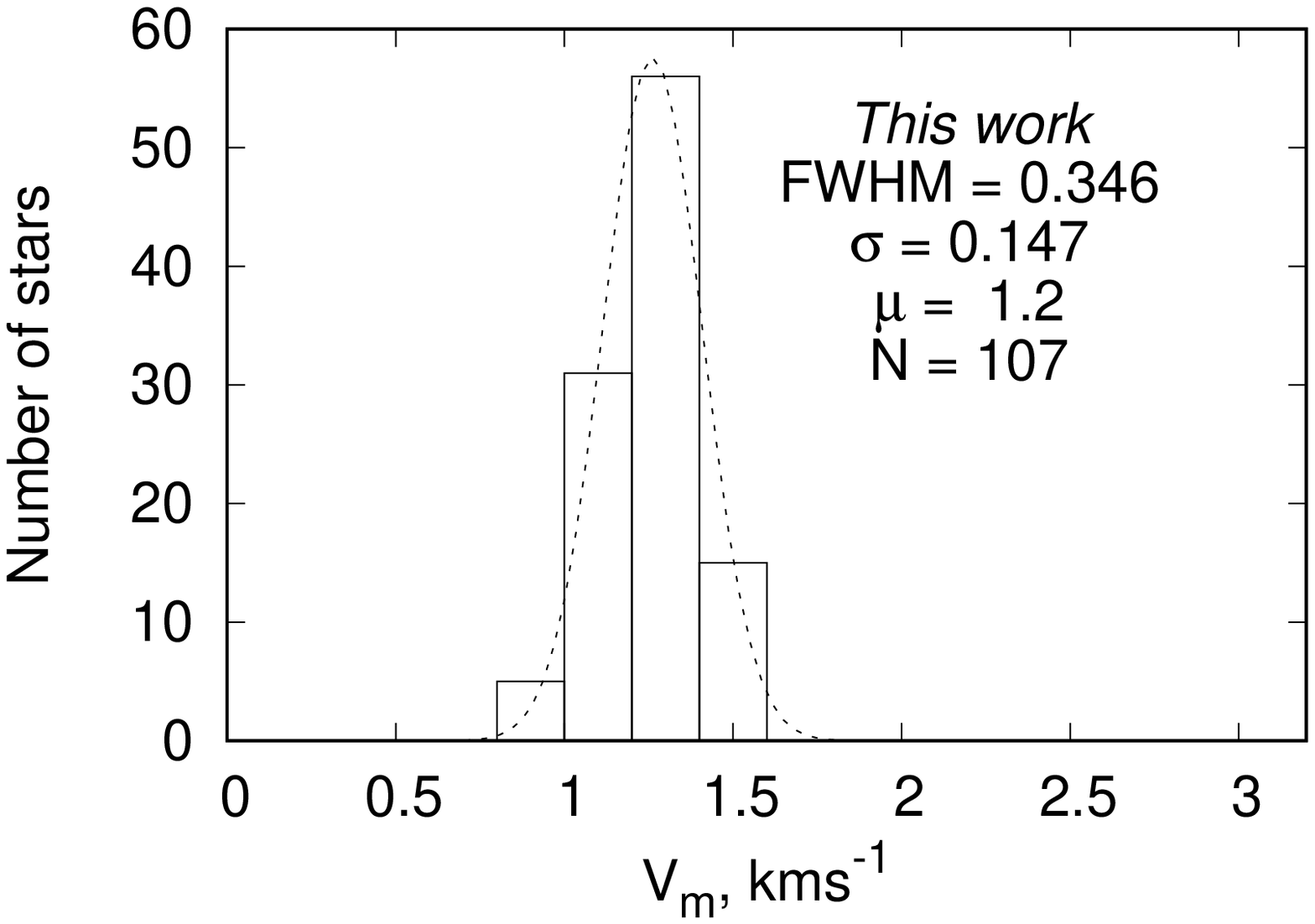}    
\includegraphics[width=55mm]{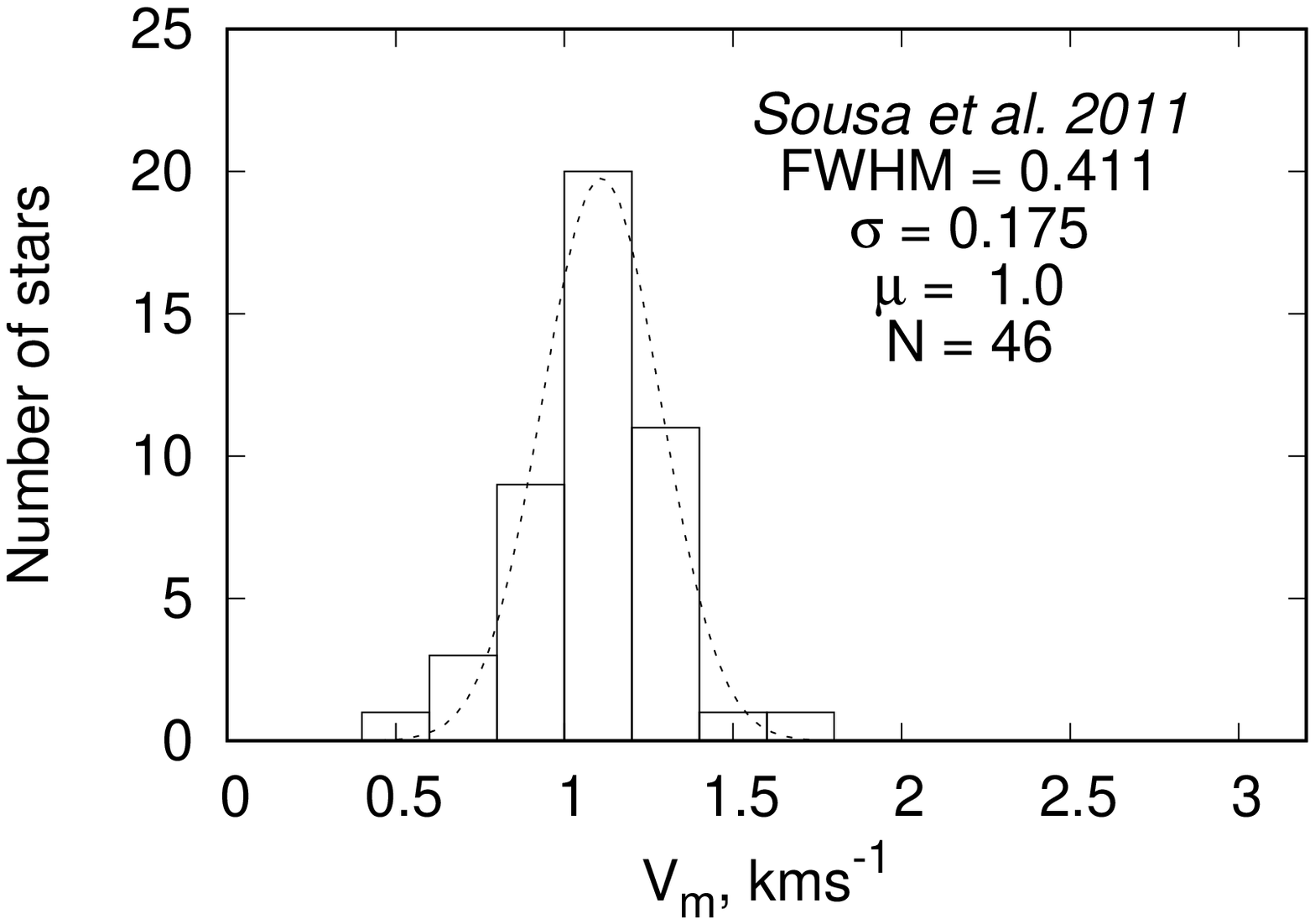} \\ 
\includegraphics[width=55mm]{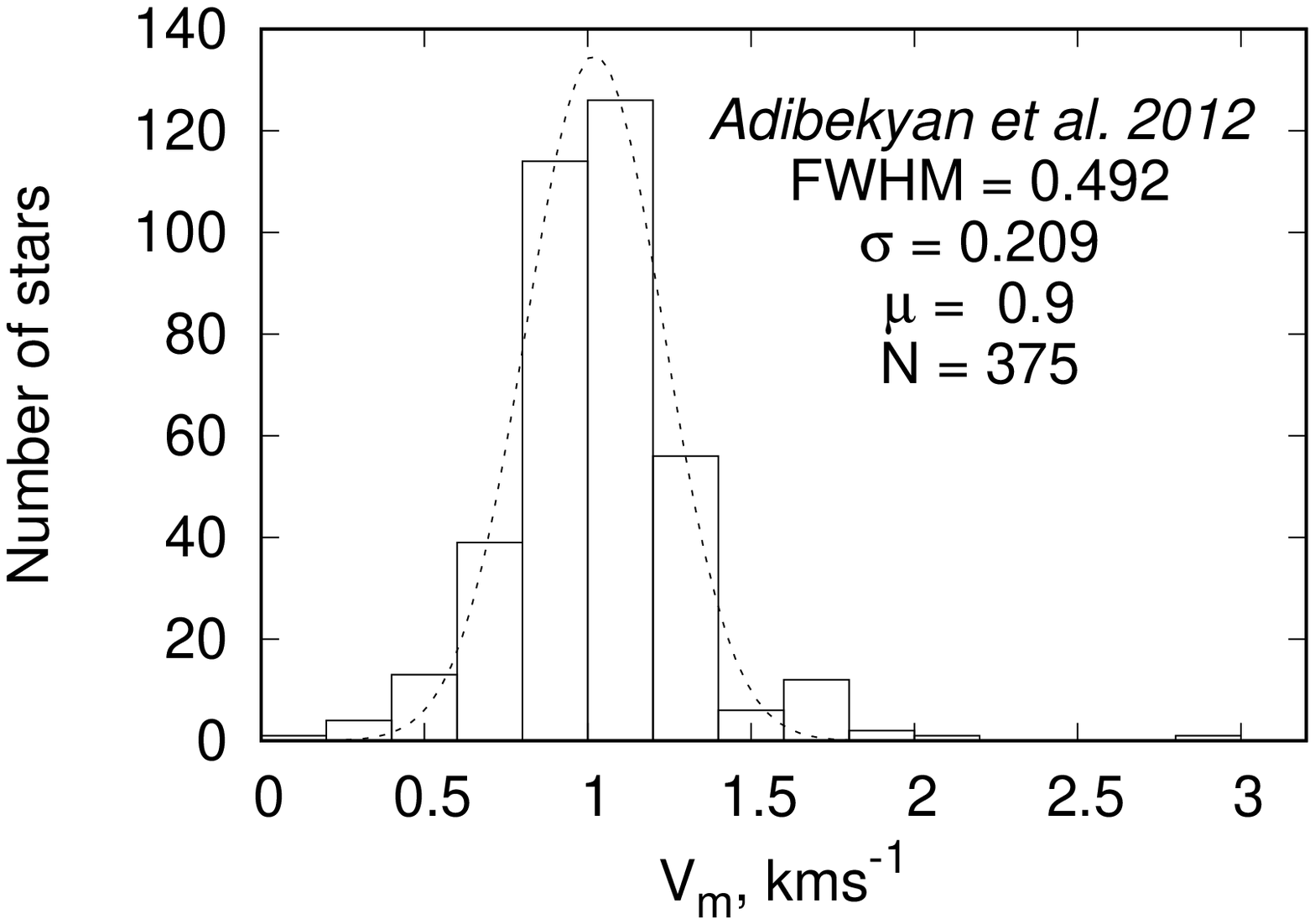}    
\includegraphics[width=55mm]{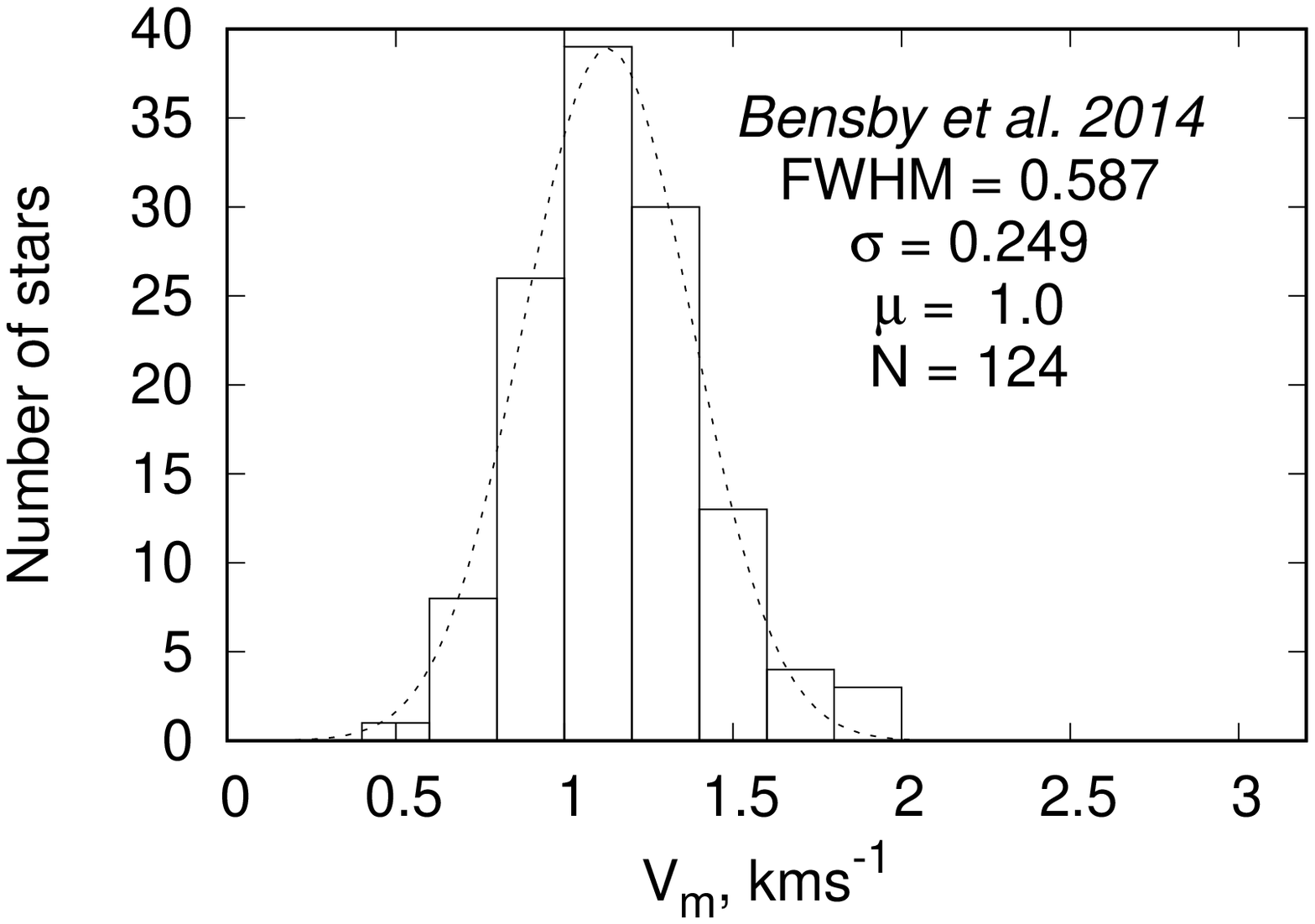}    
\caption{Distributions of microturbulent velocities for the metal rich domain ([Fe/H]$>$0.0) in this work and from the results by \protect\cite{sous11,adib12,bens14}. Parameters $\mu$ and $\sigma$ are the mean of the distribution and standard deviation, FHWM is the full width at half maximum, N is the number of stars with [Fe/H]$>$0.0\,dex in the comparison samples.}
\label{_figure_microturbulence_comparison}
\end{figure*}

The microturbulent velocity is an important parameter that must be understood with as much precision as possible to properly determine the chemical abundances. We compare our distribution of \Vm\, to those by the other authors across the metal rich domain ([Fe/H]$>$7.67\,dex) of their samples (Fig. \ref{_figure_microturbulence_comparison}). \Vm\, in the atmospheres of our stars lies in the range of 1.0--1.4\,\kmps, with the well defined peak at 1.2\,\kmps. The peaks of the distributions in comparison are at 0.8--1.0\,\kmps.

In general, all distributions are of a similar shape. At the same time, our distribution of \Vm, determined from the fits to the observed line profiles, is narrower ($\sigma$=0.147) than the similar distributions ($\sigma>$0.2) obtained from the equivalent width analyses provided by the different authors independently on the number of stars used.

\begin{figure*}
\centering
\includegraphics[width=55mm]{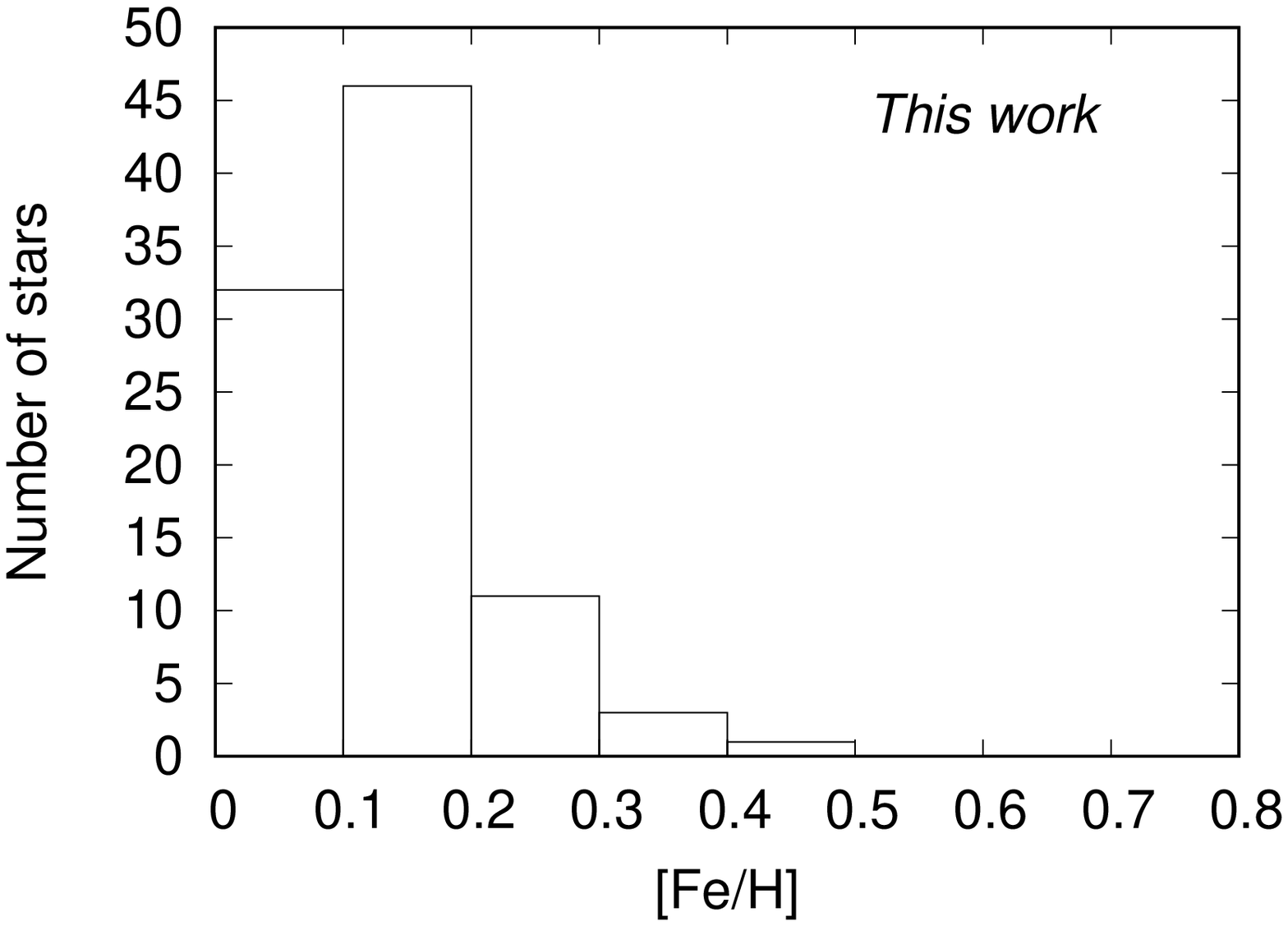}    
\includegraphics[width=55mm]{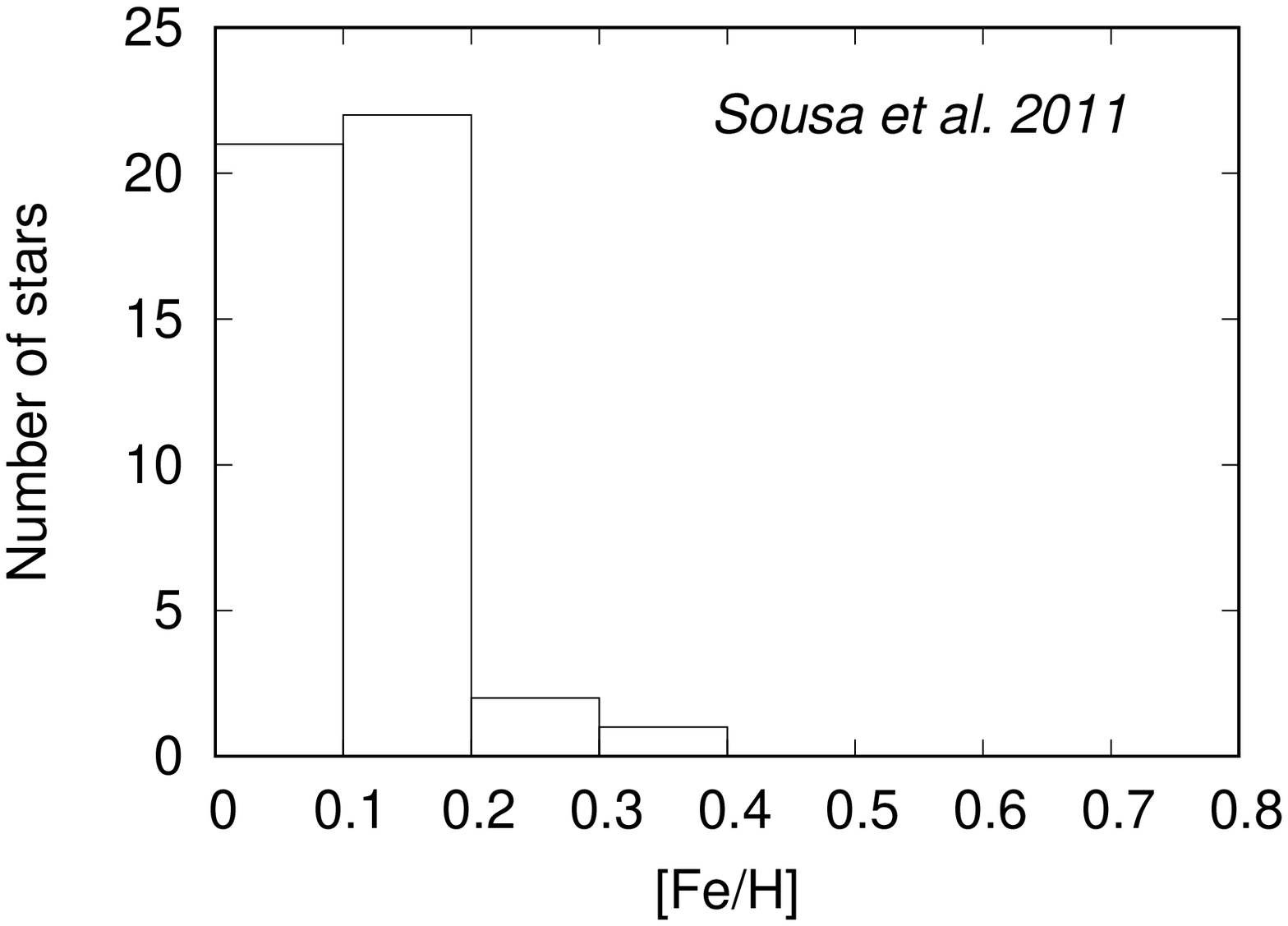} \\ 
\includegraphics[width=55mm]{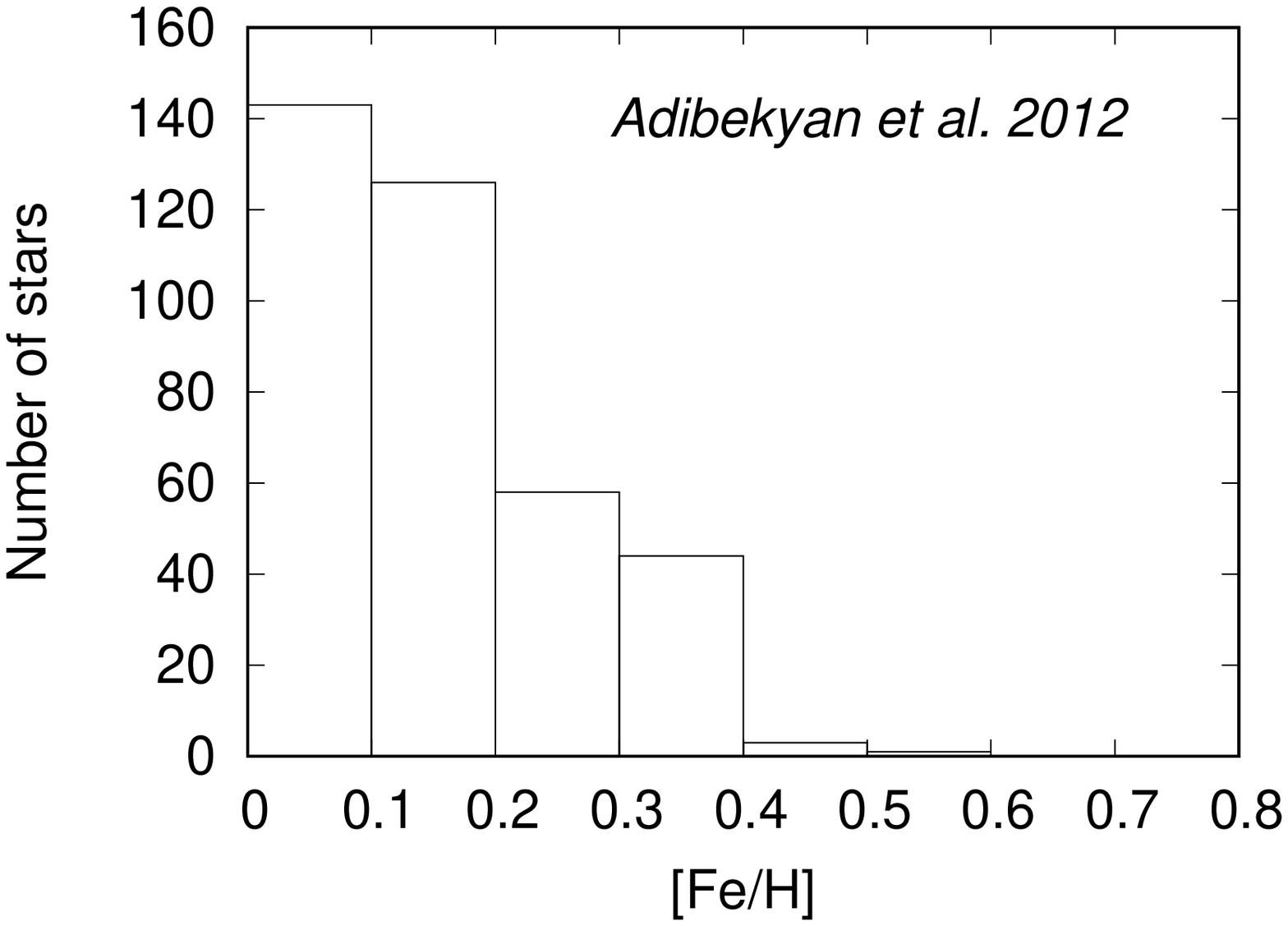}    
\includegraphics[width=55mm]{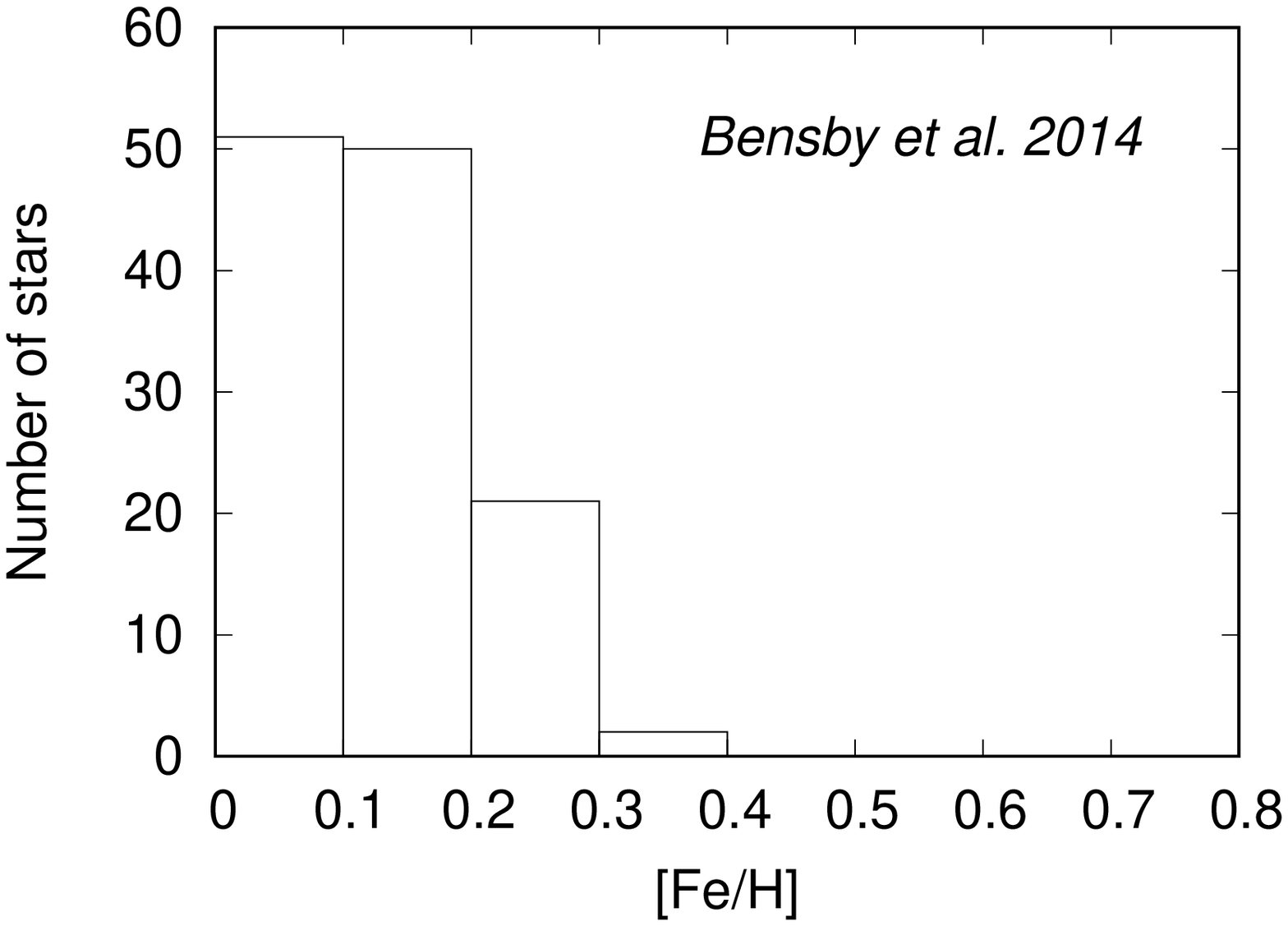}    
\caption{Distributions of metallicities in this work and from the results by \protect\cite{sous11,adib12,bens14}.}
\label{_figure_metallicity_distributions}
\end{figure*}

\subsection{The common stars}

\begin{figure*}
\centering
\includegraphics[width=55mm]{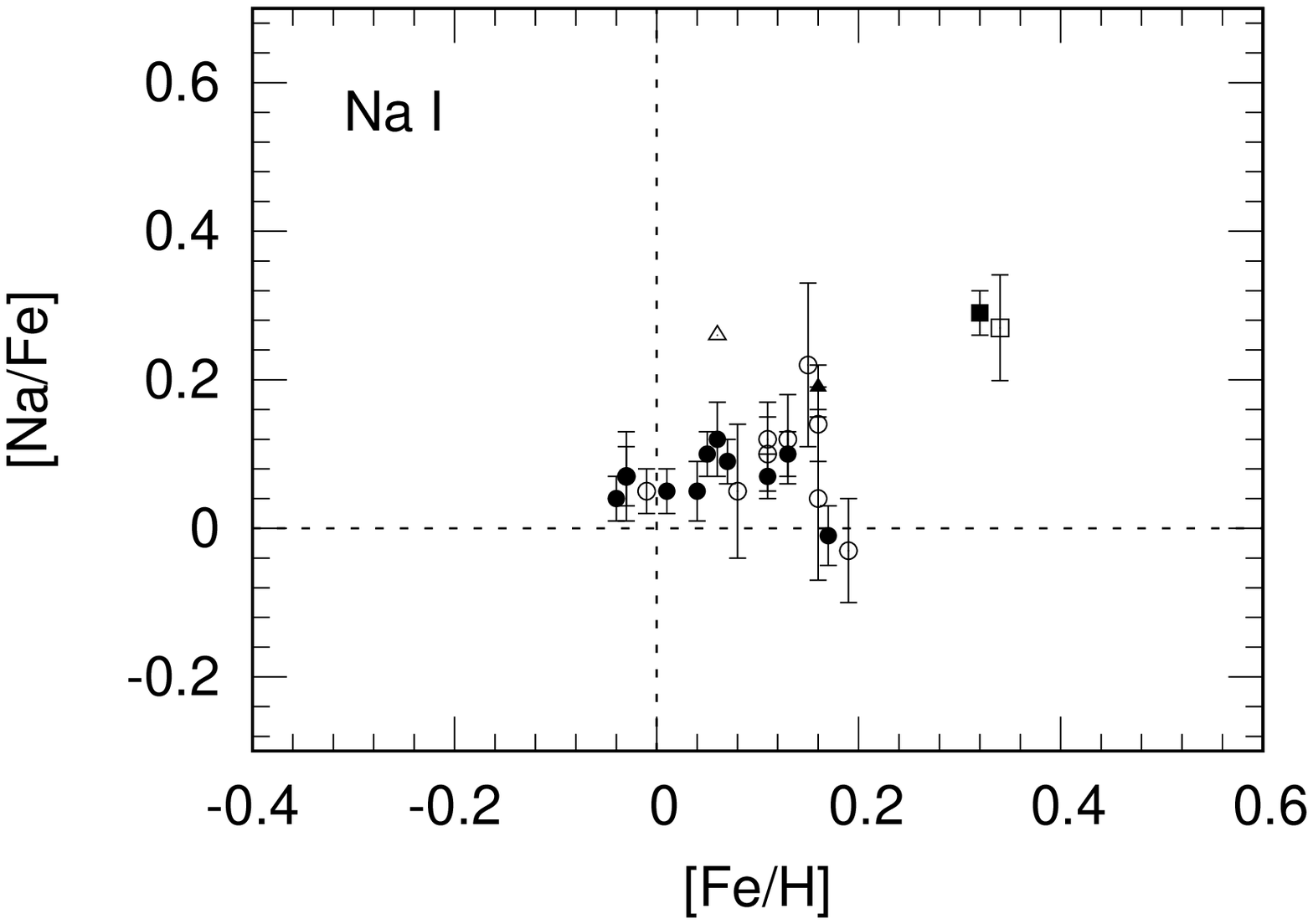} 
\includegraphics[width=55mm]{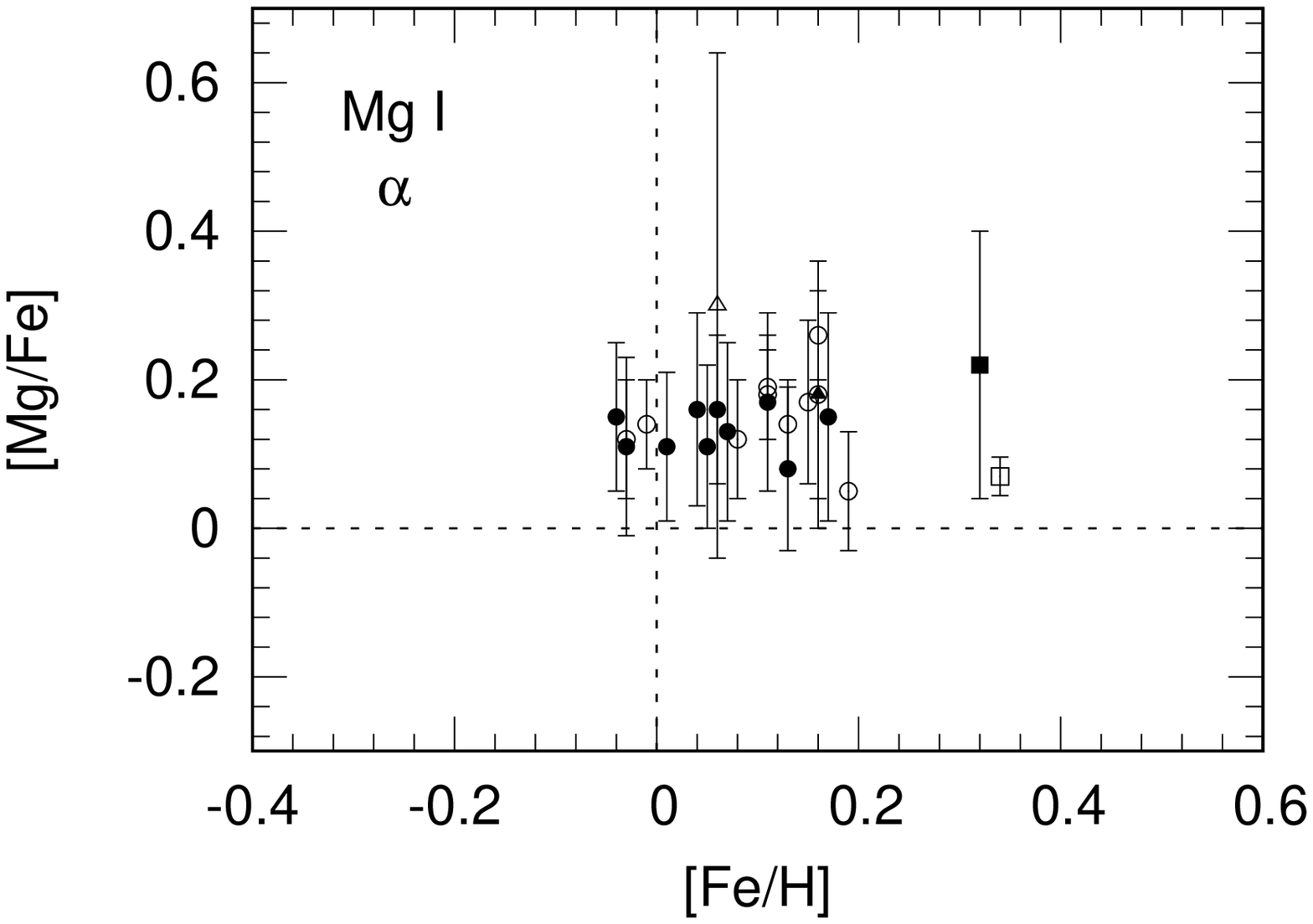} 
\includegraphics[width=55mm]{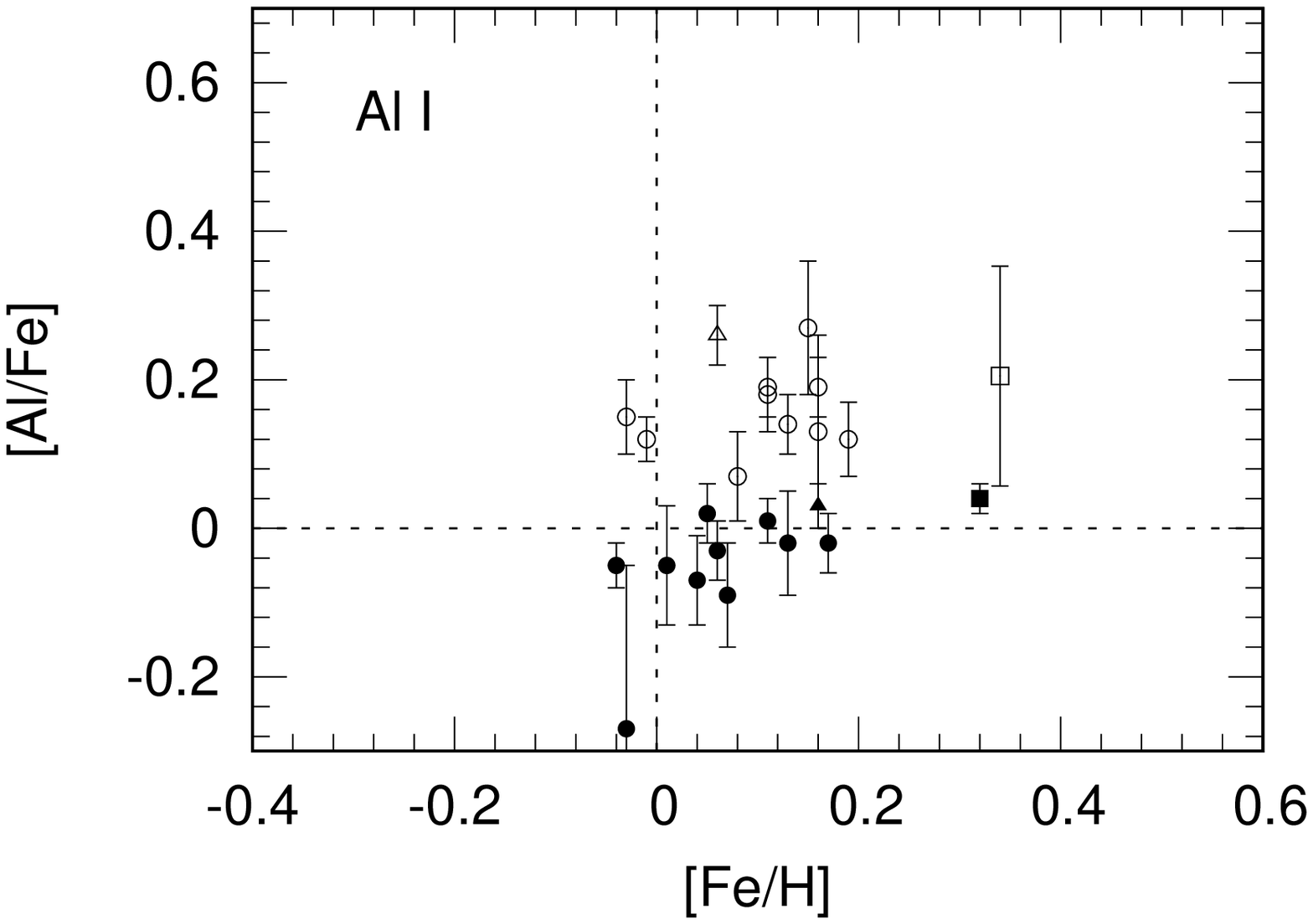} 
\includegraphics[width=55mm]{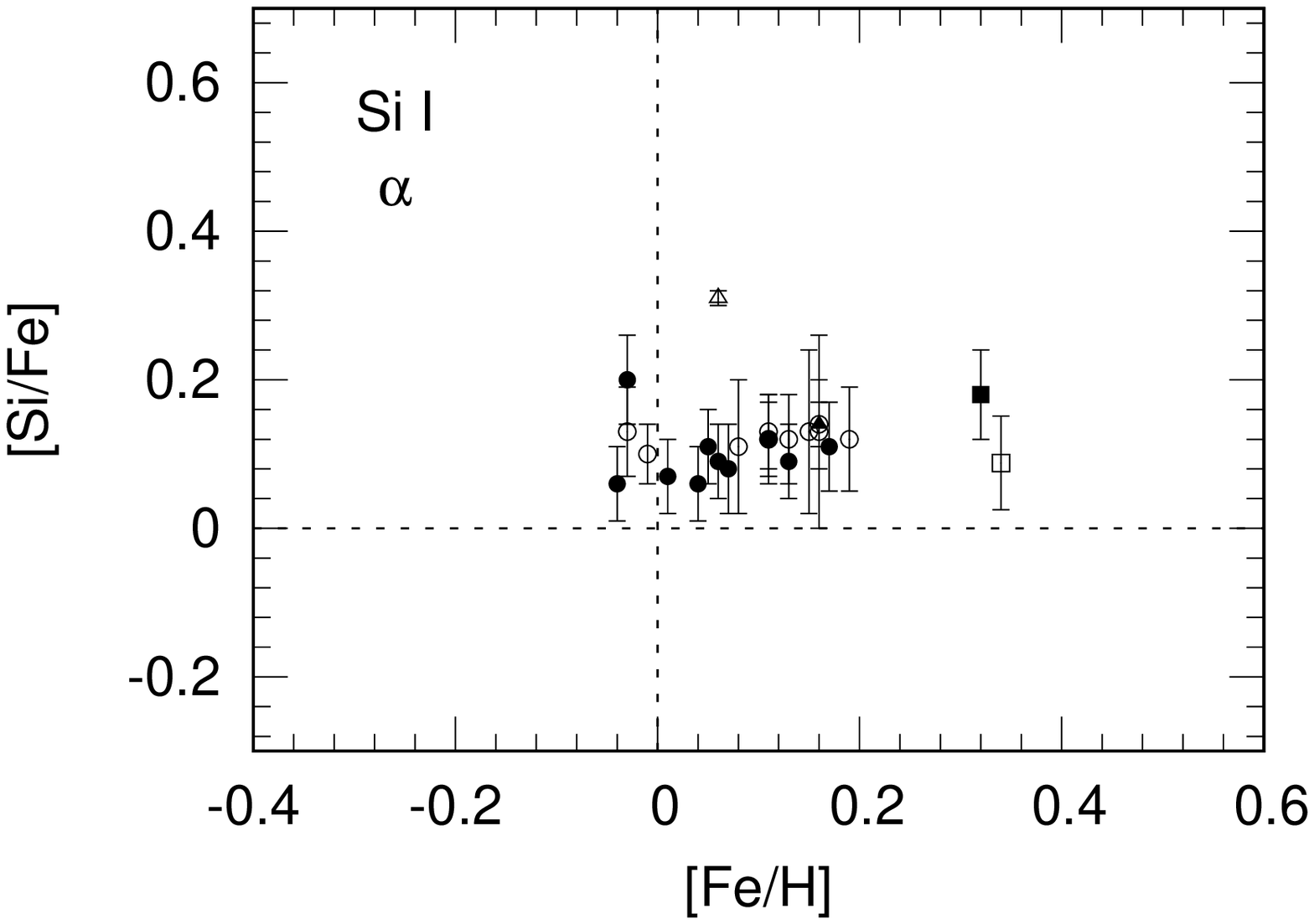} 
\includegraphics[width=55mm]{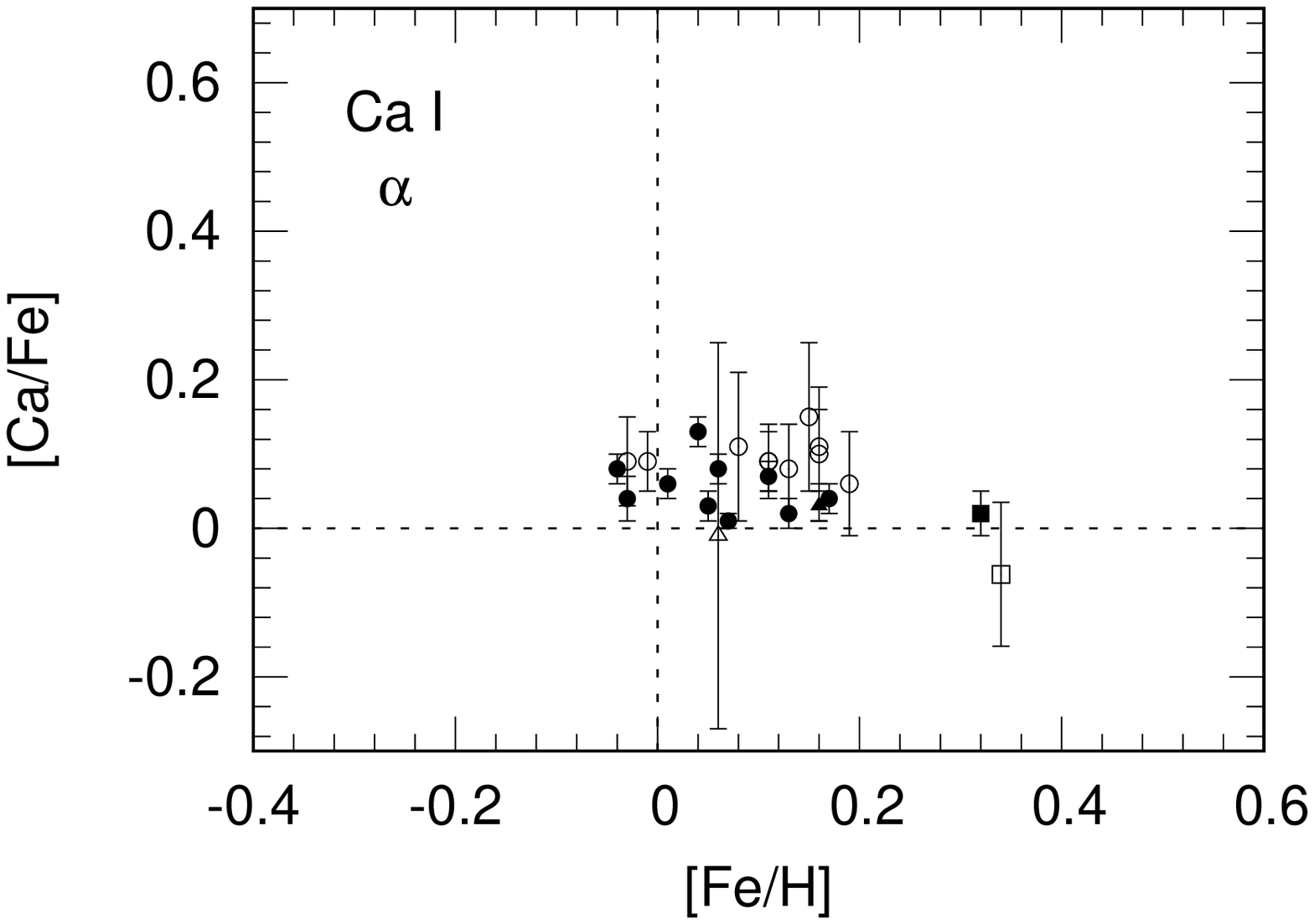} 
\includegraphics[width=55mm]{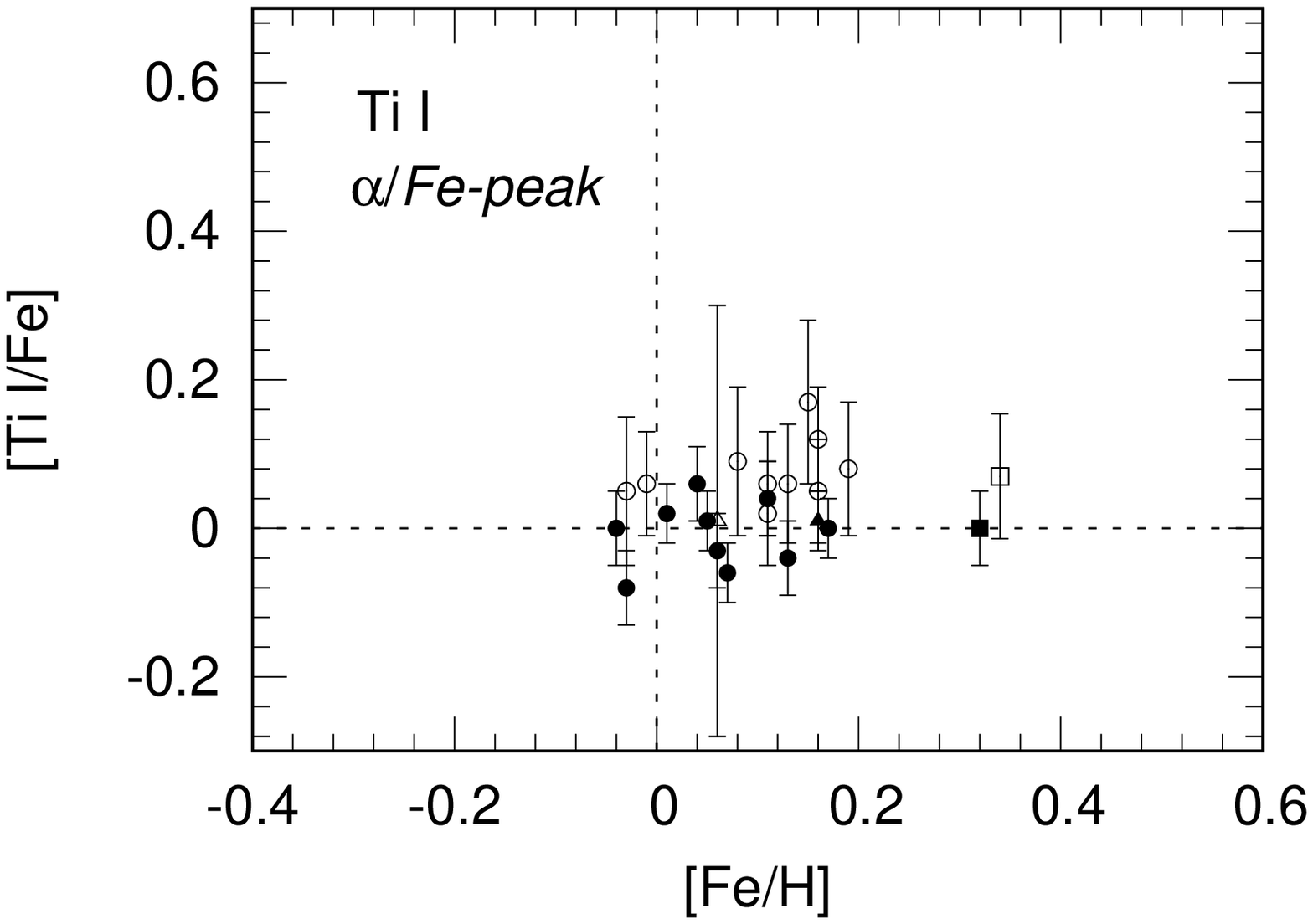} 
\includegraphics[width=55mm]{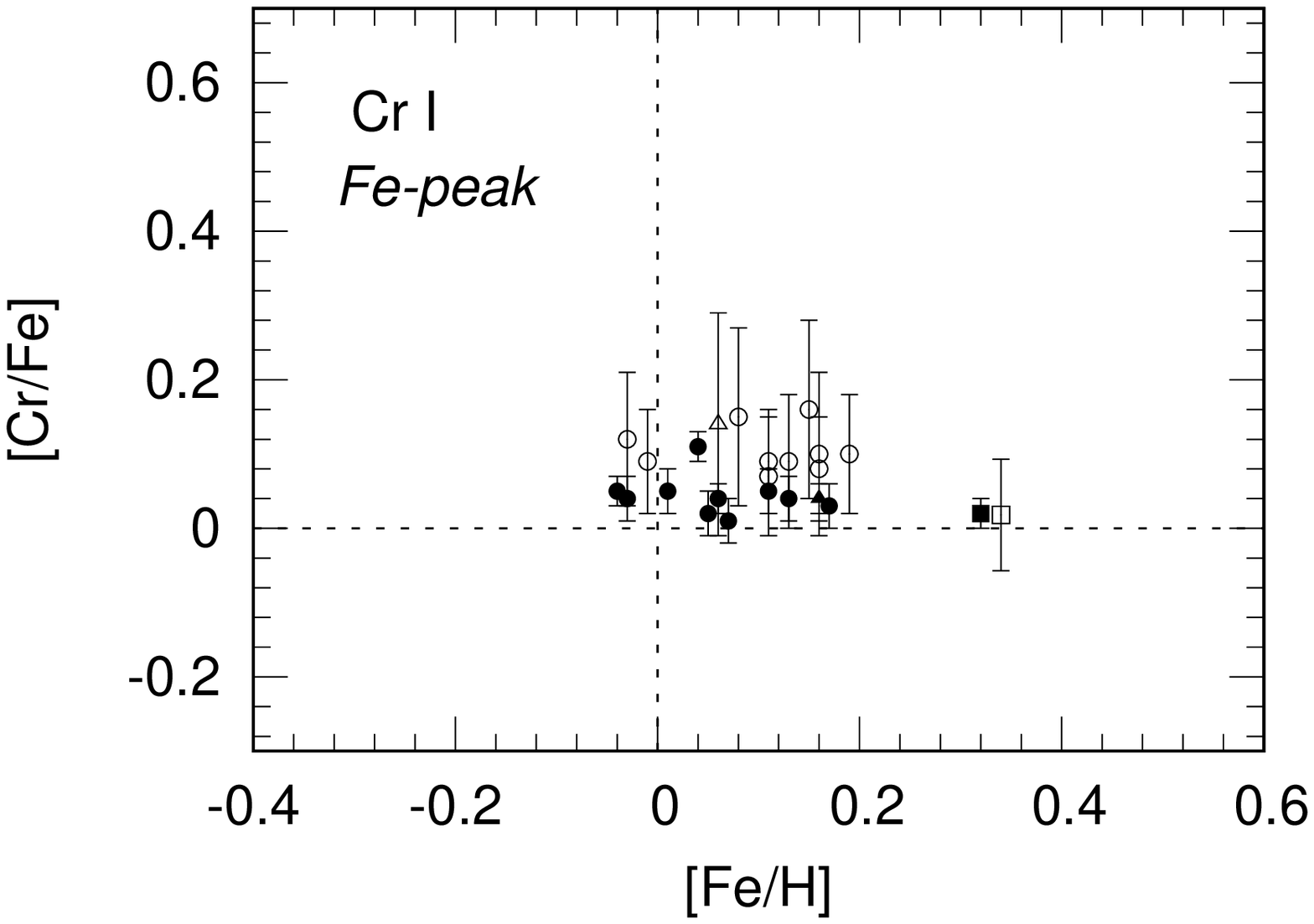} 
\includegraphics[width=55mm]{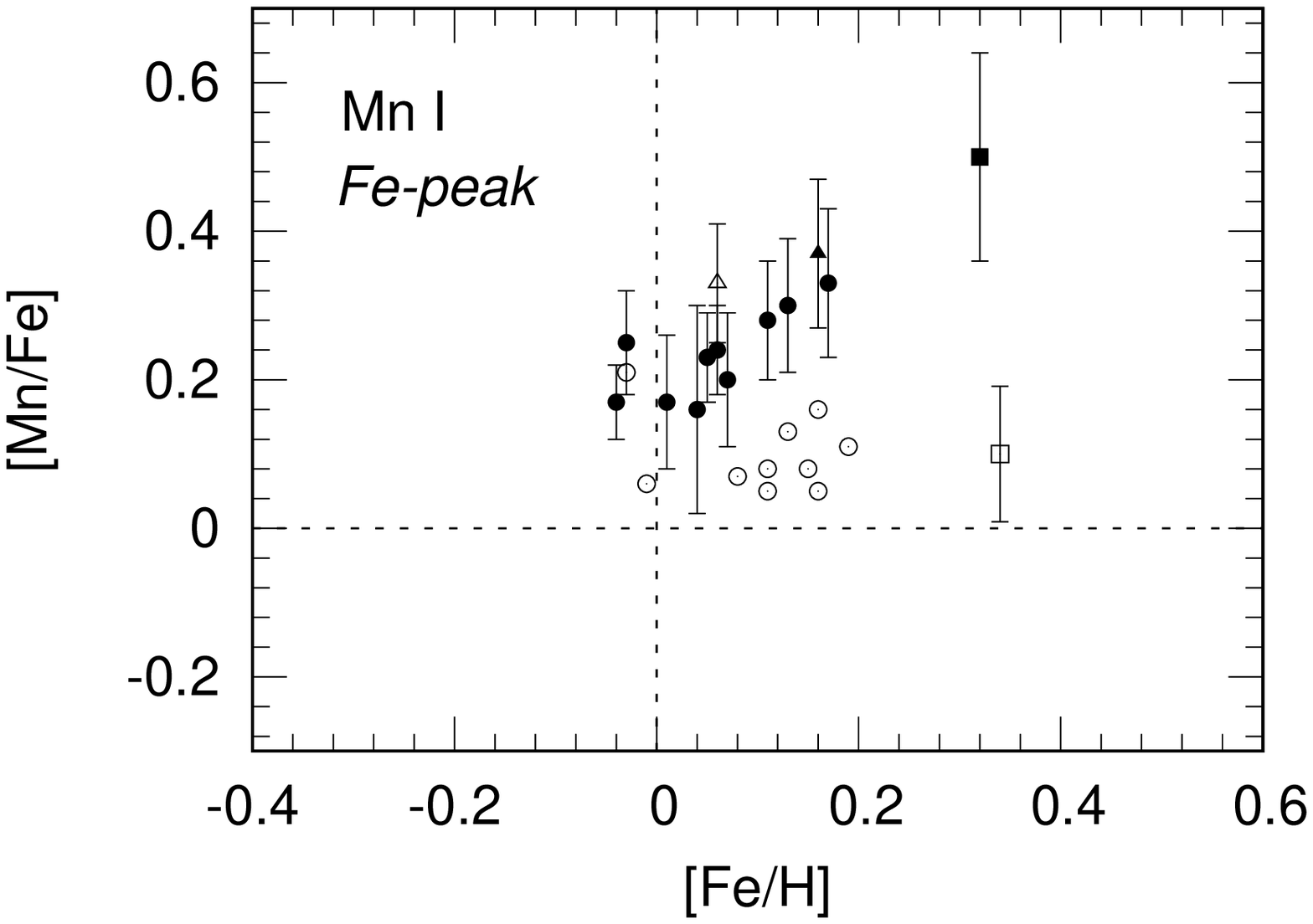} 
\includegraphics[width=55mm]{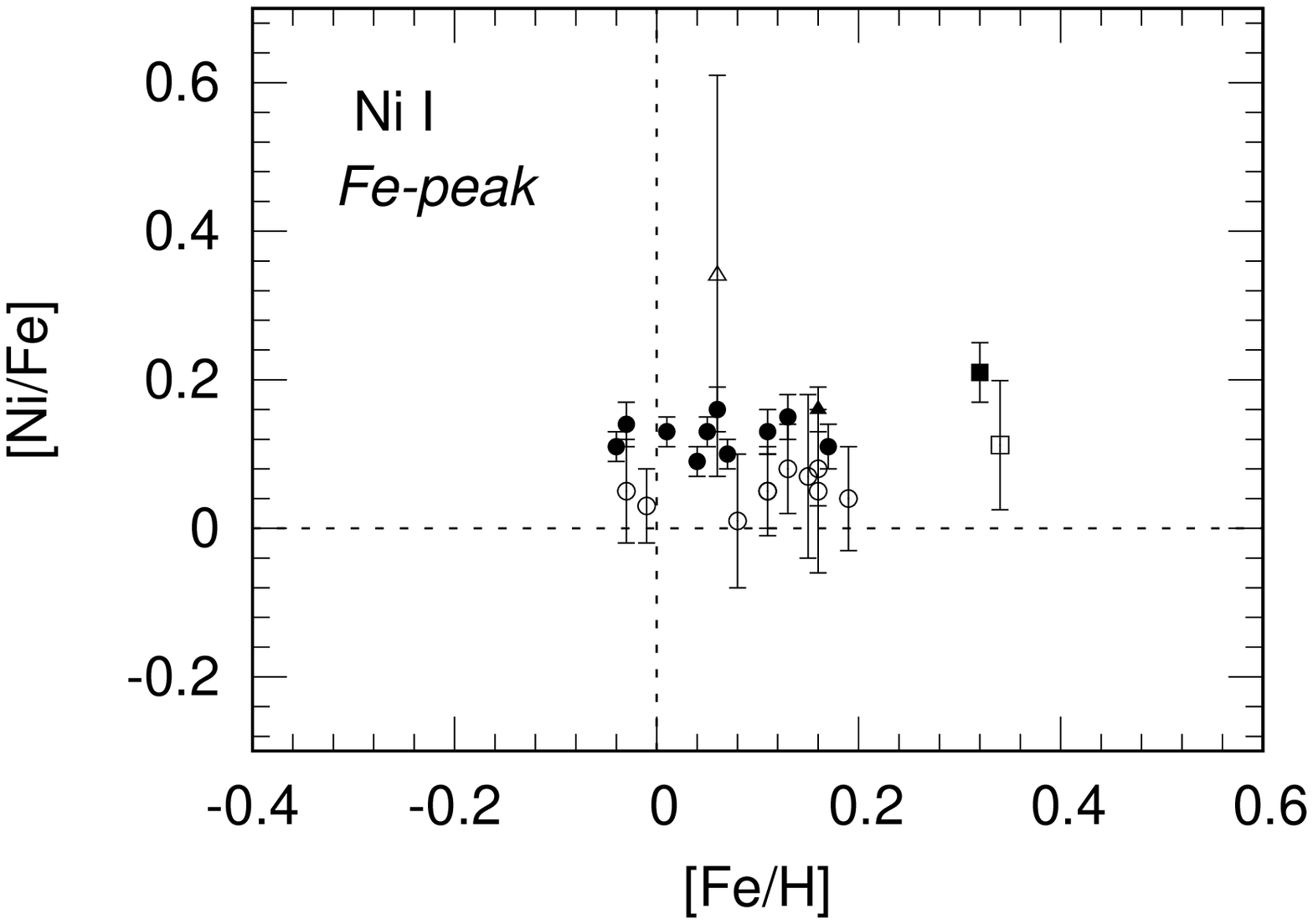} 
\includegraphics[width=55mm]{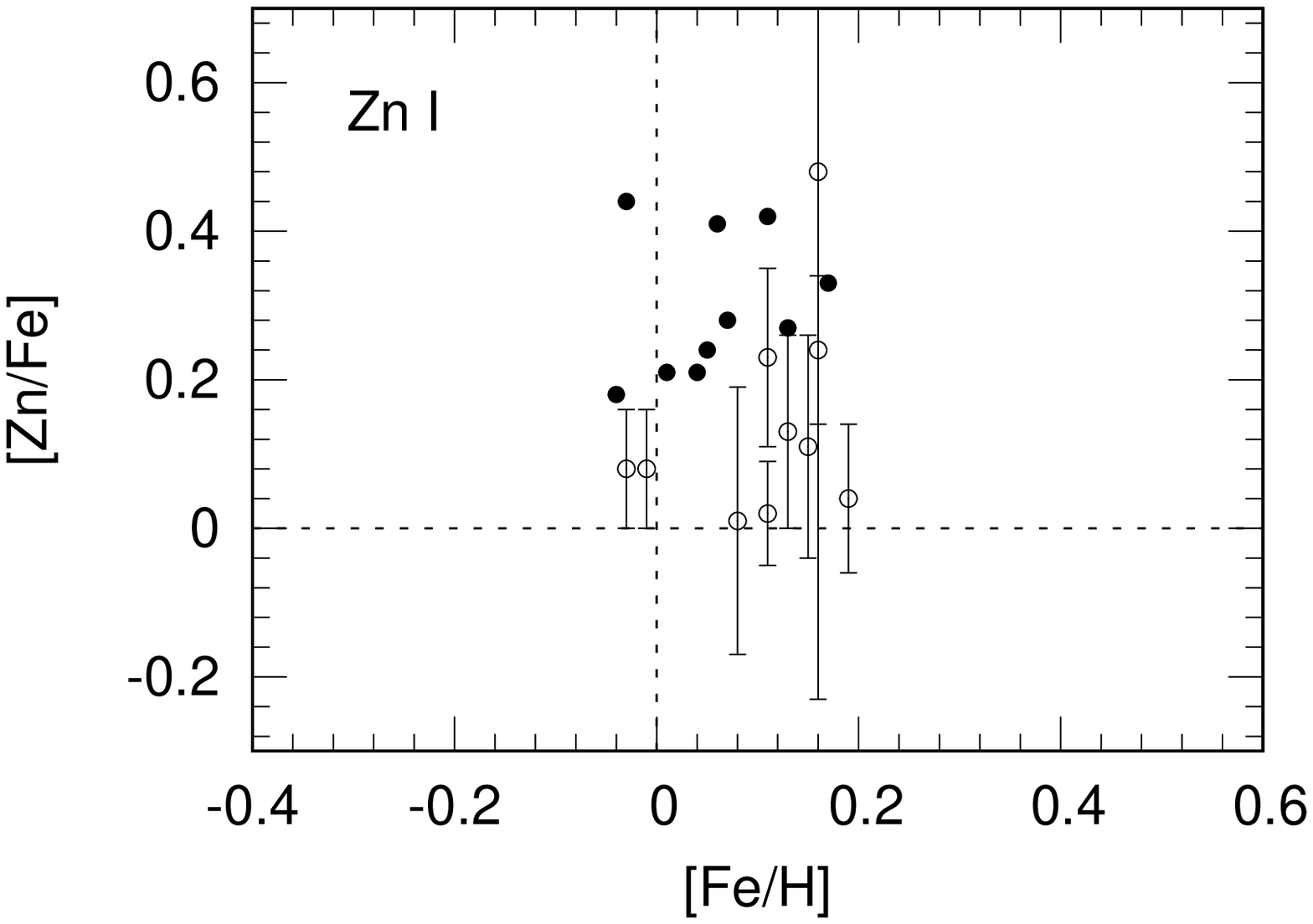} 
\caption{The dependencies of [X/Fe] versus [Fe/H] for the common stars with \protect\cite{felt98}, \protect\cite{adib12}, \protect\cite{bens14} \protect\citep[][in case of Mn]{batt15} -- shown with empty triangles, squares and circles, respectively. Our stars are drawn using the same figures, but filled. The reference solar abundances for each work are converted to the scale of \protect\cite{ande89}.}
\label{_figure_overlapping_stars}
\end{figure*}

We carried out a detailed comparison of our results to \cite{bens14} and a few other authors (Fig. \ref{_figure_overlapping_stars}). These are the only works which have stars in common and references to the solar scales in use, which are essential for a direct abundance comparison.

In total, we found 26 stars from our sample that were previously analysed. The full list is given in Table \ref{_table_all_common_stars}. Note, that the data in Table \ref{_table_all_common_stars} are given 'as is' -- with the [M/H] values provided in the solar scales adopted by the respective authors. Note the main differences between our procedures:

- The stars in common were observed using different spectrometers, and processed using different pipeline procedures;

- We used different systems of the oscillator strengths $gf$, damping constants and the line lists;

- In our analysis we reproduced the absorption line profiles, and the other works the observed equivalent widths;

- As mentioned above, the different types of analyses as well as the sample bias effects could lead to the different distributions of the microturbulent velocities (Fig. \ref{_figure_microturbulence_comparison}) and hence the abundances (Fig. \ref{_figure_method_microturbulence}).

This work as well as the comparison papers provided values that were measured relative to the Sun as a star. We do not expect large differences to be present here, and despite the various procedures and observed data we find a good agreement between the different authors. Details of this analysis and observed differences are discussed in the following subsection.

A direct comparison of abundances for the common stars (Fig. \ref{_figure_overlapping_stars}) show approximately the same values for Na, Mg, Si, and Ca relative to iron in the different works. The agreement is within 0.02\,dex.

For the Ti, Cr, and Ni we have slightly higher deviations of the order of 0.05\,dex. For the first two elements our abundances are systematically lower, and Ni in our case is over-abundant within the same range.

We found larger differences for the abundances of Al, Mn, and Zn. This might be explained by differences in the line lists used, at least the distributions of stars in the different samples (Fig. \ref{_figure_samples}) point to this conclusion. We got 0.1--0.2\,dex higher abundances of Mn and Zn, and 0.1--0.2\,dex lower for Al. However, our aluminium and manganese distributions agree with \cite{adib12} and \cite{felt98}, respectively, and our results agree with \cite{brew16} for both these elements within the error bars.

\subsection{Stability of the results}

To verify the stability of our results, we re-determined fundamental parameters and abundances for the stars in common and the Sun using the effective temperatures, atomic line data, and line lists by \cite{bens14}, and atomic line data from the 2nd and 3rd releases of \textsc{vald}.

\begin{figure*}
\centering
\includegraphics[width=55mm]{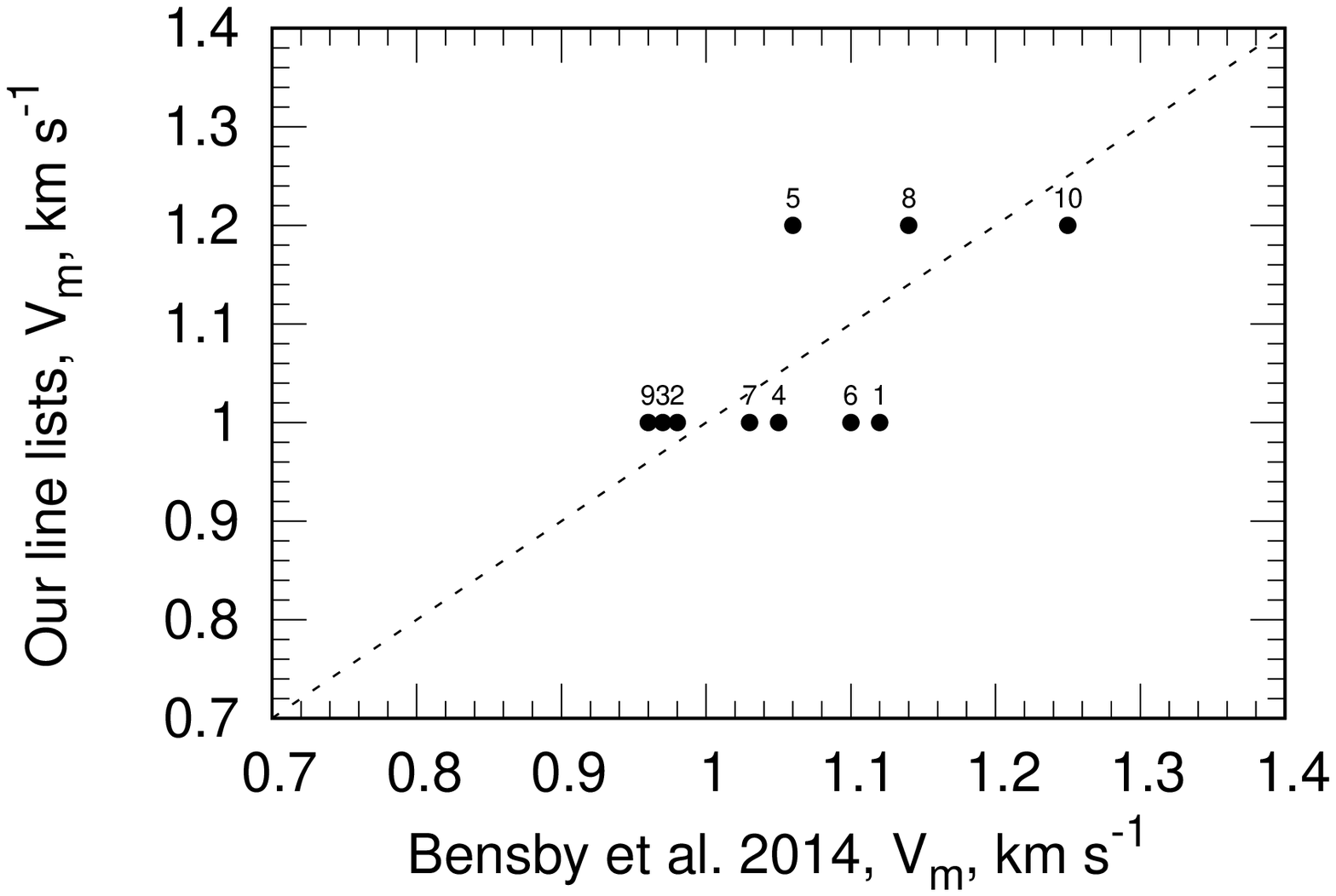}    
\includegraphics[width=55mm]{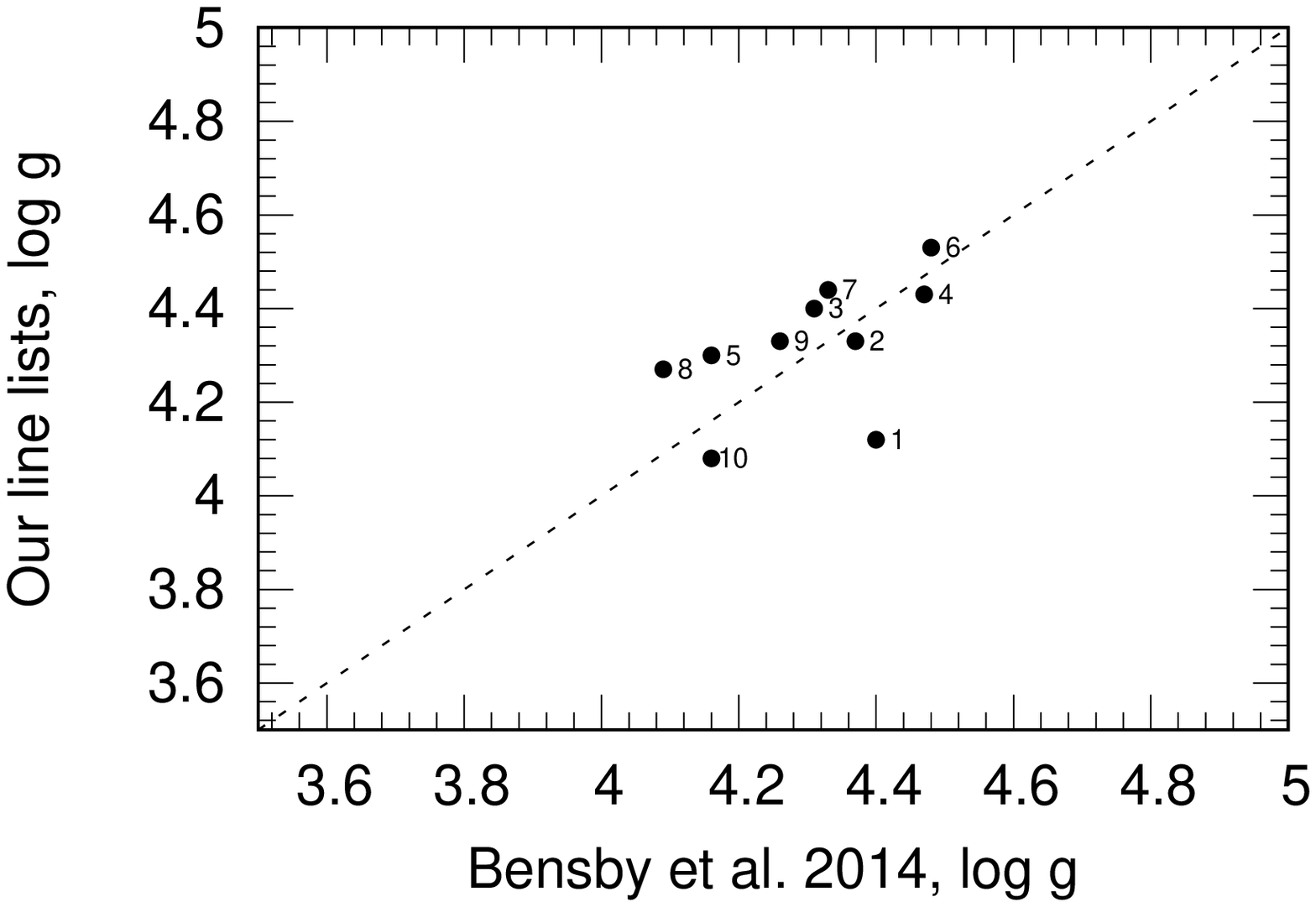} \\ 
\includegraphics[width=55mm]{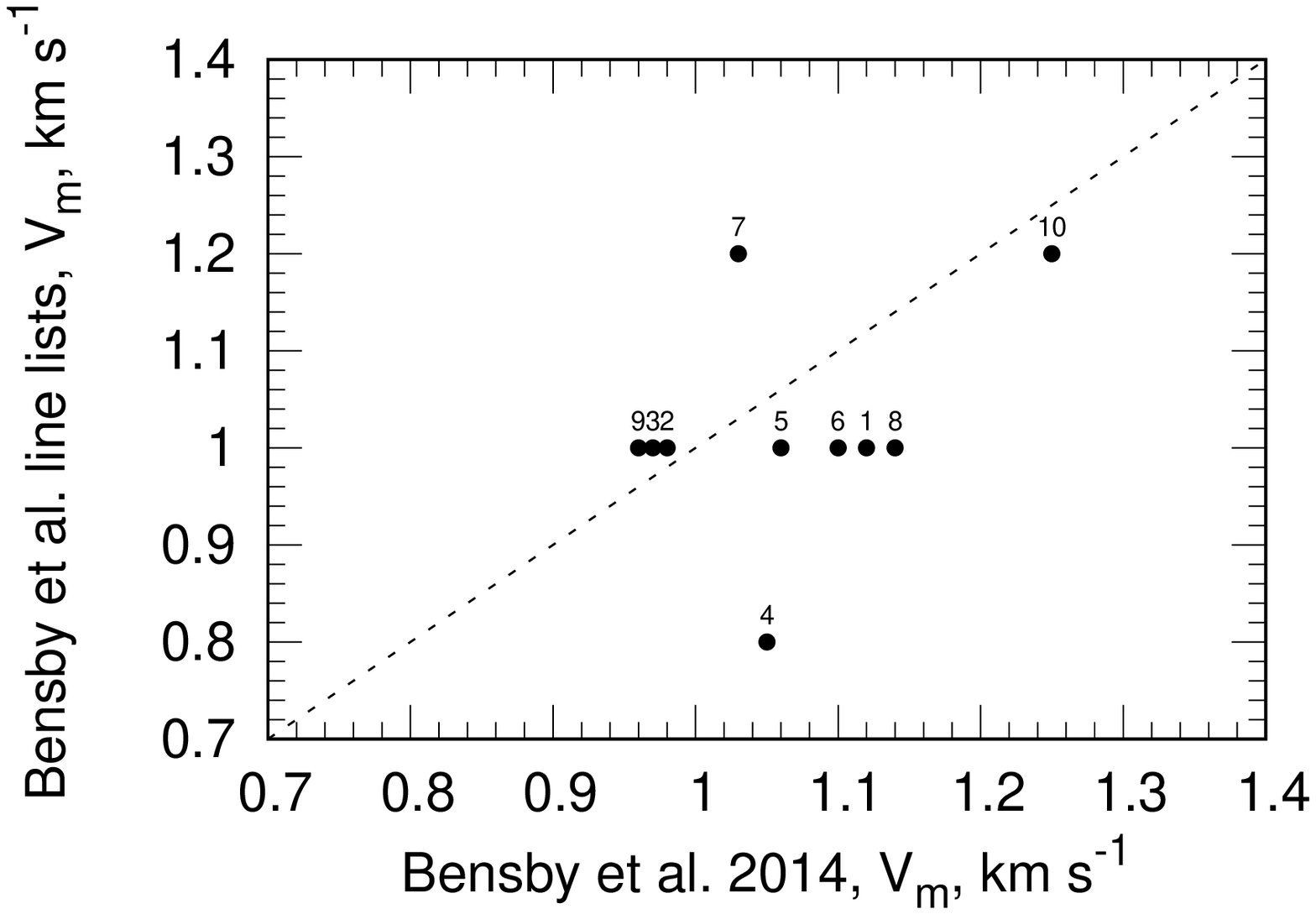}    
\includegraphics[width=55mm]{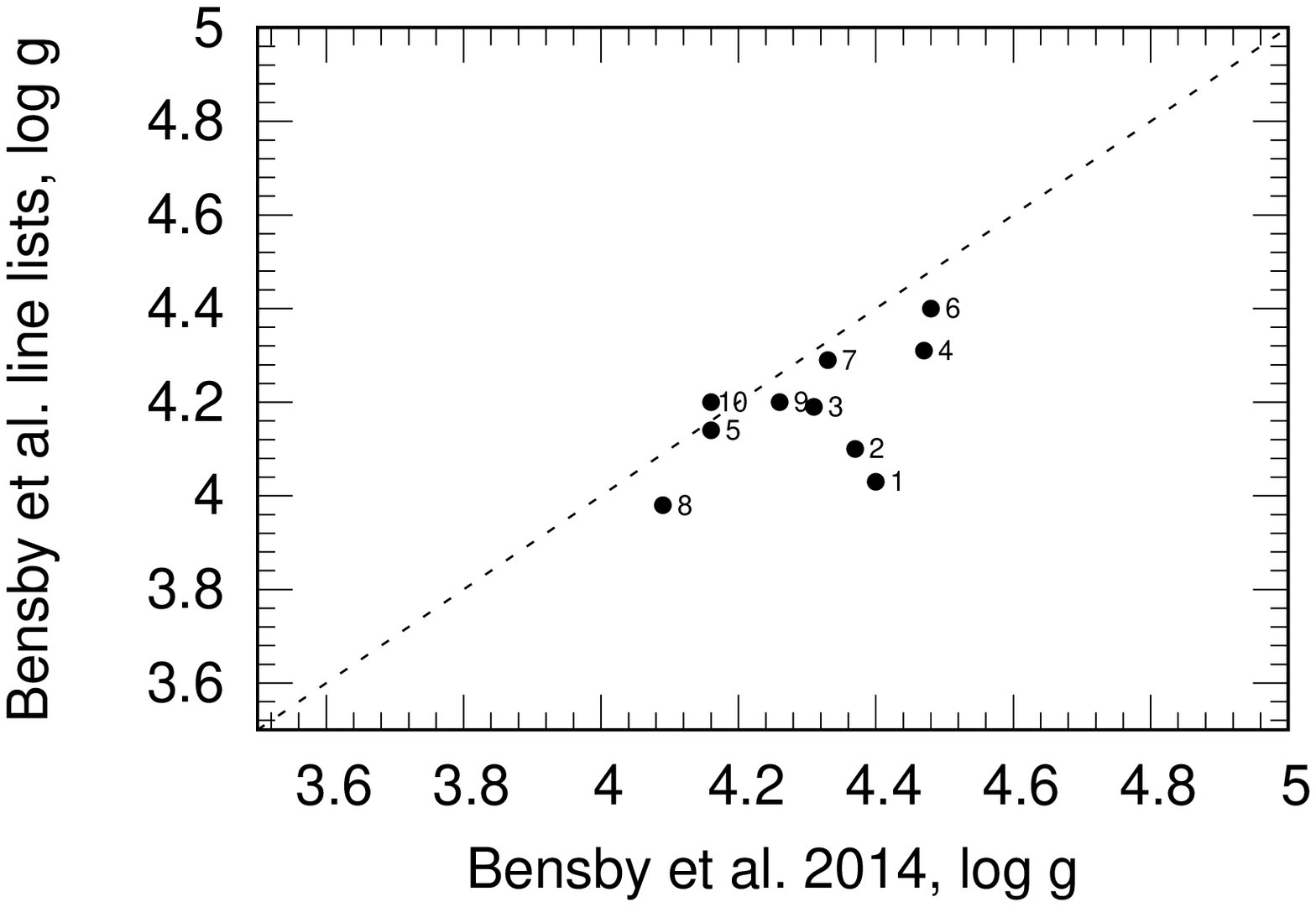}    
\caption{The comparison of microturbulence velocities and gravities for the common stars with \protect\cite{bens14}, using our ({\it top}) and their line lists and effective temperatures ({\it bottom}). Here and forth, the data point numbers correspond to HD 150936, HD 165204, HD 170706, HD 185679, HD 186194, HD 190125, HD 194490, HD 218960, HD 220981 and HD 90520, respectively.}
\label{_figure_bensby_gravity}
\end{figure*}

\begin{figure*}
\centering
\includegraphics[width=55mm]{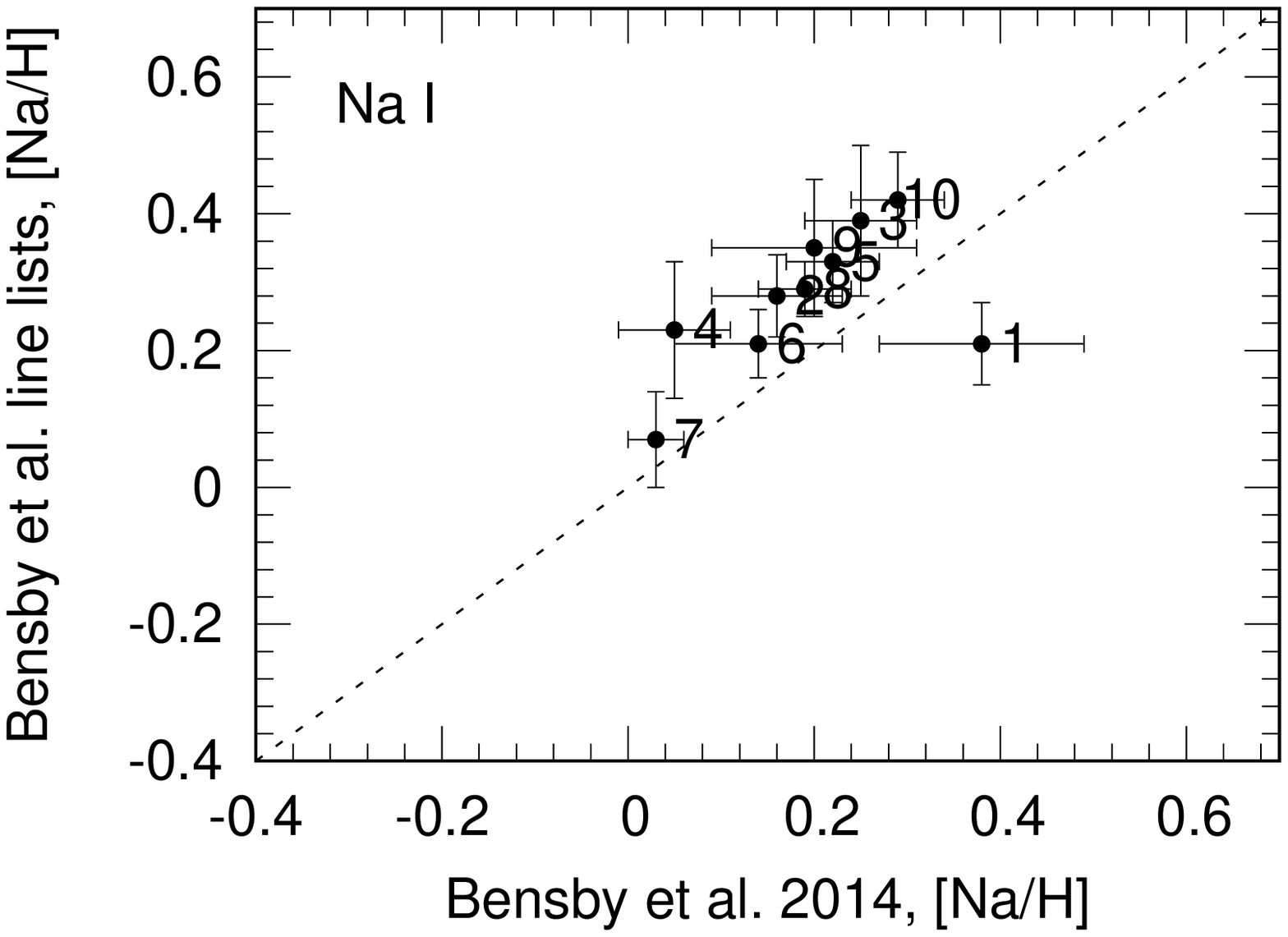} 
\includegraphics[width=55mm]{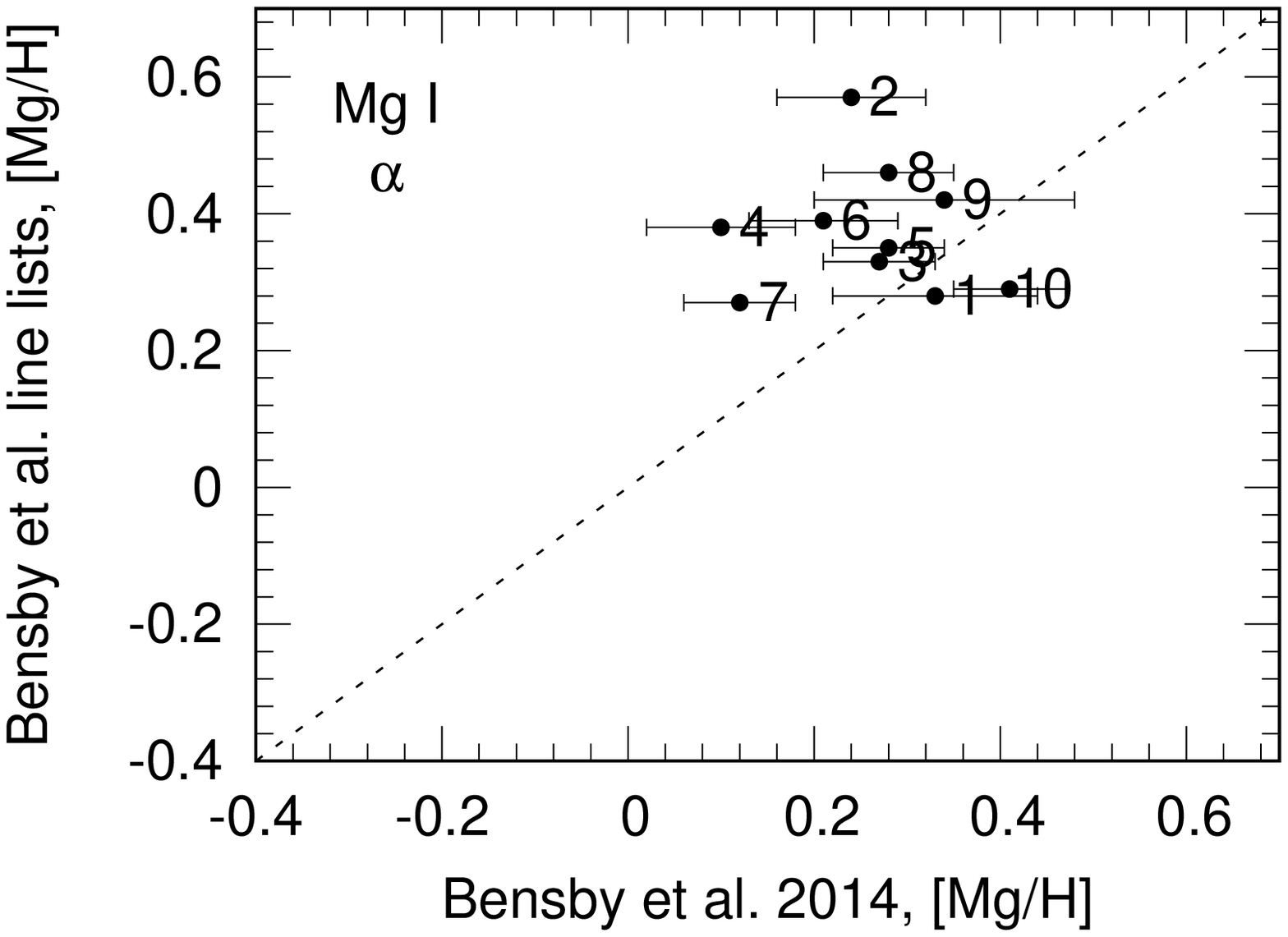} 
\includegraphics[width=55mm]{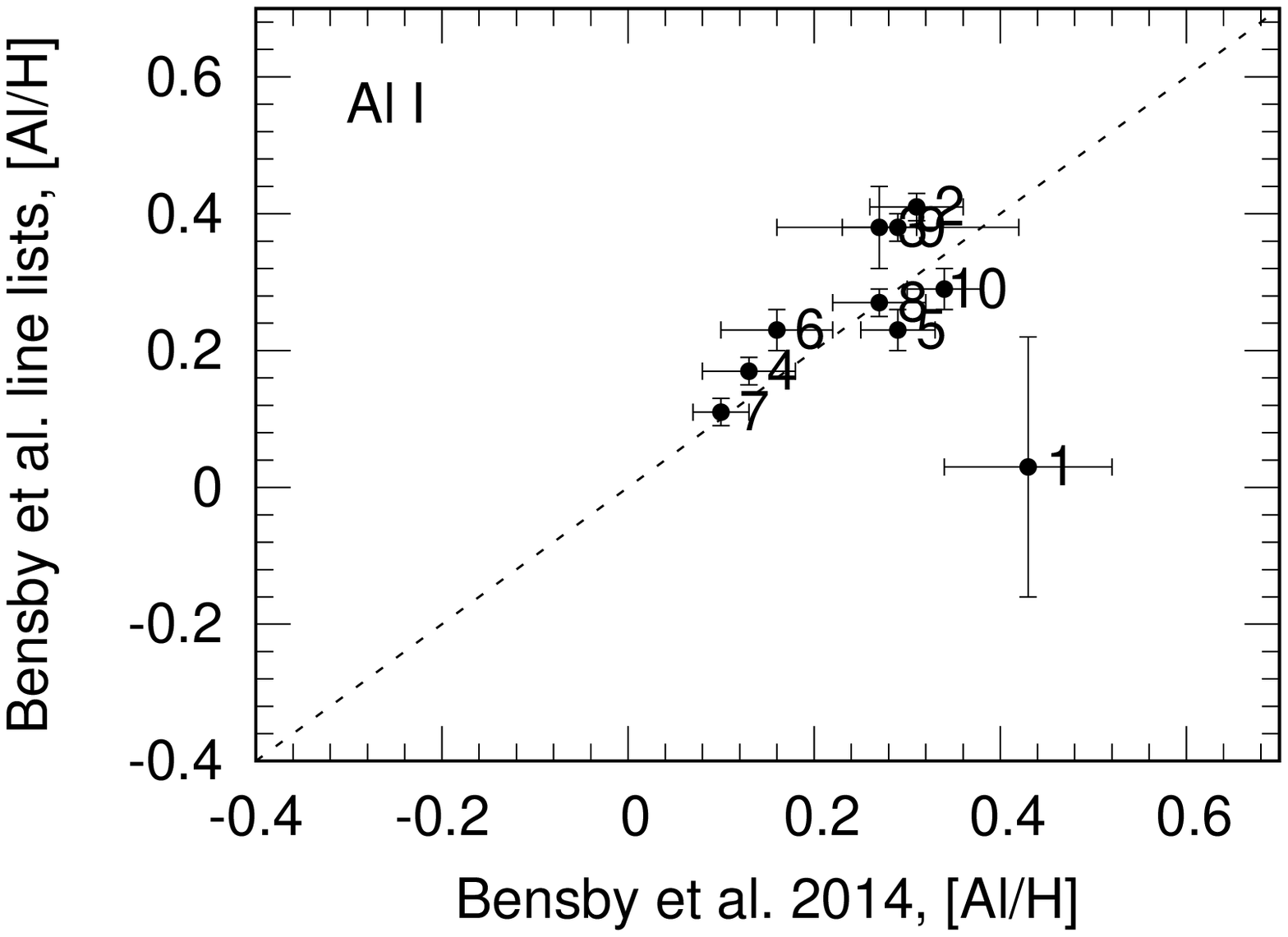} 
\includegraphics[width=55mm]{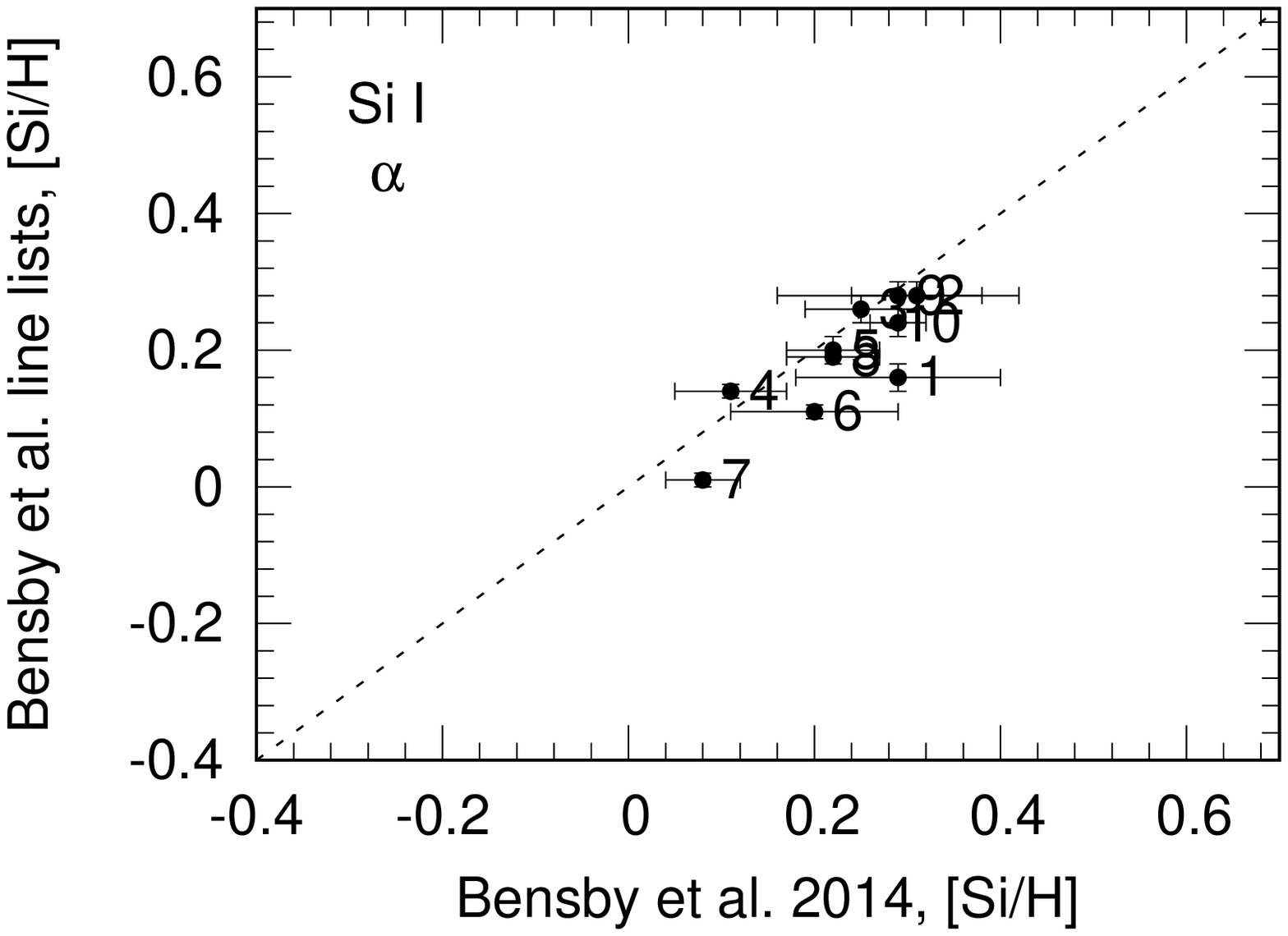} 
\includegraphics[width=55mm]{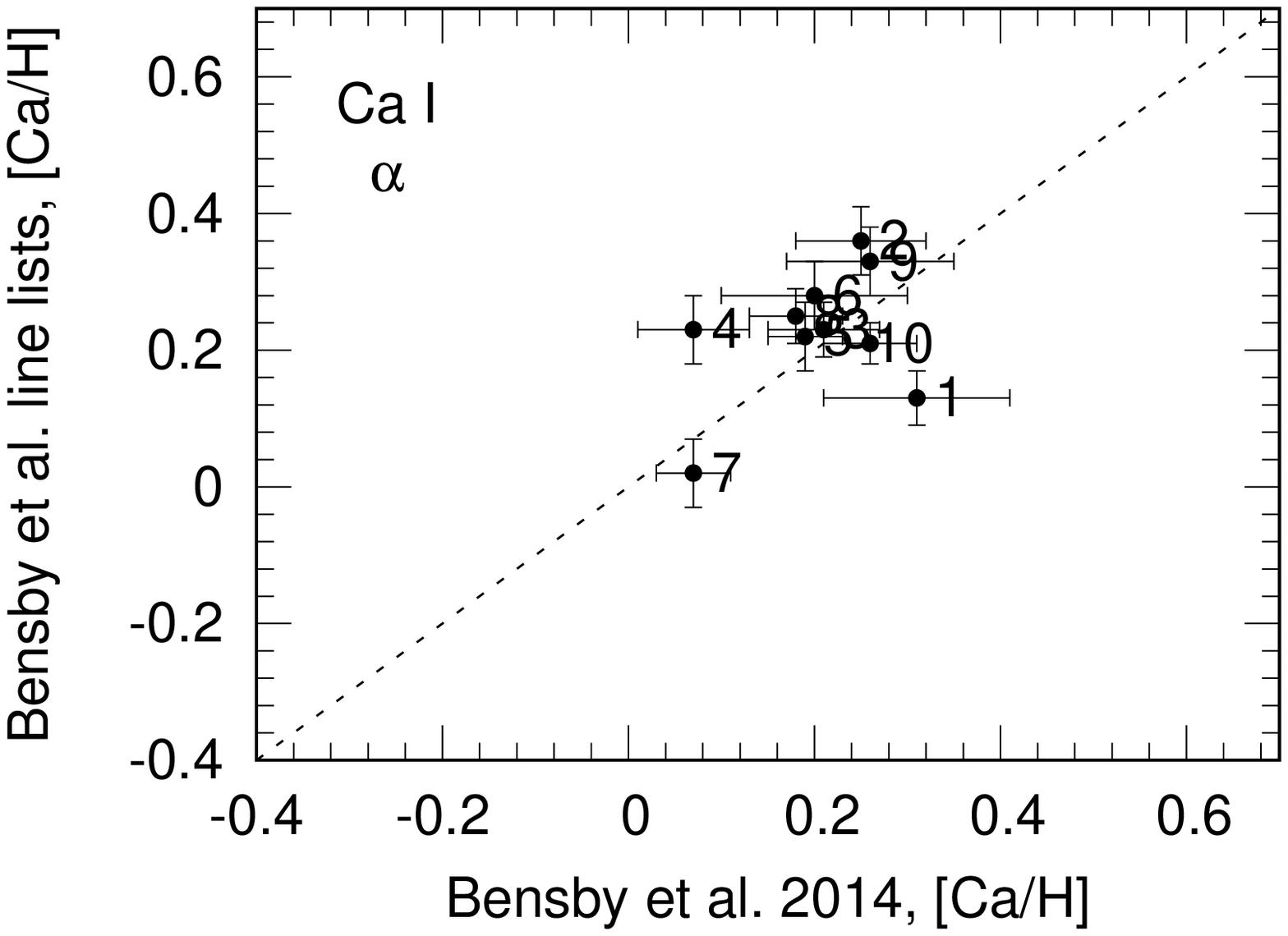} 
\includegraphics[width=55mm]{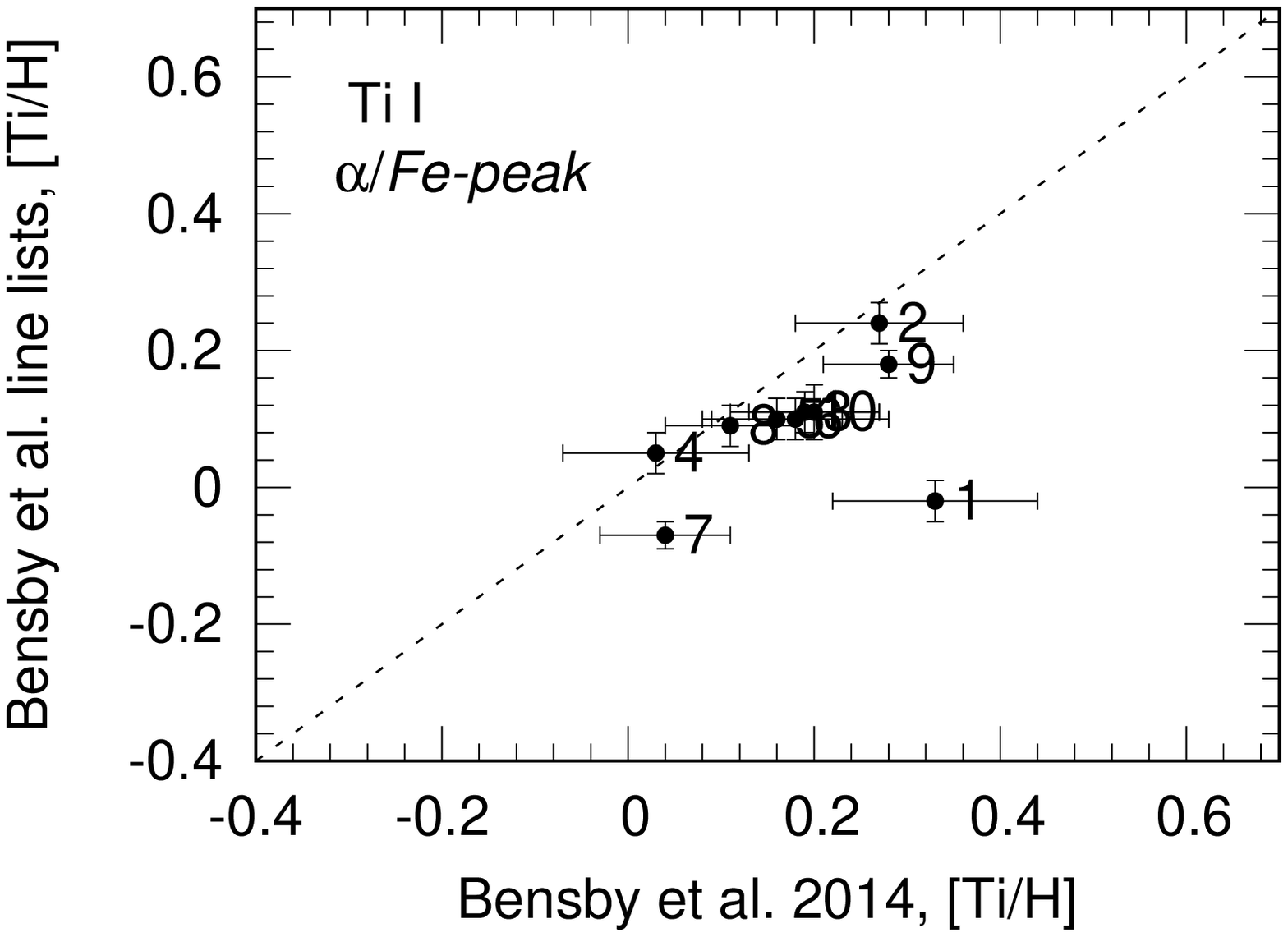} 
\includegraphics[width=55mm]{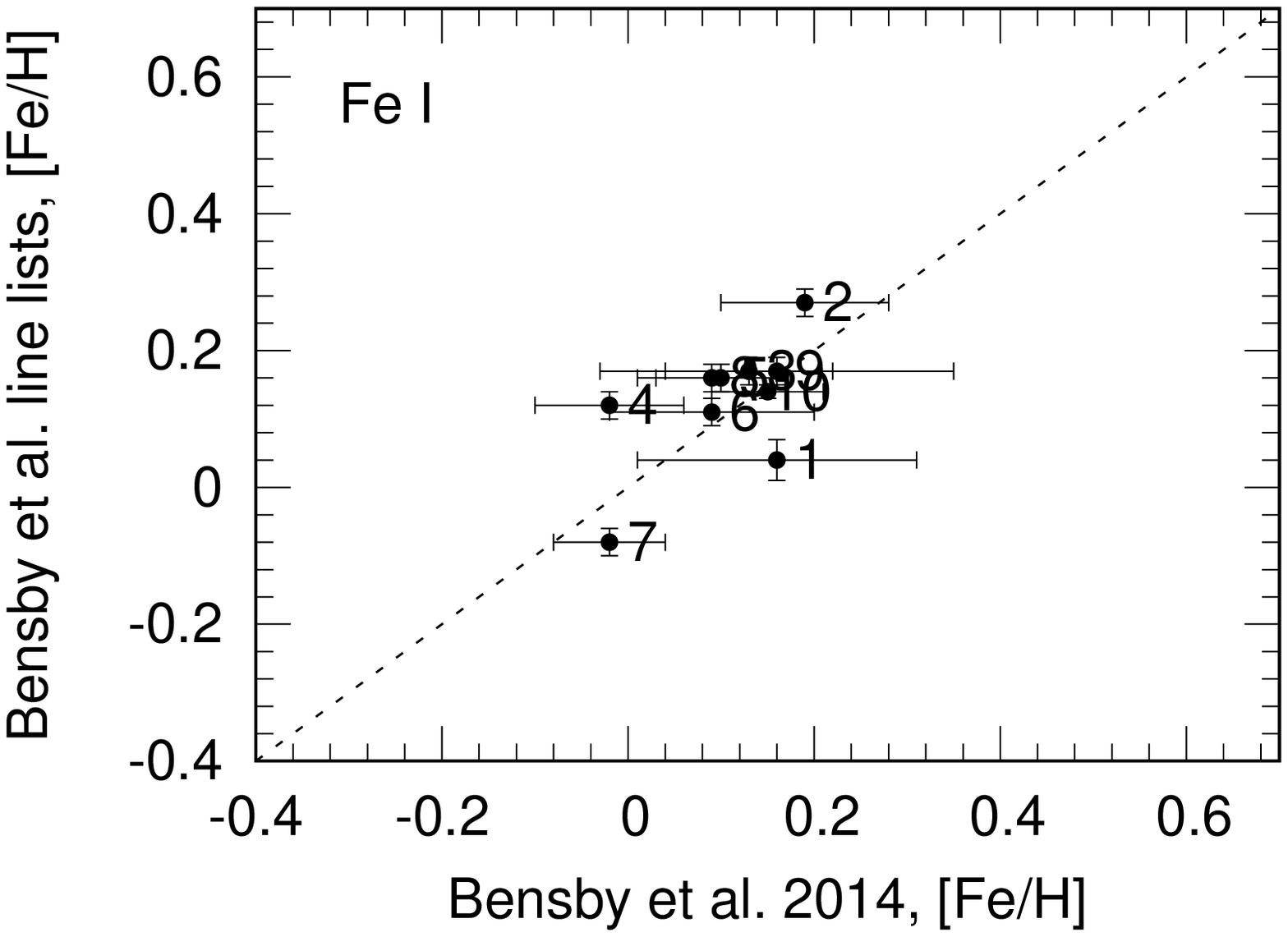} 
\includegraphics[width=55mm]{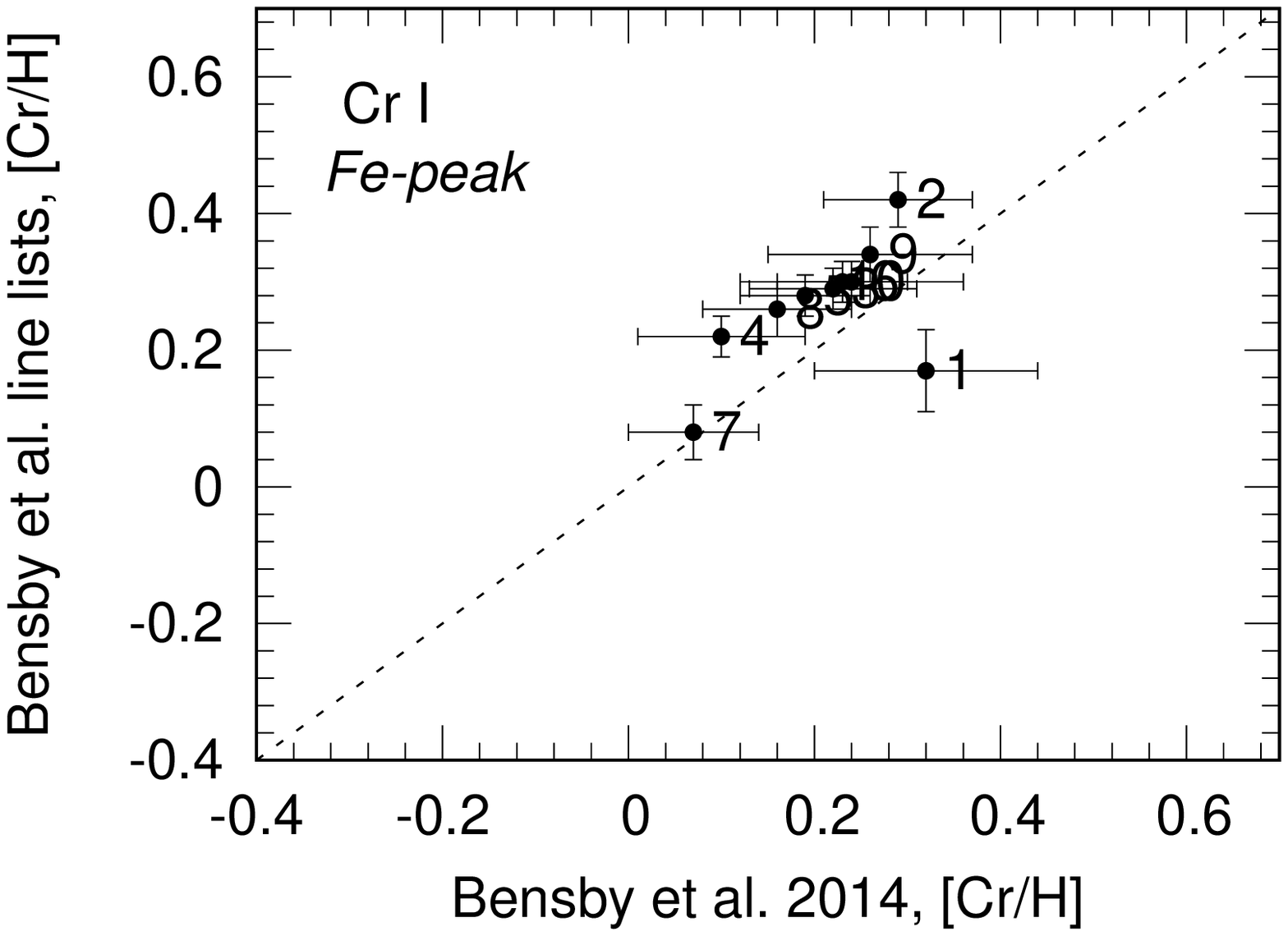} 
\includegraphics[width=55mm]{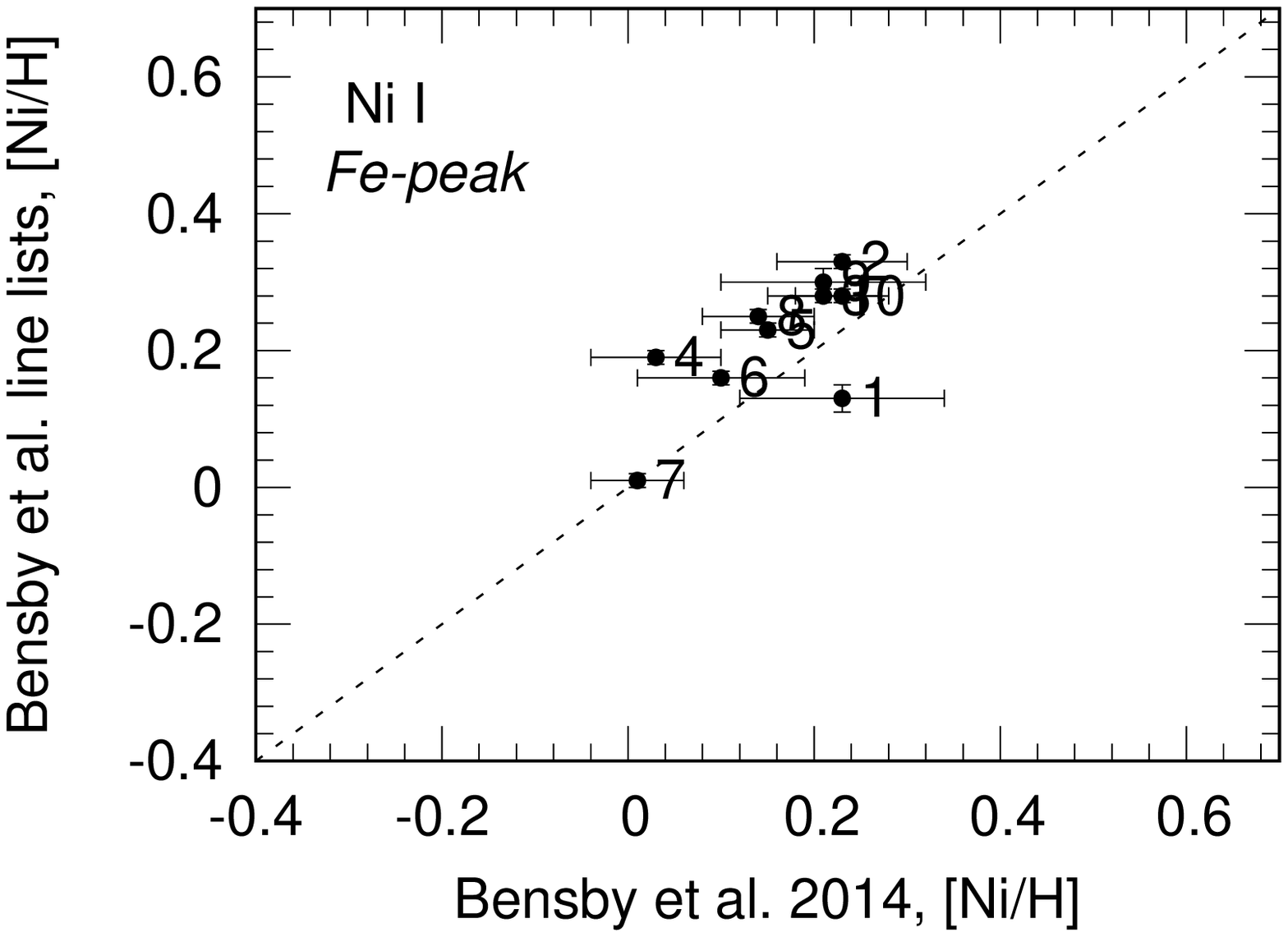} 
\caption{The comparison of abundances for the common stars with \protect\cite{bens14}, using their line lists and effective temperatures.}
\label{_figure_bensby}
\end{figure*}

\cite{bens14} analysed the spectra obtained using MIKE spectrograph with an effective resolution of R$\sim$42,000--65,000. By using their fundamental parameters and the line lists we reduced the differences between our works to the different quality spectra, the different algorithms of \logg\, and \Vm\, determination, and the algorithms of abundance computations. We fitted our theoretical spectra to the observed line profiles, \cite{bens14} carried out the equivalent width analysis. The results are summarized in Figs. \ref{_figure_bensby_gravity} and \ref{_figure_bensby}.

Our microturbulent velocities agree within $\pm$0.1\,\kmps for the eight stars out of ten, and for the two stars within $\pm$0.2\,\kmps.

The differences in \logg\, do not exceed 0.1\,dex for most stars, and only two objects show a deviation up to 0.4\,dex. The same two stars also show an abundance difference up to 0.2\,dex for the neutral iron in the worst case, but still within the error margin provided by \cite{bens14} (Fig. \ref{_figure_bensby}).

As for the other elements, one of the reasons for differences was the limited number of lines that we were able to use. Indeed, due to the wavelength range of HARPS we were able to use only a fraction of the line list provided by \cite{bens14}. The two other main factors we discuss further in the text.

The dependence on the different sets of atomic line data can be assessed using the example of the solar spectrum analysis. The results are summarized in Table \ref{_table_solar}. Note that in the cases of \textsc{vald-2} and \textsc{vald-3} we used the same line lists and fitting ranges that we used for our stars. The only difference was the atomic line data adopted in different versions of the databases. In the case of \cite{bens14}, as mentioned above, we used only a fraction of their original line list.

\textsc{vald-2} and \textsc{vald-3} show the most notable difference of 0.2\,dex for Si, which is also true for all the stars in our sample. Surface gravity is another notable issue which comes from the systematically lower \pion{Fe}{ii} abundances than those for \pion{Fe}{i} for the initial solar model atmosphere.

For the case of \cite{bens14} we also note the \logg\,issue, that comes from the higher measured abundances for the \pion{Fe}{ii} lines and the notable differences of the order of $\sim$0.1--0.3\,dex for Na, Mg, Al, and Cr.

The main problem was that we could not well reproduce the solar equivalent widths obtained by \cite{bens14} using their line list, oscillator strengths, and model atmospheres.

We found the two main reasons for these differences. Firstly, the correction factor of the van der Waals broadening that was adopted by \cite{bens14} was two times higher than in our case of using unmodified \textsc{vald} damping constants (see subsection \ref{_line_lists}). Once we applied the correction we were able to reproduce their equivalent widths. Secondly, the standard continuum levels in the solar spectrum by \cite{kuru84} that we used in our work were different to those adopted by \cite{bens14} for certain lines.

\begin{figure}
\centering
\includegraphics[width=75mm]{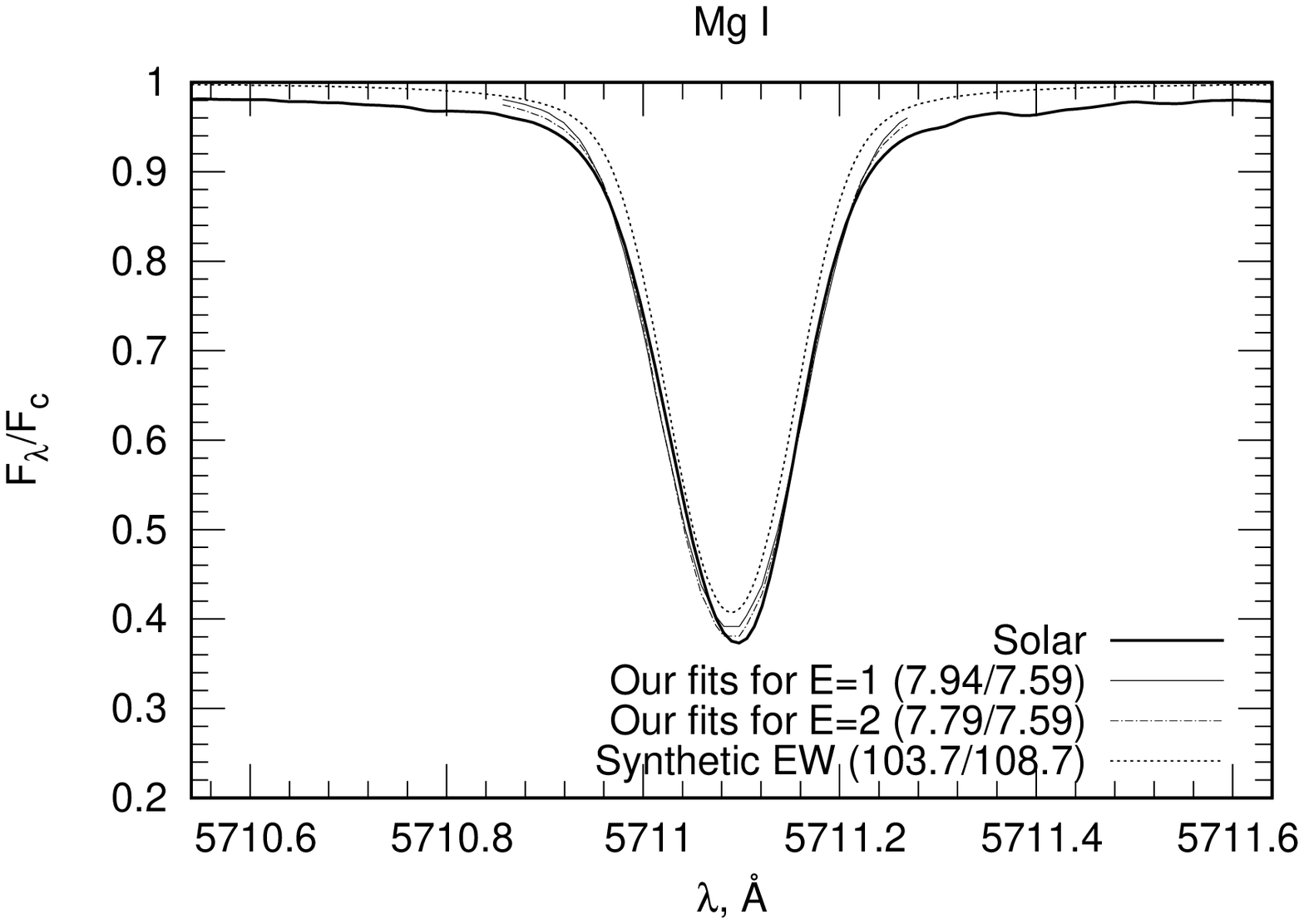} 
\includegraphics[width=75mm]{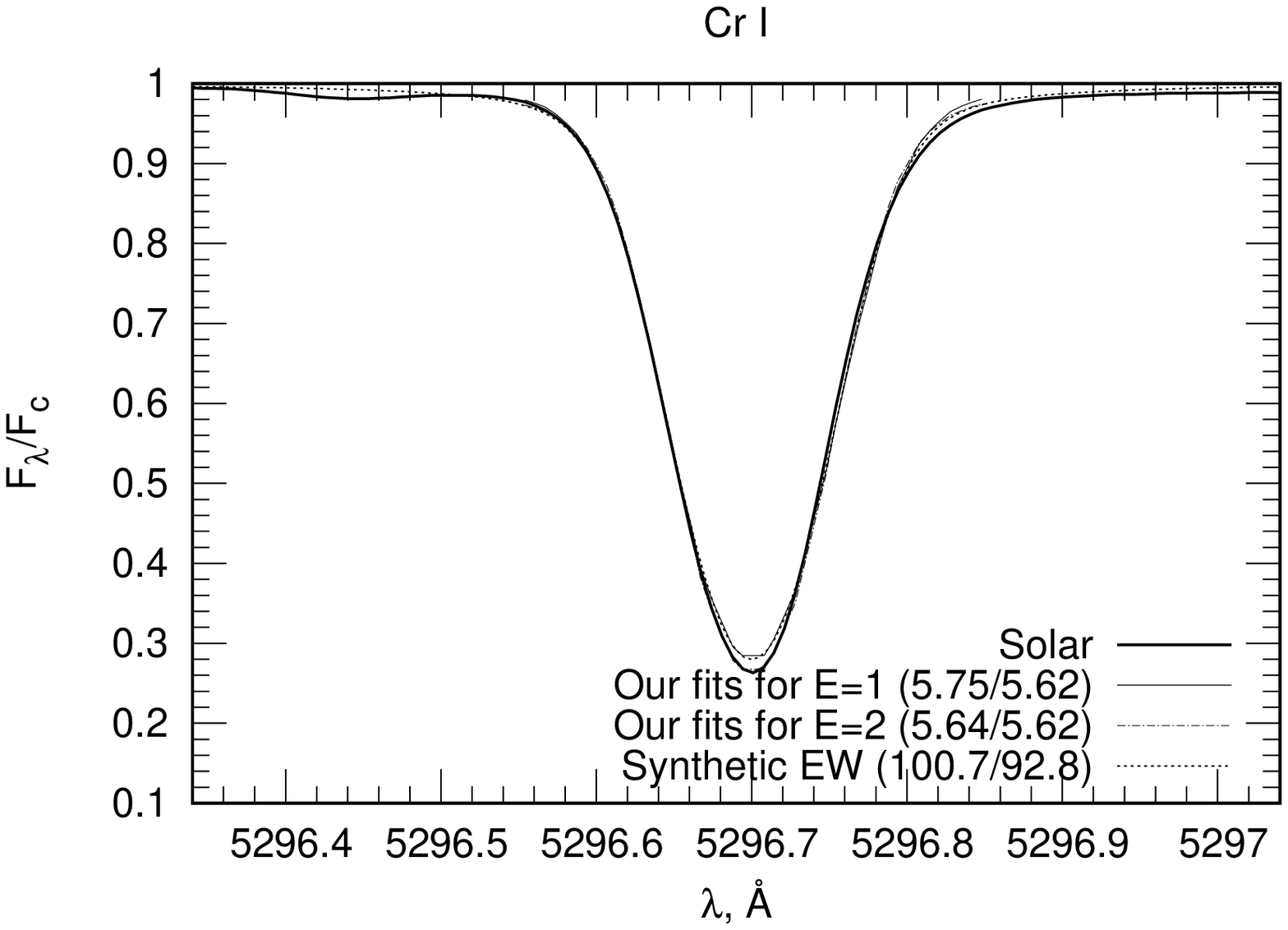} 
\caption{Fits for the different correction factors of the van der Waals broadening ($E=1$ and $E=2$). The numbers in brackets stand for the resulting abundance of the fit ({\it left}) and the reference abundance by \protect\cite{bens14} ({\it right}). 'Synthetic EW' is the theoretical spectral line for the fixed reference abundance by \protect\cite{bens14}. The numbers in brackets depict the equivalent width of the theoretical line and the reference equivalent width measured by \protect\cite{bens14}, respectively. 'Solar' is the observed solar spectrum by \protect\cite{kuru84} with their standard continuum level.}
\label{_lines}
\end{figure}

In Fig. \ref{_lines} we show two of our fits to the solar spectrum by \cite{kuru84} for the different enhancement factors of the van der Waals broadening ($E=1$ and $E=2$) \citep{full71} and their resulting abundances in comparison to the reference abundances by \cite{bens14}. We also show the theoretical spectral line that reproduces the equivalent width for the fixed abundance of \cite{bens14} to illustrate the influence of the adopted continuum level. We see that a similar continuum treatment for the case of \pion{Cr}{i} line returns the reference abundance for $E=2$ and an abundance that is 0.1\,dex higher for $E=1$. The different continuum level for the \pion{Mg}{i} line provides 0.2\,dex higher abundance even in the case of $E=2$. We should not underestimate the importance of these factors when attempting to compare with results by other authors.

For the solar lines this gave us a larger error than for the same elements in the stellar spectra (Fig. \ref{_figure_bensby}). However, the systematically lower \logg\,that we found for our stars in common (Fig. \ref{_figure_bensby_gravity}) using the line data by \cite{bens14} comes from the overabundance of \pion{Fe}{ii} lines due to the same reasons discussed above.

\begin{table*}
\centering
\caption{The solar abundance analyses using our line list based on \textsc{vald-2} and \textsc{vald-3}, and using the line list by \protect\cite{bens14} in comparison to the abundances determined by \protect\cite{bens14}. The reference solar scales by \protect\cite{ande89} and \protect\cite{aspl09} are shown in the last two columns.}
\label{_table_solar}
\begin{tabular}{ccccccc}
\hline\hline
                &  \textsc{vald-2}  &  \textsc{vald-3}  &    Bensby list    &   Bensby results  &        AG89       &        AS09        \\\hline
\logg           &        4.47       &        4.64       &        4.34       &        4.42       &         --        &         --         \\
\Vm             &        1.0        &        1.0        &        1.0        &        0.88       &         --        &         --         \\
\Vsini          &   1.73$\pm$0.12   &   1.65$\pm$0.11   &   1.84$\pm$0.12   &         --        &         --        &         --         \\
\pion{Na}{i}    &  -5.72$\pm$0.03   &  -5.76$\pm$0.02   &  -5.54$\pm$0.15   &  -5.81$\pm$0.00   &  -5.71$\pm$0.03   &  -5.80$\pm$0.04    \\
\pion{Mg}{i}    &  -4.29$\pm$0.09   &  -4.41$\pm$0.16   &       -4.30       &  -4.45$\pm$0.01   &  -4.46$\pm$0.05   &  -4.44$\pm$0.04    \\
\pion{Al}{i}    &       -5.58       &       -5.55       &       -5.51       &  -5.59$\pm$0.00   &  -5.57$\pm$0.07   &  -5.59$\pm$0.03    \\
\pion{Si}{i}    &  -4.51$\pm$0.11   &  -4.33$\pm$0.02   &  -4.51$\pm$0.02   &  -4.54$\pm$0.01   &  -4.49$\pm$0.05   &  -4.53$\pm$0.03    \\
\pion{Ca}{i}    &  -5.64$\pm$0.03   &  -5.71$\pm$0.01   &  -5.72$\pm$0.08   &  -5.71$\pm$0.01   &  -5.68$\pm$0.02   &  -5.70$\pm$0.04    \\
\pion{Ti}{i}    &  -7.01$\pm$0.04   &  -7.10$\pm$0.03   &  -7.10$\pm$0.03   &  -7.14$\pm$0.08   &  -7.05$\pm$0.02   &  -7.09$\pm$0.05    \\
\pion{Ti}{ii}   &  -6.98$\pm$0.21   &  -7.09$\pm$0.16   &  -7.14$\pm$0.08   &  -7.14$\pm$0.10   &         --        &         --         \\
\pion{Cr}{i}    &  -6.35$\pm$0.03   &  -6.44$\pm$0.02   &  -6.22$\pm$0.06   &  -6.41$\pm$0.01   &  -6.37$\pm$0.03   &  -6.40$\pm$0.04    \\
\pion{Fe}{i}    &  -4.42$\pm$0.03   &  -4.43$\pm$0.03   &  -4.46$\pm$0.02   &  -4.49$\pm$0.07   &  -4.37$\pm$0.03   &  -4.54$\pm$0.04    \\
\pion{Fe}{ii}   &  -4.46$\pm$0.02   &  -4.47$\pm$0.04   &  -4.47$\pm$0.02   &  -4.48$\pm$0.08   &         --        &         --         \\
\pion{Ni}{i}    &  -5.73$\pm$0.03   &  -5.77$\pm$0.03   &  -5.77$\pm$0.02   &  -5.81$\pm$0.03   &  -5.79$\pm$0.04   &  -5.82$\pm$0.04    \\\hline
\end{tabular}                                                     
\end{table*}

\section{Discussion}
\label{_discussion}

\subsection{Results and comparisons}

Our CHEPS sample is quite diverse in terms of surface gravity and effective temperature, and yet it maintains homogeneity in various senses. For instance, the selected stars generally have low chromospheric activity, meaning the spectra of our sample should be minimally affected by spotty activity caused by magnetic field phenomena. Also, all of our stars are from the nearby solar neighbourhood, the distance does not exceed 170 pc, and most of them belong to the thin disc population (Fig. \ref{_figure_general_properties}).

Over the last years a few extended studies of the atmospheric abundances of stars with F-G spectral classes were performed using various approaches. It is worth noting, that in our analysis we used a different procedure in comparison to other authors. The use of line profiles allows us to develop an advanced procedure of the abundance analysis:

- We determined the distribution of microturbulent velocities in the atmospheres of our stars using the known condition $D_a$\,=\,0, the distribution of \Vm\, found in Fig. \ref{_figure_microturbulence_comparison} covers the range of 0.8--1.4\,\kmps, with a maximum at 1.2\,\kmps.

- To minimize the effects of blending we used only parts of the absorption line profiles that are less affected by other lines. The use of our procedure is better suited to the case of metal rich stars due to the increased line blending effects.

- Our analysis allowed us to determine rotational velocities (Fig. \ref{_figure_rotation}). Furthermore, we showed that our fast rotating stars have lower \logg, as expected for a population of younger stars.

- We used only the effective photometric temperatures from \cite{jenk08} as an input parameter to provide the complete analysis of our stellar spectra. The other parameters, i.e. [Fe/H], \Vsini, \Vm, \logg\, were found iteratively from the fits to observed absorption line profiles in the spectra.

In the framework of this work, we performed a direct comparison between our results and those of the other authors. We find the following:

1) The abundance distributions of the metal rich samples of stars from our work and those of the other authors are shown in Fig. \ref{_figure_samples}. We see notable differences between all of the different samples, and hence we neglect to analyse the mean sample abundances giving preference to the trends of abundances with metallicities, since they are generally the same for the different authors.

2) In Fig. \ref{_figure_microturbulence_comparison} we show the microturbulent velocity distributions of our sample and the samples of the other authors. The adopted or determined microturbulent velocity is of crucial importance for the determination of accurate chemical abundances (Fig. \ref{_figure_method_microturbulence}). We can see that our distribution of \Vm\, is narrower in comparison to the samples of \cite{sous11,adib12,bens14}. A notable fraction of their stars show lower \Vm, which formally should convert into larger measured abundances. In some sense, the shape of the \Vm\, distribution reflects some of the uncertainties in the procedures of abundance determination: fits to the 'true' continuum, line blending, differences in $gf$, etc. Basically, we would expect a rather narrow distribution of \Vm, due to the fact that the convective envelopes of solar like stars on the main sequence, and those just moving off onto the subgiant branch, should be very similar.

3) The abundance distributions of common elements for the samples of different authors are shown in Fig. \ref{_figure_samples}. We can see here that the internal agreement is better than the absolute agreement. In other words, comparing relative results between stars using a given methodology is more robust than comparing absolute results from different procedures directly. On the other hand, different authors also generally use different samples, therefore, hypothetically, absolute differences can be explained by differences in the local abundance distributions in the Galaxy, particularly, when comparing small numbers of stars. The comparison of common stars (Fig. \ref{_figure_overlapping_stars}) partially proves this, as we see a little bit better agreement for them, even despite the differences in the adopted input parameters, i.e. \Teff, \logg, \Vm, except for a few elements with a limited number of 'good' lines. However, we reiterate that it is still likely that we are seeing here the effects of the differences in the procedures adopted by each author, like the use of different line lists, adjusting parameters, quality of spectra, different interpretations of line blending effects, as well as systematic errors in the effective temperatures and/or gravities.

4) Modern procedures aim to analyse huge sets of spectral data using automated routines or manually \citep[e.g.][]{vale05,adib12,bens14,brew16,jenk17b}. In our analysis we also used the comparatively old classical results by \cite{edva93} and \cite{felt98} that were carried out in the framework of the classical approach. In Fig. \ref{_figure_samples} we show only a few of the known abundance distributions obtained by different authors using different procedures. \cite{sant04,sous08,neve09,sous11} with \cite{adib12} and \cite{vale05} with \cite{brew16} obtained extended abundance determination sets in the framework of the same procedures (the type of analysis, model atmospheres, line lists selection, and the spectra). As expected, their results agree well, and therefore the similarities between these works is yet another reason to use independent procedures, like the one we used in our paper, for verification purposes. Various approaches provide notable differences between authors that exceed the internal accuracy of each method e.g. \cite{vale05} with \cite{brew16} and \cite{bens14} with \cite{batt15}.

5) We obtained notable differences at the level of 0.2\,dex for Mn abundances in the atmospheres of metal rich stars between our work and those computed by \cite{batt15} and \cite{adib12}. In most cases, this level of offset could be explained by differences in the adopted line lists. On the other hand, slopes of the dependence between [Mn/Fe] and [Fe/H] agree well between all comparison works (Table \ref{_table_abundance_slopes}). It is interesting to note that \cite{felt98} and \cite{brew16} obtained very similar results for this element (Fig. \ref{_figure_samples}).

6) Despite the different mean abundances in the samples, all comparison works show mostly the same trends with metallicity. Even if the computed numbers might be different (Table \ref{_table_abundance_slopes}), the distributions in Fig. \ref{_figure_samples} point to the fact that differences in numbers might originate from the sample bias and border effects, as well as from a relatively large spread of abundances for certain elements like Mg, Al, Ti, Mn, and Zn. Despite these differences, general trends remain the same.

7) Only 12 stars from our sample have the confirmed planetary systems to date \citep[see][]{jenk13b,jenk17a}. Similarly to the other studies, e.g. \cite{vale05,neve09}, we found that the abundance distributions for hosts and stars without planets show no significant differences in the solar and metal rich domains. \cite{adib12} found the over-abundance of $\alpha$-elements for planet-hosting stars at low metallicities, though since we only have metal rich stars we cannot verify this result.

\subsection{Differences between the samples}

We can suggest a few possible explanations of the apparent differences in Fig. \ref{_figure_samples} between different authors:

1) Astrophysics -- different samples draw stars that reach us from different parts of the Galaxy. We understand stellar nucleosynthesis, whereby metal rich stars that were formed at distances of a few kpc from the Sun, were formed in the epoch after the birth of the Sun. However, most of our stars belong to the thin disc which is well mixed due to the high stellar density. Eventually, we hope to measure all the abundances for stars near the Sun, such that we can say something more definite about homogeneity of the local population.

2) Differences in the details of procedures. To get matched abundances, it is necessary to perform fits to observed spectra, i.e. line profiles or equivalent widths, using the same procedures as those that were used in the literature. Three main sets of input data should be used: observed spectra, model atmospheres, and line lists. Overall, we can describe the differences in the abundance distribution of an element $\Delta_{total}$ by the formula: $$\Delta_{total}=\Delta_{ma}+\Delta_{Teff}^{logg}+\Delta_{gf}+\Delta_X,$$ where:

- $\Delta_{ma}$ is the difference due to model atmosphere structures computed by different authors. We do not expect large differences here. Our experience shows that in most cases $\Delta_{ma}<0.1$\,dex. Indeed, well known programs that employ 1D model atmosphere computations, e.g. \textsc{atlas}, \textsc{marcs}, \textsc{nextgen}, \textsc{sam12}, use practically the same equation of state, opacity sources and convection treatments. Some differences here may be caused by abundance peculiarities, but it is not the case for solar-like dwarfs.

- $\Delta_{Teff}^{logg}$ is the difference of the adopted or computed \Teff\, and \logg\, converted into abundances. The determination of effective temperatures for stars remains a problem for these kinds of investigations. Different methods of \Teff\, determination could be characterised as photometric, spectroscopic and bolometric. Unfortunately, only the last one has a simple physical meaning. The spectroscopic \Teff\, really corresponds to the temperatures in line forming regions of model atmospheres, i.e. above the photosphere layers where the main source of energy originates. In our work, we used photometric \Teff, which is more similar to bolometric \Teff. Differences between the \Teff\, of the star may be of the order of 50-100\,K, which translates into abundance differences up to 0.2\,dex for neutral iron.

- $\Delta_{gf}$ represents the difference caused by uncertainties in the atomic line data for the absorption lines. There are no conventional line lists selected and approved for this kind of work, and every group uses their own pre-selected list of 'good' lines. Moreover, often oscillator strengths of some lines are adjusted for the best fits to solar spectrum. The cumulative effect of these updates can be seen in Fig. \ref{_figure_samples}. As we noted above, the internal agreement of the results is better than the agreement between the different authors. The effects of solely oscillator strengths and different adoption of the same line list can be seen on the solar spectrum analysis in Table \ref{_table_solar}: we found that even \textsc{vald-2} and \textsc{vald-3} provide notable differences in the spectroscopic parameters of the same lines of some elements. The damping constants treated differently also provide the abundance differences up to 0.2\,dex.

There is also a contribution to the $\Delta_{gf}$ due to a particular choice of the solar abundances at the line pre-selection stage. It can result in the different mean abundances shown by different authors, too. Our comparison works relied mostly on the scales by \cite{ande89} (most of the works shown in Fig. \ref{_figure_samples}), \cite{aspl09} (oscillator strengths in \cite{bens14}), and their derivatives. These scales are not that much different in terms of the abundances, except Fe. What is more important is that our main selection criteria for our line list was a good fit to the observed solar line profiles -- we did not aim to specifically describe our reference scale. Therefore, our resulting solar abundances are more defined by the atomic line data we took from the \textsc{vald}, than by the reference scale (Table \ref{_table_solar}).

- $\Delta_{X}$ represents the possible detrimental effects caused by all other factors. For example, usually authors use pre-selected line lists that have been drawn from the analysis of a solar spectrum. To be certain in the line list output it is important to adopt the similar continuum level. The ideal way would be to process the solar spectra using the same pipeline procedures as those used for the stars in the sample. This will enhance the determination accuracy of the stellar fundamental properties in relation to the Sun.

Although this is sensible, we should still account for the fact that blending effects increase in the spectra of metal-rich stars, or stars that rotate faster, which is one of the principal reasons why we used synthetic spectra fits to observed spectra to determine our abundances. Finally, we should also note a few other possible effects that can contribute to the $\Delta_{X}$ parameter. For instance, NLTE effects should be fully considered, more sophisticated models of micro- and macroturbulent velocities, stellar variability, and different kinds of stellar activity can all contribute also.

\section{Summary}
\label{_summary}

We carried out a spectral analysis of 107 metal rich ([Fe/H]$\ge$7.67\,dex) target stars from the Calan-Hertfordshire Extrasolar Planet Search program observed with HARPS to determine the abundances of Na, Mg, Al, Si, Ca, Ti, Cr, Mn, Fe, Ni, Cu, and Zn in their atmospheres. We used the independent procedure of finding the best fit to the absorption line profiles in high resolution spectra. The abundances, rotational velocities, microturbulence velocities, and surface gravities were found using this iterative process.

Our analysis allowed us to determine the positively skewed normal distribution of projected rotational velocities with a maximum peaking at 3\,\kmps. We obtained a Gaussian distribution of microturbulent velocities that has a maximum at 1.2\,\kmps and a full width at half maximum $\Delta v_{1/2}$=0.35\,\kmps, indicating that metal rich dwarfs and subgiants in our sample have a very restricted range in microturbulent velocities in comparison to samples of other authors, and independent of the number of stars. We also confirm that the abundance distributions for planet hosts and stars without planets show no significant differences, at least for our sample.

For most of the elements our abundances agree up to $\pm$0.05\,dex or better for the stars in common with other works. However, we do find systematic differences between the mean abundances in general samples. Our analysis suggests that the selection of line lists and atomic line data along with the adopted continuum level influence these differences the most.

The observed trends with metallicity remain as a reliable marker for the study of chemical evolution of the Galaxy, as they remain the same across all comparison works despite differences in the spectra, methods and input data. We confirm the positive trends of abundances versus metallicity for Na, Mn, Ni, Zn, and to a lesser degree, Al. A slight negative trend is observed for Ca, whereas Si and Cr tend to follow iron.

\section*{Acknowledgements}

We acknowledge funding by EU PF7 Marie Curie Initial Training Networks (ITN) RoPACS project (GA N 213646) and the special support by the NAS of Ukraine under the Main Astronomical Observatory GRAPE/GPU/GRID computing cluster project. We also acknowledge funding by CATA-Basal grant (PB06, Conicyt), from Fondecyt through grants 1161218 and 3110004, partial support from Centro de Astrof\'\i sica FONDAP 15010003, the GEMINI-CONICYT FUND and from the Comit\'e Mixto ESO-GOBIERNO DE CHILE and support from the UK STFC via grants ST/M001008/1 and Leverhulme Trust RPG-2014-281. This research has made use of the SIMBAD database, operated at CDS, Strasbourg, France. We thank the anonymous referee for her/his thorough review and highly appreciate the comments and suggestions, which significantly contributed to improving the quality of the publication.

\input{ms.bbl}

\appendix
\section{Additional tables and plots}

We provide here the table with the mean abundances and slopes found in this and comparison works. The tables with the abundances and fundamental properties for the stars in our sample, and the spectral fitting ranges used in our work. We also include a full list of the common stars and their fundamental properties found by different authors. And finally, the plots for the abundance dependencies on \Teff\, and \logg. The latter are discussed in subsection \ref{_dependencies}.

\begin{table*}
\centering
\caption{The mean abundances [X/Fe] and their slopes ($X_f=\Delta[X/Fe]/\Delta[Fe/H]$) in this work, in \protect\cite{adib12} (A), and in \protect\cite{bens14} (B). Only the stars with [Fe/H]$\ge$7.67\,dex were accounted. $N_l$ is the total number of lines in the list, $\rho$(\Teff) and $\rho$(\logg) are the Pearson product-moment correlation coefficients for [X/H] versus \Teff\, and \logg.}
\begin{tabular}{ccrrrrrrrrr}
\hline\hline
\multicolumn{1}{c}{\multirow{2}{*}{Element}}                         &                                       \multicolumn{9}{c}{Abundances and slopes}                                       \\
\multicolumn{1}{c}{} & ~$N_l$  & ~~~$\rho$(\Teff) & ~~~$\rho$(\logg) &      [X/Fe]~~     &       [X/Fe]A~    &       [X/Fe]B~    &       $X_f$~~~~   &       $X_f$A~~~~  &       $X_f$B~~~~  \\\hline
\pion{Na}{i}         &    9    &      -0.20       &      -0.14       &   0.10$\pm$0.08   &   0.09$\pm$0.11   &   0.07$\pm$0.07   &   0.30$\pm$0.07   &   0.28$\pm$0.10   &   0.20$\pm$0.06   \\
\pion{Mg}{i}         &    7    &      -0.43       &      -0.02       &   0.13$\pm$0.05   &   0.01$\pm$0.07   &   0.16$\pm$0.07   &   0.10$\pm$0.05   &   0.02$\pm$0.07   &  -0.23$\pm$0.06   \\
\pion{Al}{i}         &    8    &      -0.36       &       0.01       &  -0.03$\pm$0.05   &   0.05$\pm$0.12   &   0.15$\pm$0.05   &   0.13$\pm$0.05   &   0.01$\pm$0.12   &  -0.01$\pm$0.05   \\
\pion{Si}{i}         &   28    &      -0.43       &      -0.02       &   0.10$\pm$0.05   &   0.02$\pm$0.03   &   0.11$\pm$0.04   &  -0.07$\pm$0.05   &   0.06$\pm$0.03   &  -0.04$\pm$0.04   \\
\pion{Ca}{i}         &   23    &      -0.21       &       0.03       &   0.05$\pm$0.05   &  -0.00$\pm$0.04   &   0.09$\pm$0.04   &  -0.15$\pm$0.04   &  -0.26$\pm$0.03   &  -0.14$\pm$0.04   \\
\pion{Ti}{i}         &   23    &      -0.28       &       0.29       &  -0.01$\pm$0.04   &   0.06$\pm$0.08   &   0.06$\pm$0.04   &  -0.03$\pm$0.04   &  -0.22$\pm$0.07   &   0.10$\pm$0.04   \\
\pion{Ti}{ii}        &    3    &       0.26       &      -0.08       &  -0.05$\pm$0.08   &   0.02$\pm$0.05   &        --~~~~~~   &  -0.21$\pm$0.08   &   0.05$\pm$0.05   &        --~~~~~~   \\
\pion{Cr}{i}         &   52    &      -0.20       &       0.11       &   0.03$\pm$0.03   &   0.01$\pm$0.02   &   0.10$\pm$0.03   &  -0.07$\pm$0.03   &  -0.05$\pm$0.02   &   0.09$\pm$0.03   \\
\pion{Mn}{i}         &   20    &      -0.46       &       0.16       &   0.25$\pm$0.09   &   0.04$\pm$0.06   &        --~~~~~~   &   0.53$\pm$0.07   &   0.36$\pm$0.05   &        --~~~~~~   \\
\pion{Fe}{i}         &   63    &      -0.28       &       0.08       &        --~~~~~~   &        --~~~~~~   &        --~~~~~~   &        --~~~~~~   &        --~~~~~~   &        --~~~~~~   \\
\pion{Fe}{ii}        &   15    &      -0.29       &       0.15       &   0.00$\pm$0.04   &        --~~~~~~   &  -0.00$\pm$0.01   &   0.05$\pm$0.04   &        --~~~~~~   &   0.01$\pm$0.01   \\
\pion{Ni}{i}         &   25    &      -0.25       &       0.00       &   0.13$\pm$0.03   &   0.02$\pm$0.04   &   0.05$\pm$0.04   &   0.13$\pm$0.03   &   0.18$\pm$0.03   &   0.22$\pm$0.04   \\
\pion{Cu}{i}         &    9    &      -0.30       &      -0.01       &   0.02$\pm$0.09   &        --~~~~~~   &        --~~~~~~   &   0.28$\pm$0.08   &        --~~~~~~   &        --~~~~~~   \\
\pion{Zn}{i}         &    1    &      -0.56       &       0.11       &   0.29$\pm$0.14   &        --~~~~~~   &   0.13$\pm$0.11   &   0.29$\pm$0.14   &        --~~~~~~   &   0.04$\pm$0.11   \\\hline
\end{tabular}
\label{_table_abundance_slopes}
\end{table*}

\begin{table*}
\centering
\caption{The abundances and fundamental properties found for the stars in our sample. Full version available online.}
\begin{tabular}{cccccccccc}
\hline\hline
Name         &   Distance   &   \Teff  &   Sp. \logg  &   Ph. \logg  &  Ph. [M/H] &       \Vm       &      \Vsini     &     \pion{Fe}{i}     &   ...   \\\hline
HD 6790      &    105.93    &   6012   &     4.40     &     4.36     &    0.20    &   0.8$\pm$0.2   &   4.7$\pm$0.2   &    -0.06$\pm$0.02    &   ...   \\
HD 7950      &    117.37    &   5426   &     3.94     &     3.83     &    0.17    &   1.2$\pm$0.2   &   2.7$\pm$0.1   &     0.11$\pm$0.02    &   ...   \\
HD 8389      &     30.51    &   5243   &     4.52     &     4.46     &    0.40    &   1.2$\pm$0.2   &   1.4$\pm$0.1   &     0.32$\pm$0.03    &   ...   \\
HD 8446      &     73.64    &   5819   &     4.14     &     4.15     &    0.28    &   1.2$\pm$0.2   &   3.9$\pm$0.1   &     0.13$\pm$0.02    &   ...   \\
HD 9174      &     78.93    &   5577   &     4.05     &     4.03     &    0.10    &   1.2$\pm$0.2   &   2.9$\pm$0.1   &     0.26$\pm$0.01    &   ...   \\
...          &      ...     &   ...    &     ...      &     ...      &    ...     &       ...       &       ...       &          ...         &   ...   \\\hline
\end{tabular}
\label{_table_abundances}
\end{table*}

\begin{table*}
\centering
\caption{The fitting spectral ranges used in our work. In the last column we show the number of lines for that specific element in that range found in \textsc{vald-2} along the lines of other elements. Full version available online.}
\begin{tabular}{ccc}
\hline\hline
Element      &       Fitting range      &    $N_l$    \\\hline
11.00        &   5148.790 -- 5148.890   &      1      \\
11.00        &   5682.540 -- 5682.720   &      1      \\
11.00        &   5688.110 -- 5688.330   &      2      \\
11.00        &   6154.140 -- 6154.300   &      2      \\
11.00        &   6160.650 -- 6160.860   &      3      \\
...          &           ...            &     ...     \\\hline
\end{tabular}
\label{_table_lines}
\end{table*}

\begin{table*}
\centering
\caption{Iron abundances in the atmospheres of common stars. The metallicity values are given in the solar scales adopted by the respective authors, therefore cannot be compared directly. We also included the overlapping stars from the results of the automatic pipeline procedure by \protect\cite{kord13}.}
\label{_table_all_common_stars}
\begin{tabular}{ccc|ccc}
\hline\hline
Name       &     \Teff/\logg/[Fe/H]   &     Reference     &    Name      &     \Teff/\logg/[Fe/H]   &     Reference     \\\hline
HD 8389    &       5378/4.50/0.47     &   \cite{sous06}   &   HD 150936  &       5542/4.13/0.16     &   \cite{jenk08}   \\
           &       5283/4.37/0.34     &   \cite{sous08}   &              &       5692/4.40/0.24     &   \cite{bens14}   \\
           &       5243/4.46/0.40     &   \cite{jenk08}   &              &       5692/4.40/0.24     &   \cite{batt15}   \\
           &       5283/4.37/0.34     &   \cite{adib12}   &              &       5542/4.12/-0.03    &    This paper     \\
           &       5182/4.33/0.36     &   \cite{tsan13}   &              &                          &                   \\
           &       5222/4.44/0.45     &   \cite{mann13}   &   HD 152079  &       5785/4.38/0.32     &   \cite{jenk08}   \\
           &       5283/4.37/0.34     &   \cite{delg15}   &              &       5785/4.38/0.29     &   \cite{sant13}   \\
           &       5224/4.31/0.43     &   \cite{brew16}   &              &       5726/4.35/0.16     &    This paper     \\
           &       5243/4.52/0.32     &    This paper     &              &                          &                   \\
HD 13147   &       5502/3.99/0.16     &   \cite{jenk08}   &   HD 154672  &       5655/4.15/0.21     &   \cite{jenk08}   \\
           &       5352/3.47/-0.17    &   \cite{kord13}   &              &       5743/4.27/0.25     &   \cite{sant13}   \\
           &       5502/3.94/0.03     &    This paper     &              &       5655/4.16/0.10     &    This paper     \\
HD 23398   &       5592/4.09/0.44     &   \cite{jenk08}   &   HD 165204  &       5557/4.35/0.30     &   \cite{jenk08}   \\
           &       5438/4.03/0.31     &   \cite{kord13}   &              &       5637/4.37/0.28     &   \cite{bens14}   \\
           &       5592/4.10/0.38     &    This paper     &              &       5637/4.40/0.28     &   \cite{batt15}   \\
           &                          &                   &              &       5557/4.33/0.17     &    This paper     \\
HD 38467   &       5721/4.18/0.22     &   \cite{jenk08}   &   HD 170706  &       5698/4.17/0.24     &   \cite{jenk08}   \\
           &       5753/4.15/0.24     &   \cite{brew16}   &              &       5718/4.31/0.22     &   \cite{bens14}   \\
           &       5721/4.18/0.10     &    This paper     &              &       5718/4.33/0.22     &   \cite{batt15}   \\
           &                          &                   &              &       5698/4.40/0.13     &    This paper     \\
HD 40293   &       5549/4.33/0.13     &   \cite{jenk08}   &   HD 185679  &       5681/4.34/0.14     &   \cite{jenk08}   \\
           &       5518/4.39/0.03     &   \cite{kord13}   &              &       5724/4.08/0.12     &   \cite{kord13}   \\
           &       5549/4.51/0.00     &    This paper     &              &       5710/4.47/0.06     &   \cite{bens14}   \\
HD 42719   &       5809/3.96/0.24     &   \cite{jenk08}   &              &       5710/4.50/0.06     &   \cite{batt15}   \\
           &       5962/4.36/0.23     &   \cite{kord13}   &              &       5681/4.43/0.01     &    This paper     \\
           &       5809/4.08/0.11     &    This paper     &              &                          &                   \\
HD 48265   &       5651/3.92/0.29     &   \cite{jenk08}   &   HD 186194  &       5668/4.09/0.18     &   \cite{jenk08}   \\
           &       5798/3.95/0.36     &   \cite{sant13}   &              &       5713/4.16/0.20     &   \cite{bens14}   \\
           &       5789/4.09/0.38     &   \cite{jofr15}   &              &       5713/4.20/0.20     &   \cite{batt15}   \\
           &       5651/3.92/0.17     &    This paper     &              &       5668/4.30/0.07     &    This paper     \\
HD 66653   &       5771/4.40/0.15     &   \cite{jenk08}   &   HD 190125  &       5644/4.20/0.22     &   \cite{jenk08}   \\
           &       5809/4.42/0.09     &   \cite{dats15}   &              &       5682/4.48/0.17     &   \cite{bens14}   \\
           &       5771/4.42/-0.05    &    This paper     &              &       5682/4.50/0.17     &   \cite{batt15}   \\
           &                          &                   &              &       5644/4.53/0.04     &    This paper     \\
HD 77338   &       5290/4.90/0.22     &   \cite{felt98}   &   HD 193690  &       5558/4.58/0.24     &   \cite{jenk08}   \\
           &       5290/4.60/0.30     &   \cite{thor00}   &              &       5542/4.41/0.20     &   \cite{brew16}   \\
           &       5315/4.55/0.10     &   \cite{jenk08}   &              &       5558/4.48/0.15     &    This paper     \\
           &       5300/4.30/0.36     &   \cite{pruq11}   &   HD 194490  &       5854/4.46/0.28     &   \cite{jenk08}   \\
           &       5440/4.36/0.28     &   \cite{sant13}   &              &       5857/4.33/0.08     &   \cite{bens14}   \\
           &       5315/4.42/0.16     &    This paper     &              &       5857/4.30/0.08     &   \cite{batt15}   \\
           &                          &                   &              &       5854/4.44/-0.04    &    This paper     \\
HD 90520   &       5870/4.02/0.17     &   \cite{jenk08}   &   HD 201757  &       5597/4.02/0.15     &   \cite{jenk08}   \\
           &       6008/4.16/0.25     &   \cite{bens14}   &              &       5566/3.99/0.11     &   \cite{kord13}   \\
           &       6008/4.20/0.25     &   \cite{batt15}   &              &       5597/4.23/0.05     &    This paper     \\
           &       5870/4.08/0.06     &    This paper     &              &                          &                   \\
HD 107181  &       5581/4.01/0.26     &   \cite{jenk08}   &   HD 218960  &       5732/4.24/0.21     &   \cite{jenk08}   \\
           &       5628/4.05/0.31     &   \cite{brew16}   &              &       5796/4.09/0.20     &   \cite{bens14}   \\
           &       5581/4.17/0.22     &    This paper     &              &       5796/4.10/0.20     &   \cite{batt15}   \\
HD 126535  &       5284/4.61/0.13     &   \cite{jenk08}   &              &       5732/4.27/0.05     &    This paper     \\
           &       5305/4.46/0.07     &   \cite{tabe12}   &   HD 220981  &       5567/4.34/0.18     &   \cite{jenk08}   \\
           &       5284/4.65/0.10     &    This paper     &              &       5618/4.26/0.25     &   \cite{bens14}   \\
HD 143361  &       5505/4.44/0.06     &   \cite{jenk08}   &              &       5618/4.30/0.25     &   \cite{batt15}   \\
           &       5503/4.36/0.22     &   \cite{sant13}   &              &       5567/4.33/0.11     &    This paper     \\
           &       5505/4.42/0.18     &    This paper     &              &                          &                   \\\hline
\end{tabular}                                              
\end{table*}

\begin{figure*}
\centering
\includegraphics[width=55mm]{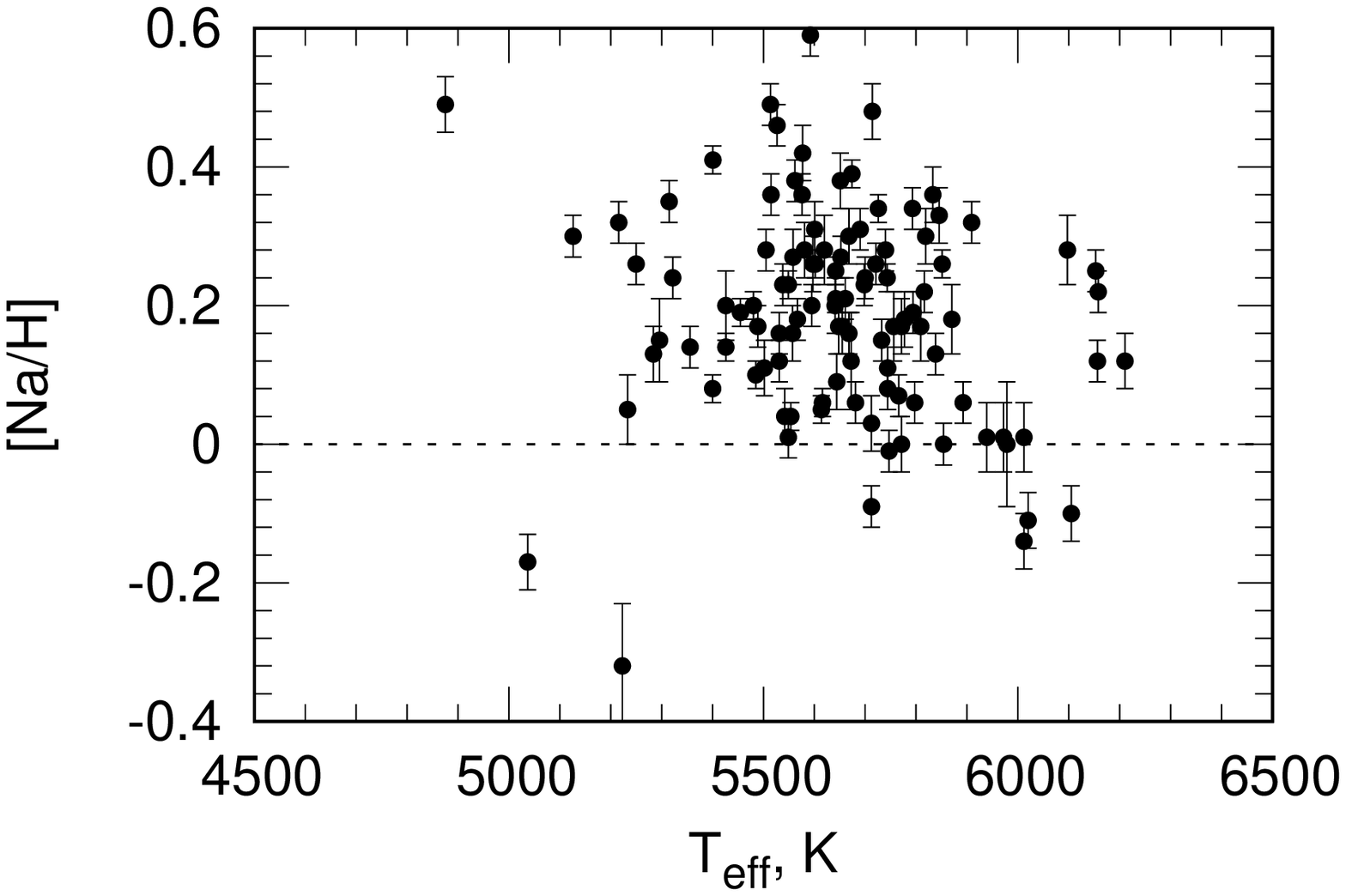} 
\includegraphics[width=55mm]{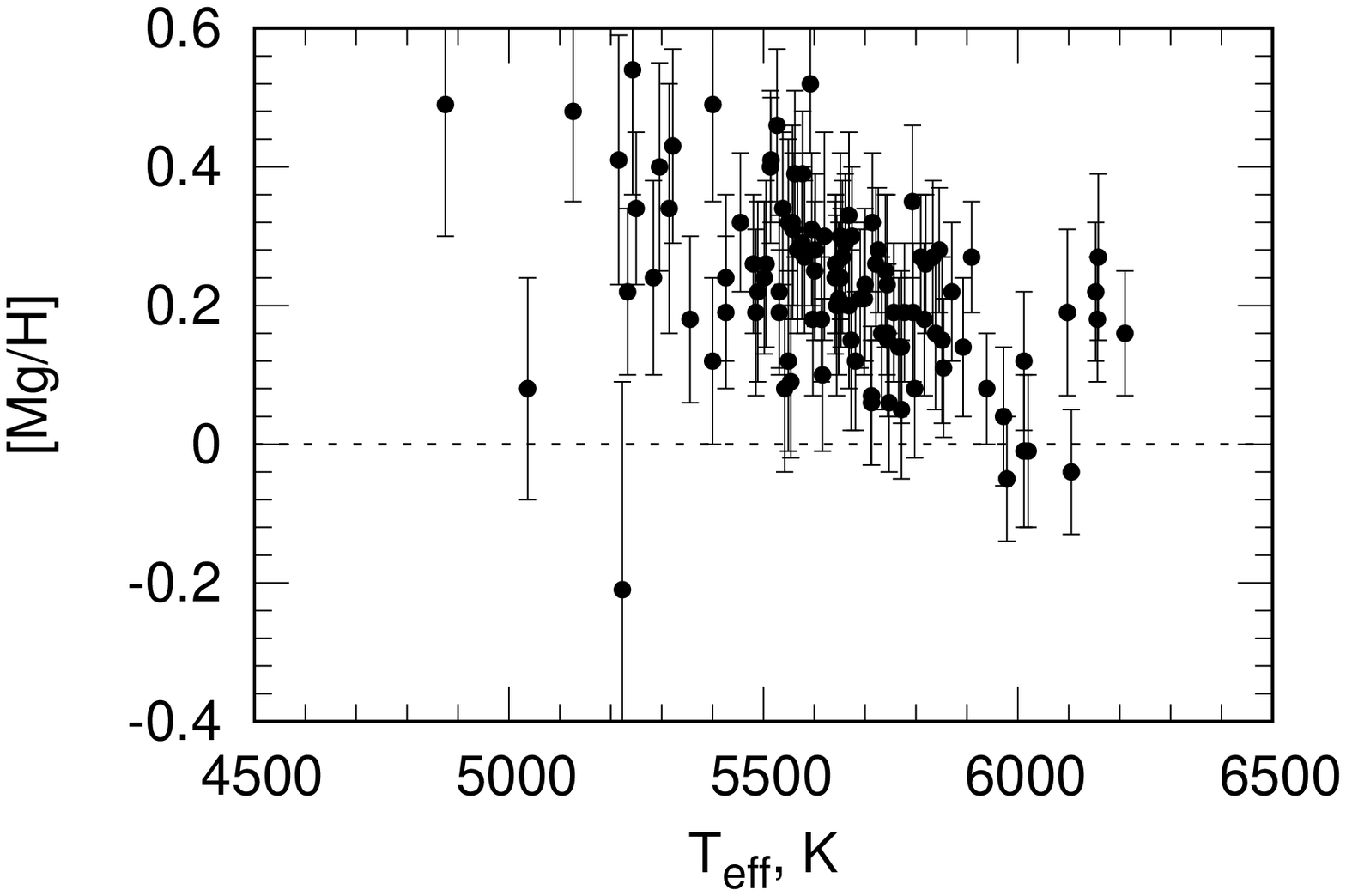} 
\includegraphics[width=55mm]{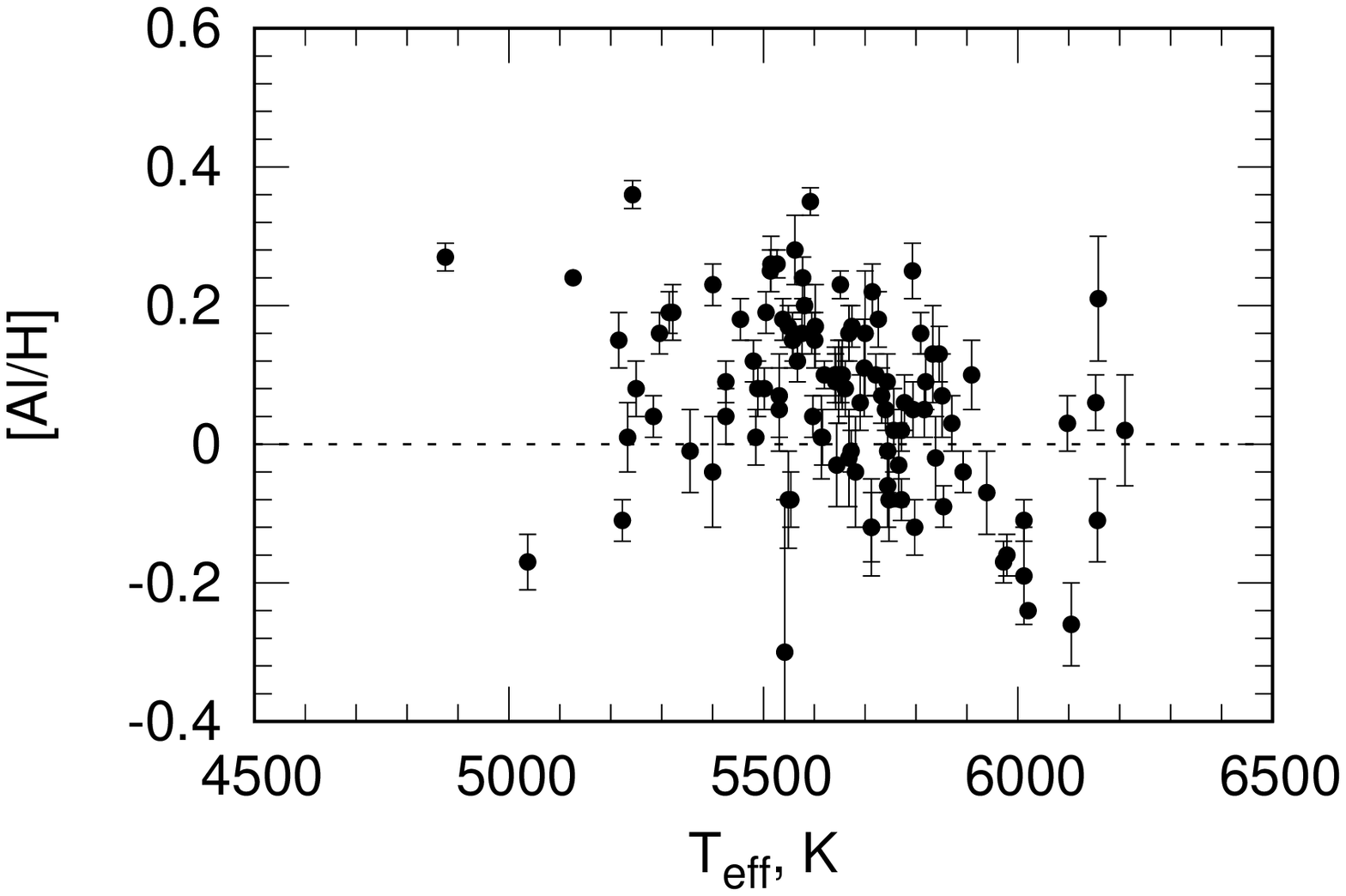} 
\includegraphics[width=55mm]{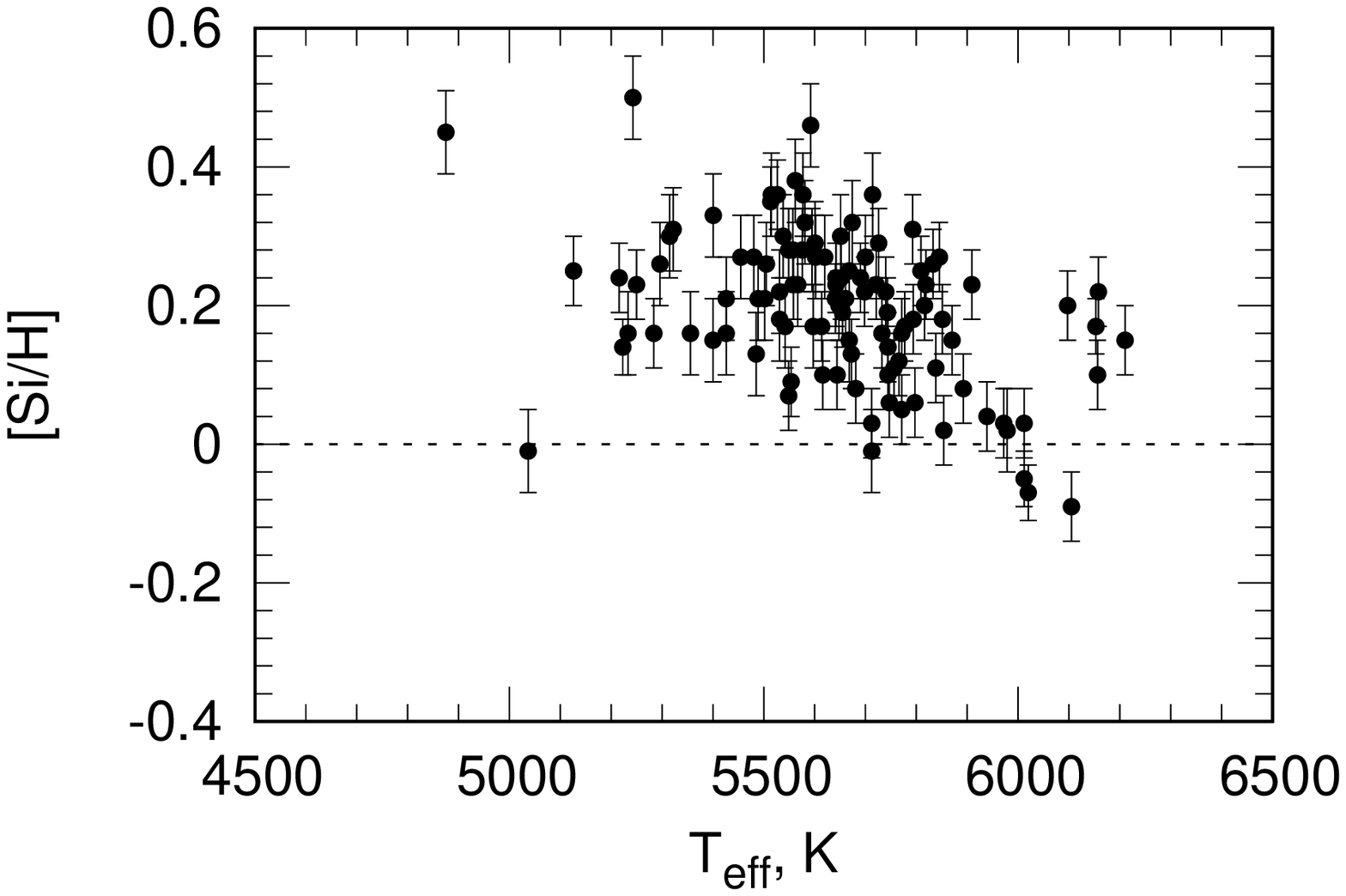} 
\includegraphics[width=55mm]{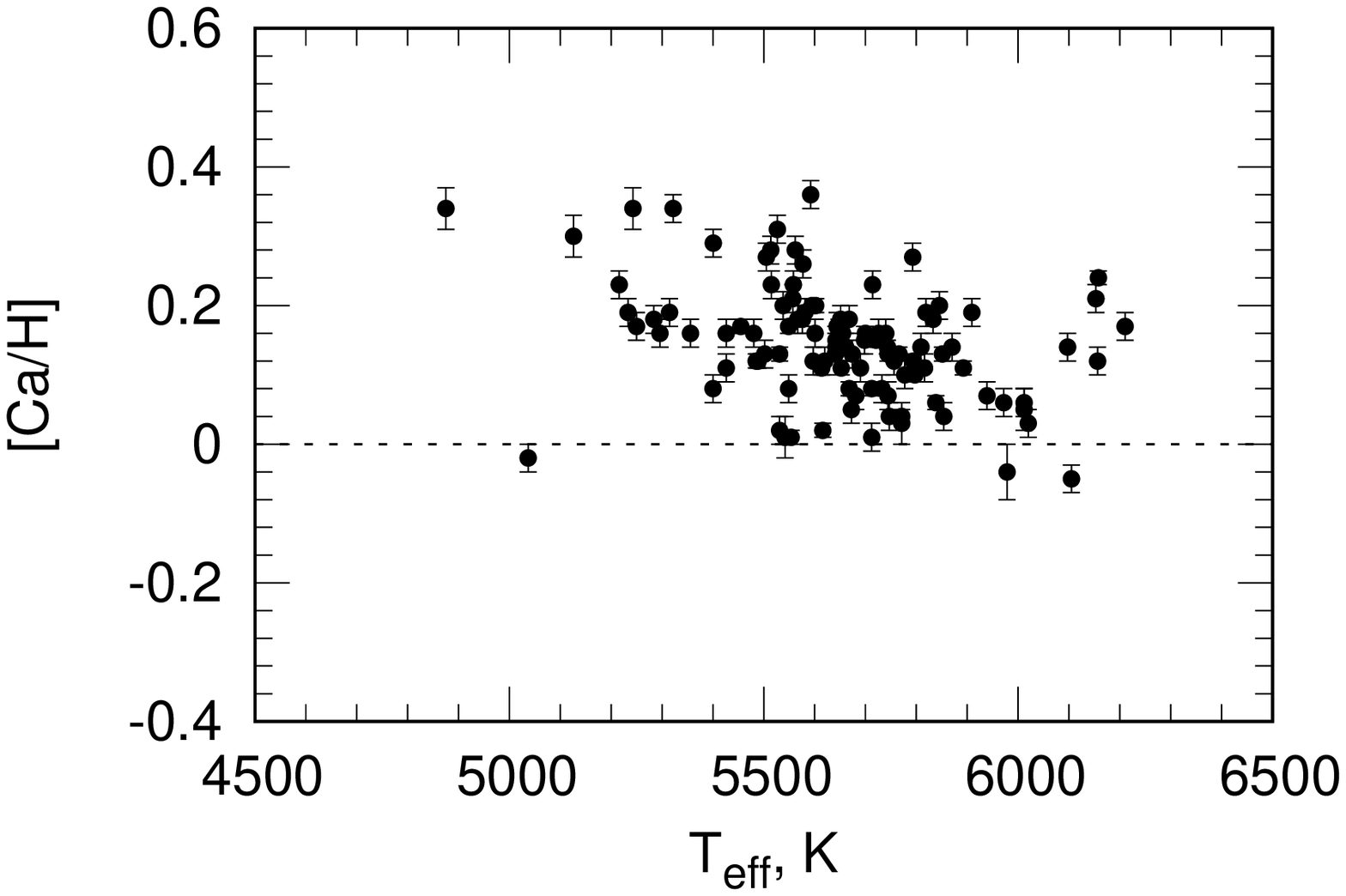} 
\includegraphics[width=55mm]{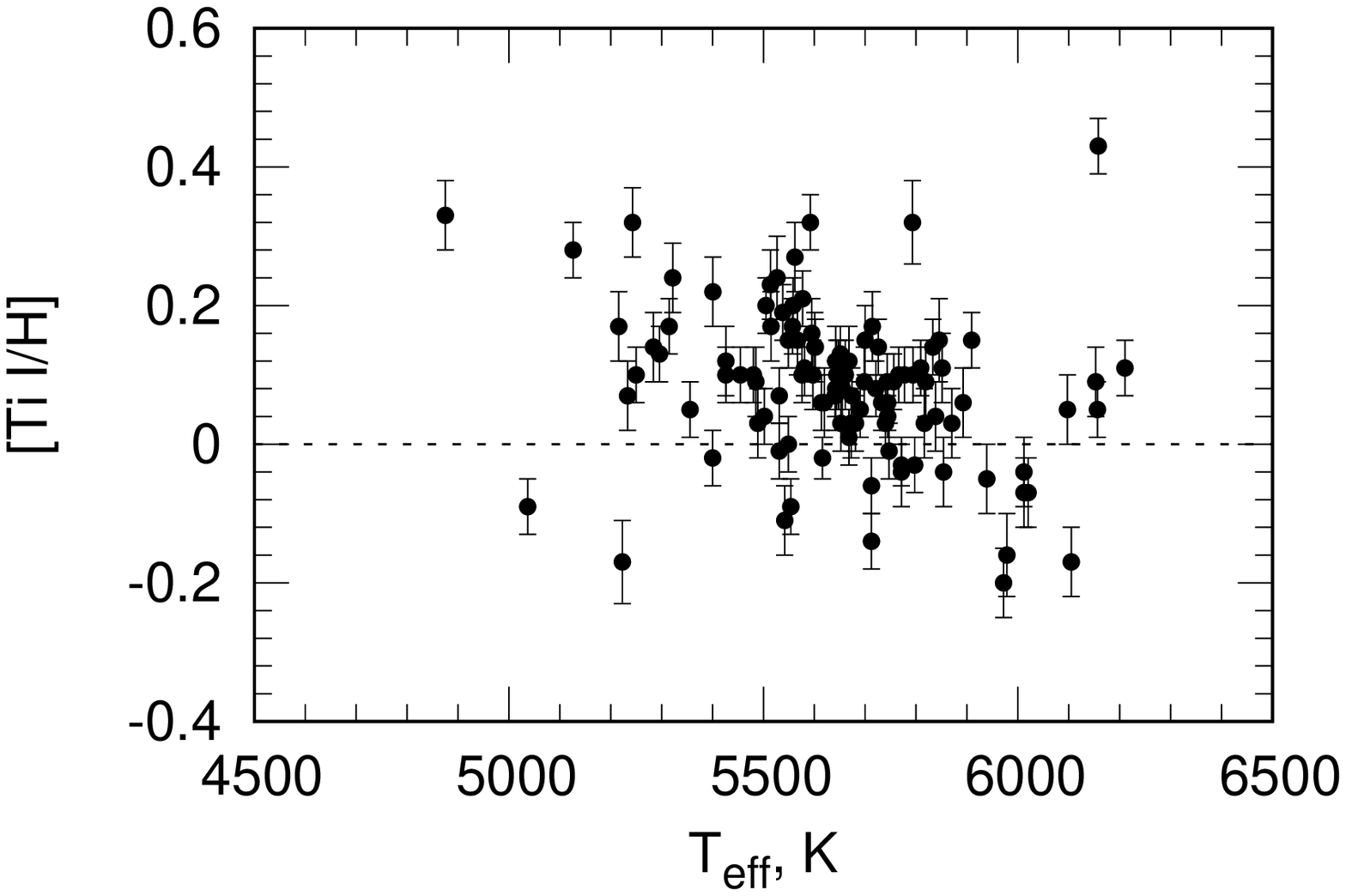} 
\includegraphics[width=55mm]{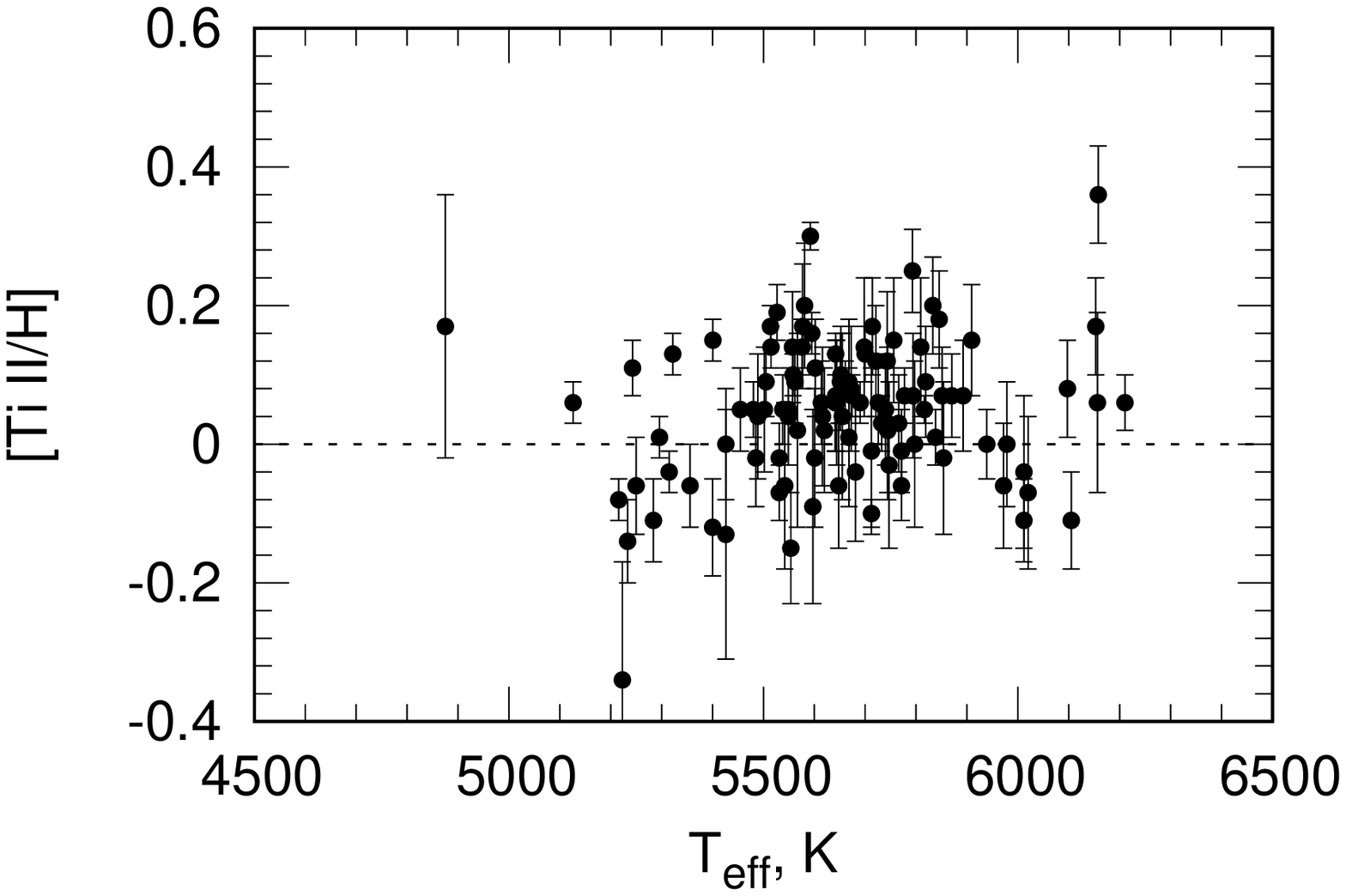} 
\includegraphics[width=55mm]{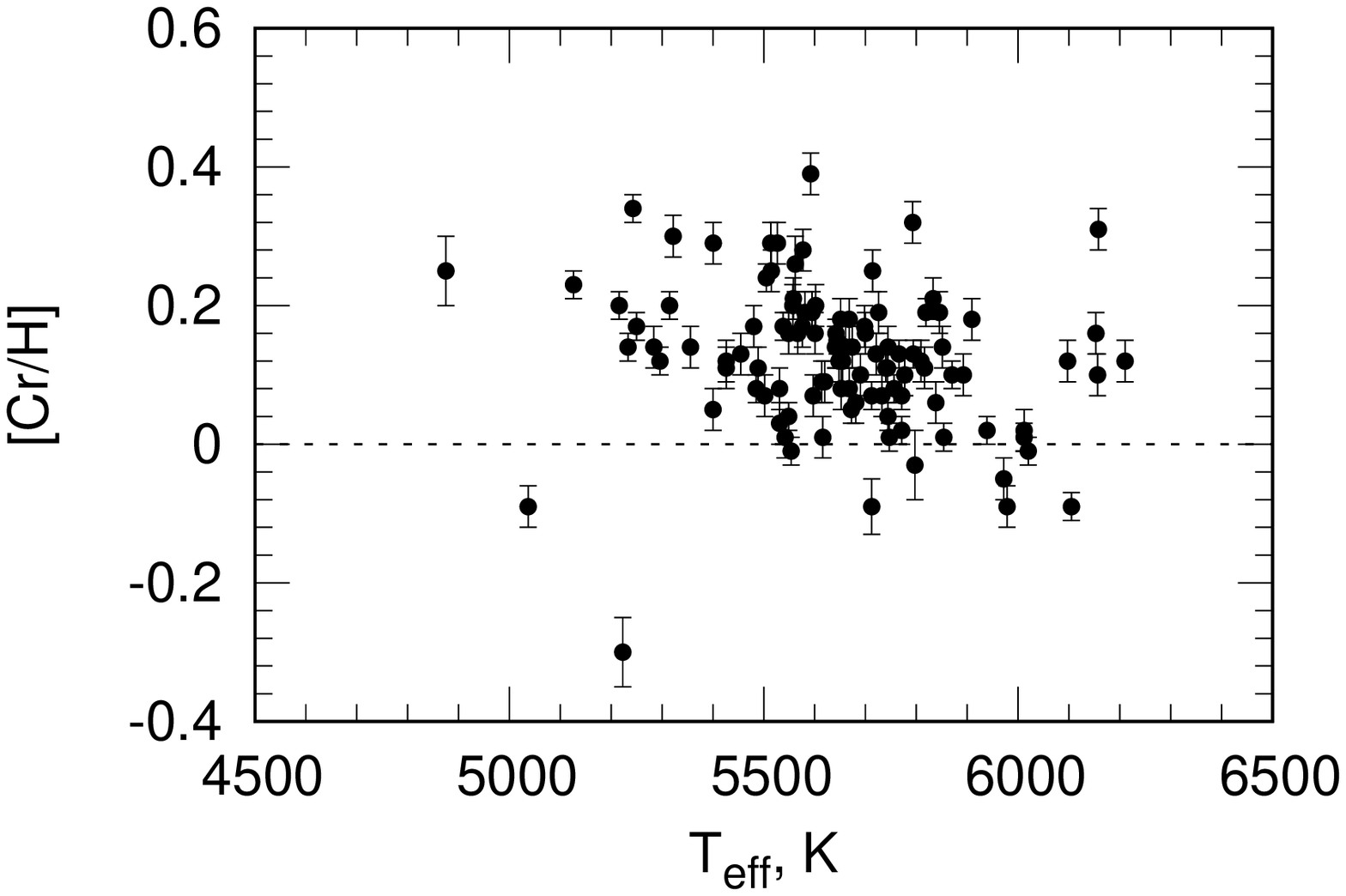} 
\includegraphics[width=55mm]{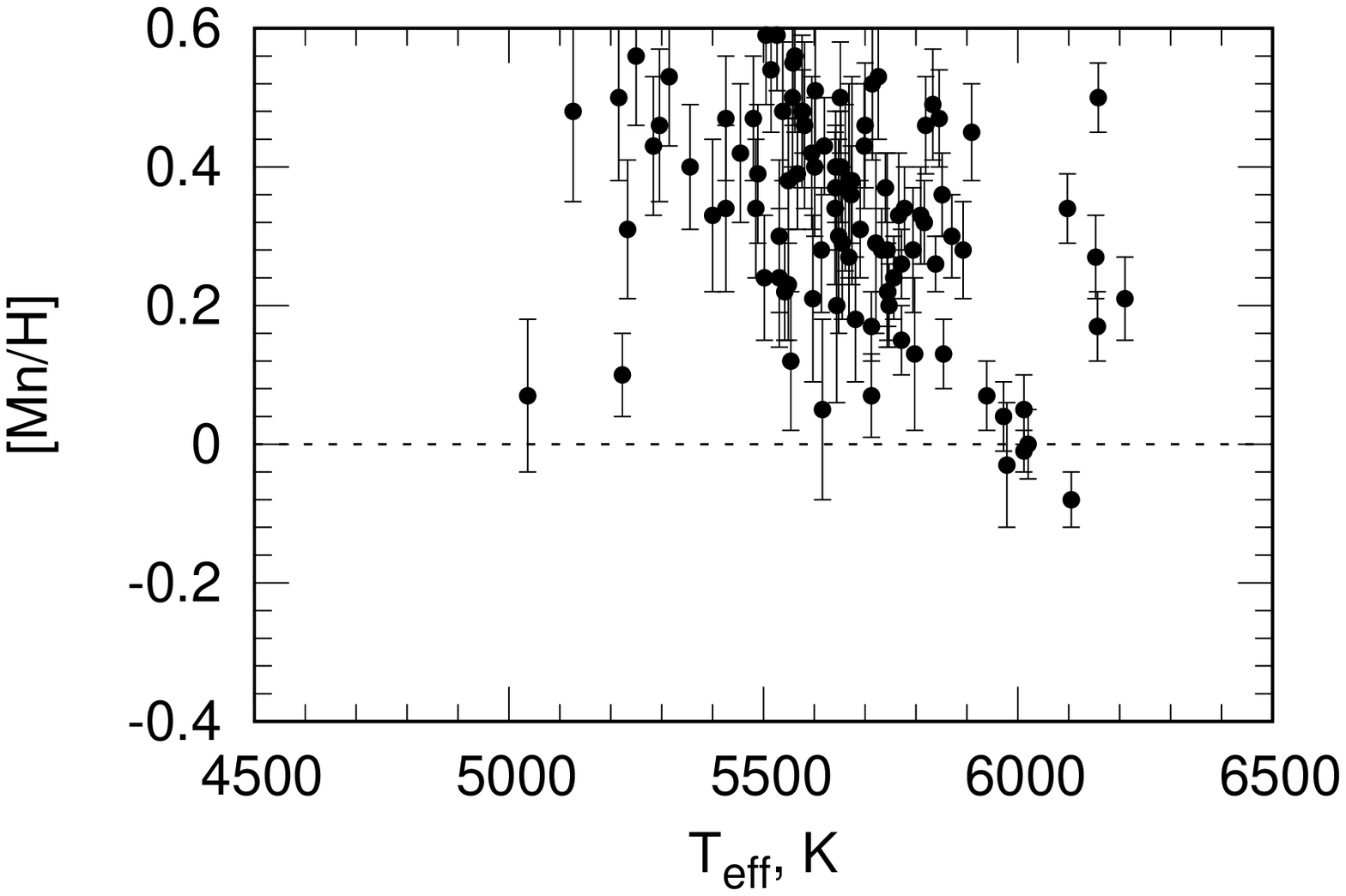} 
\includegraphics[width=55mm]{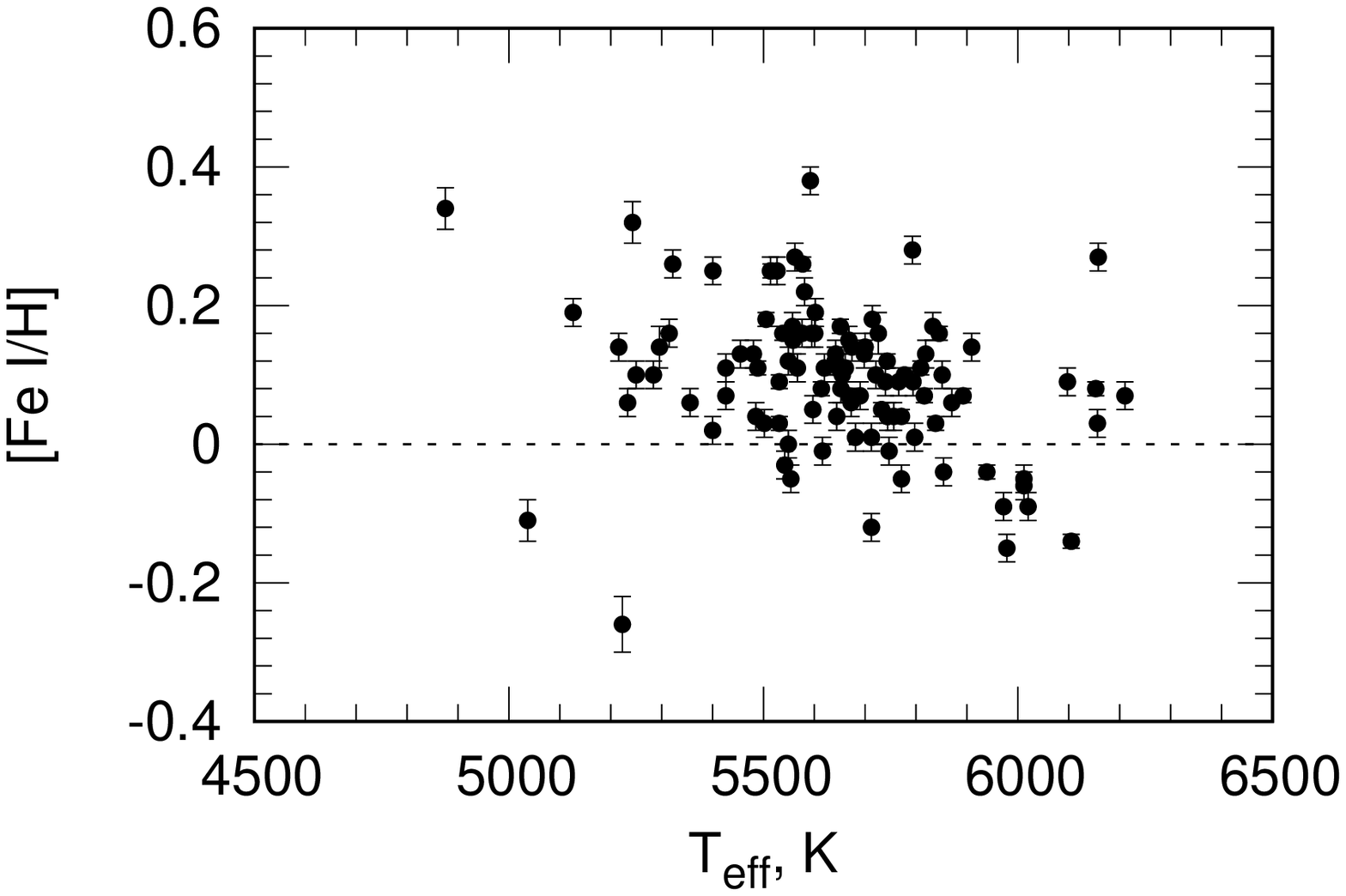} 
\includegraphics[width=55mm]{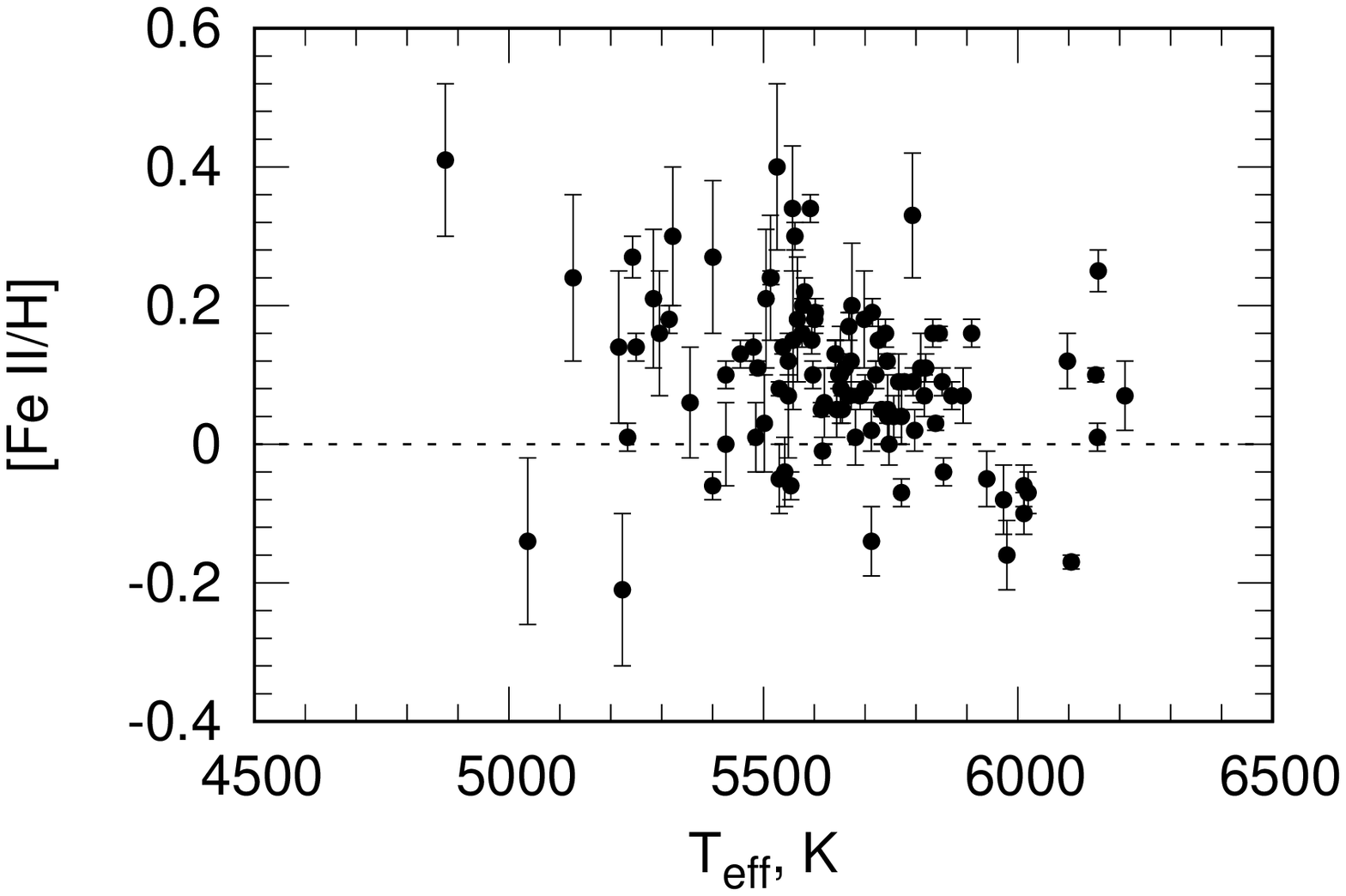} 
\includegraphics[width=55mm]{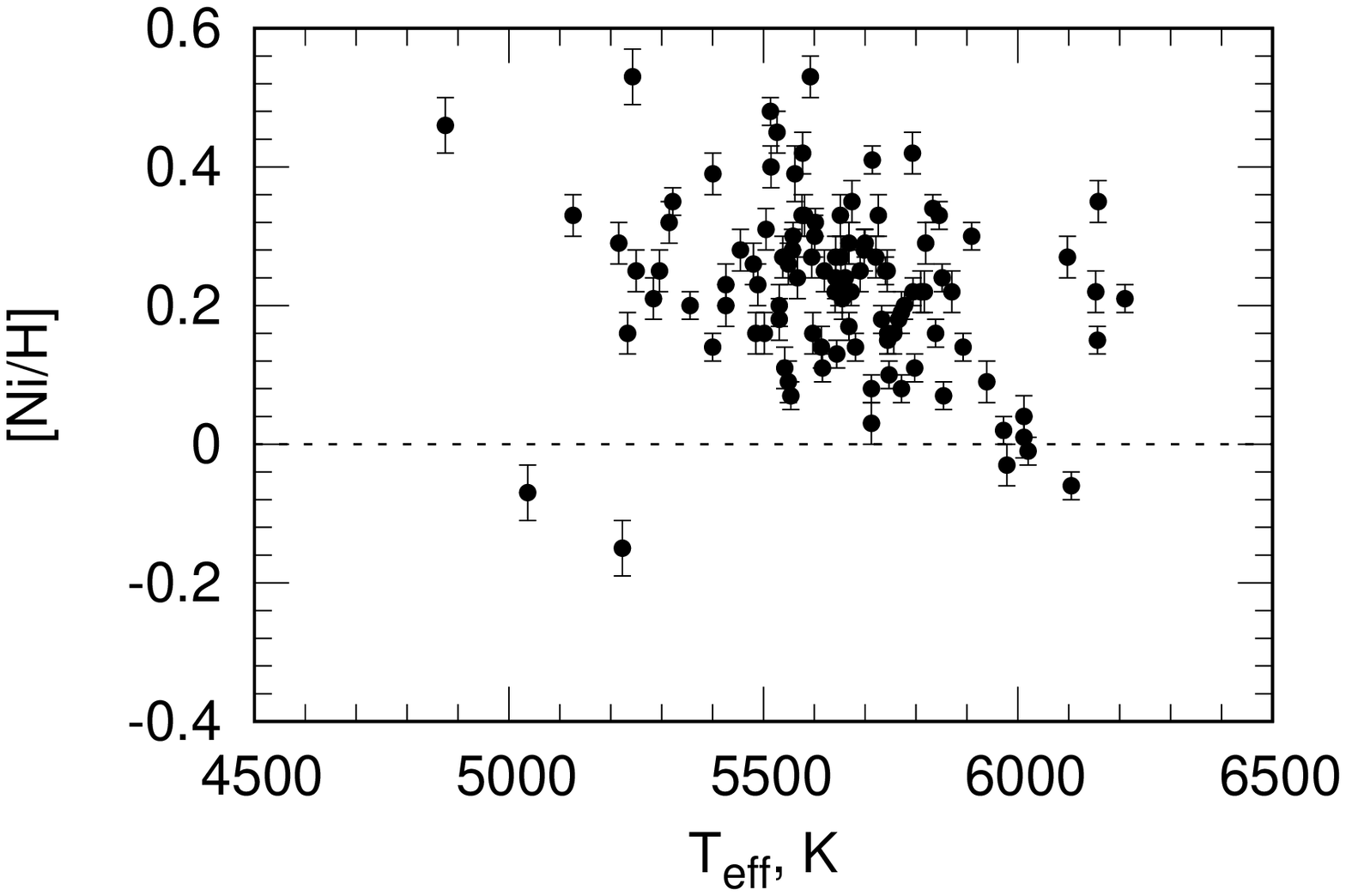} 
\includegraphics[width=55mm]{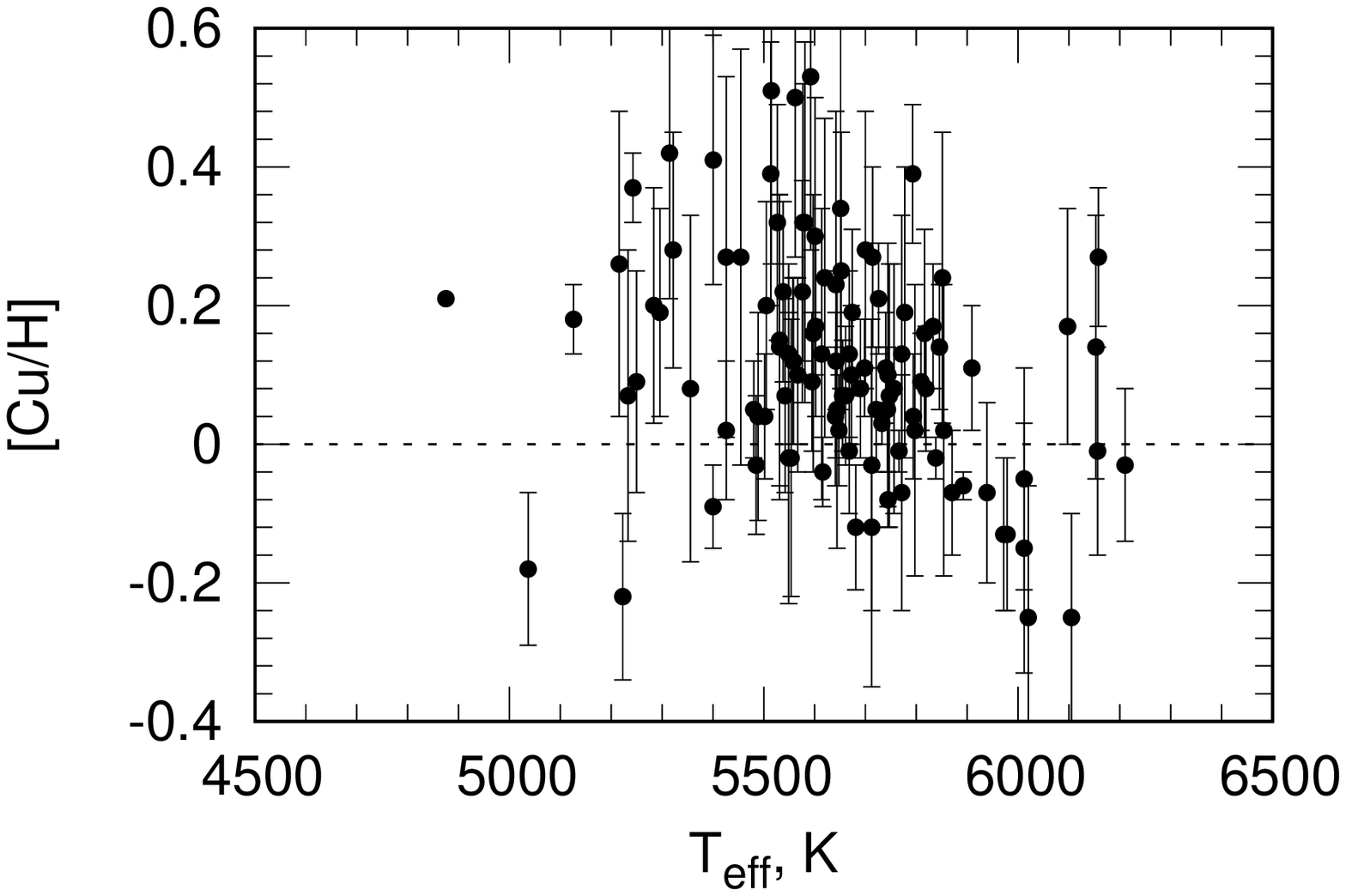} 
\includegraphics[width=55mm]{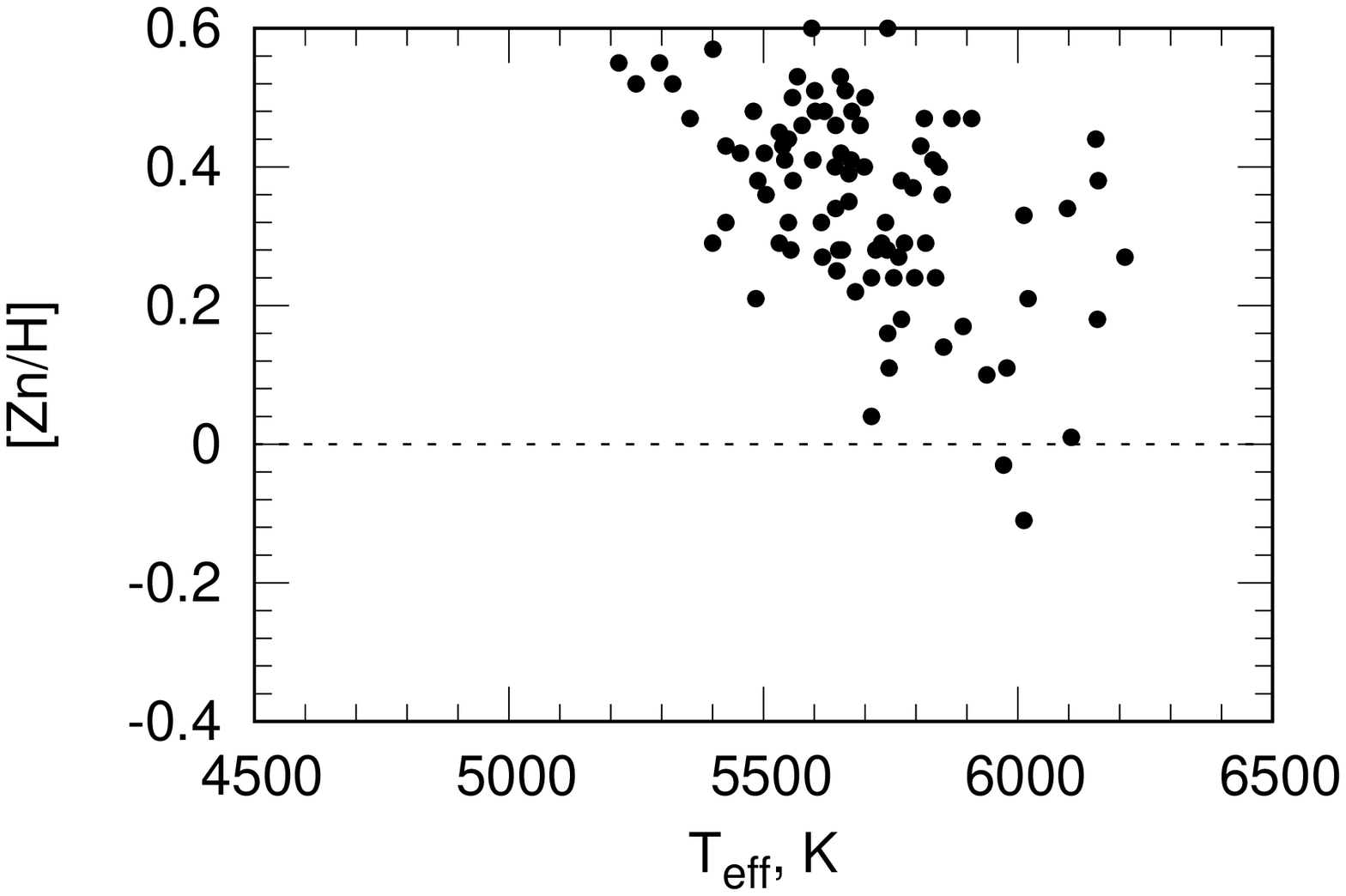} 
\caption{The dependencies of [X/H] versus \Teff\, in our sample.}
\label{_figure_temperature_dependence}
\end{figure*}

\begin{figure*}
\centering
\includegraphics[width=55mm]{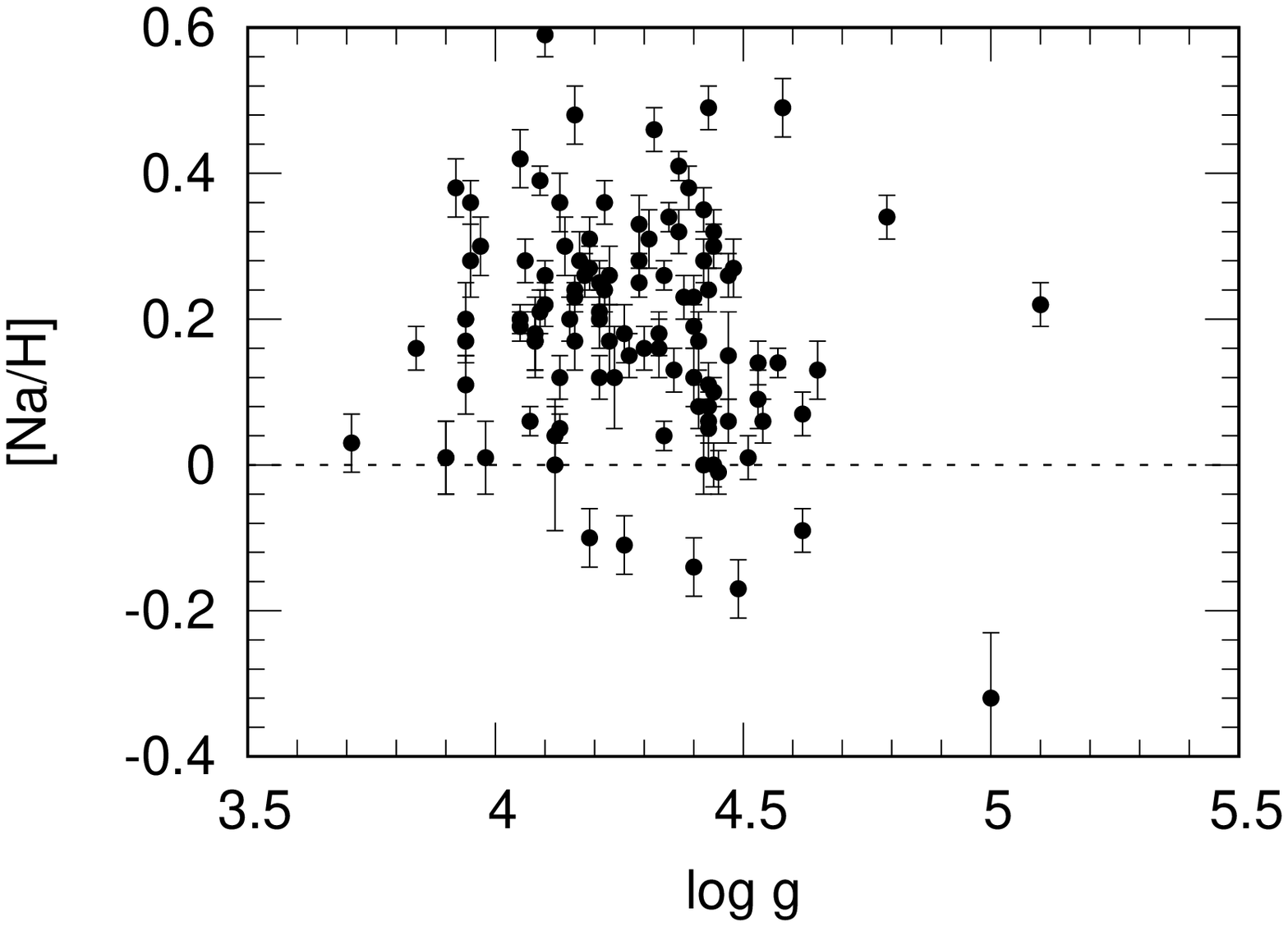} 
\includegraphics[width=55mm]{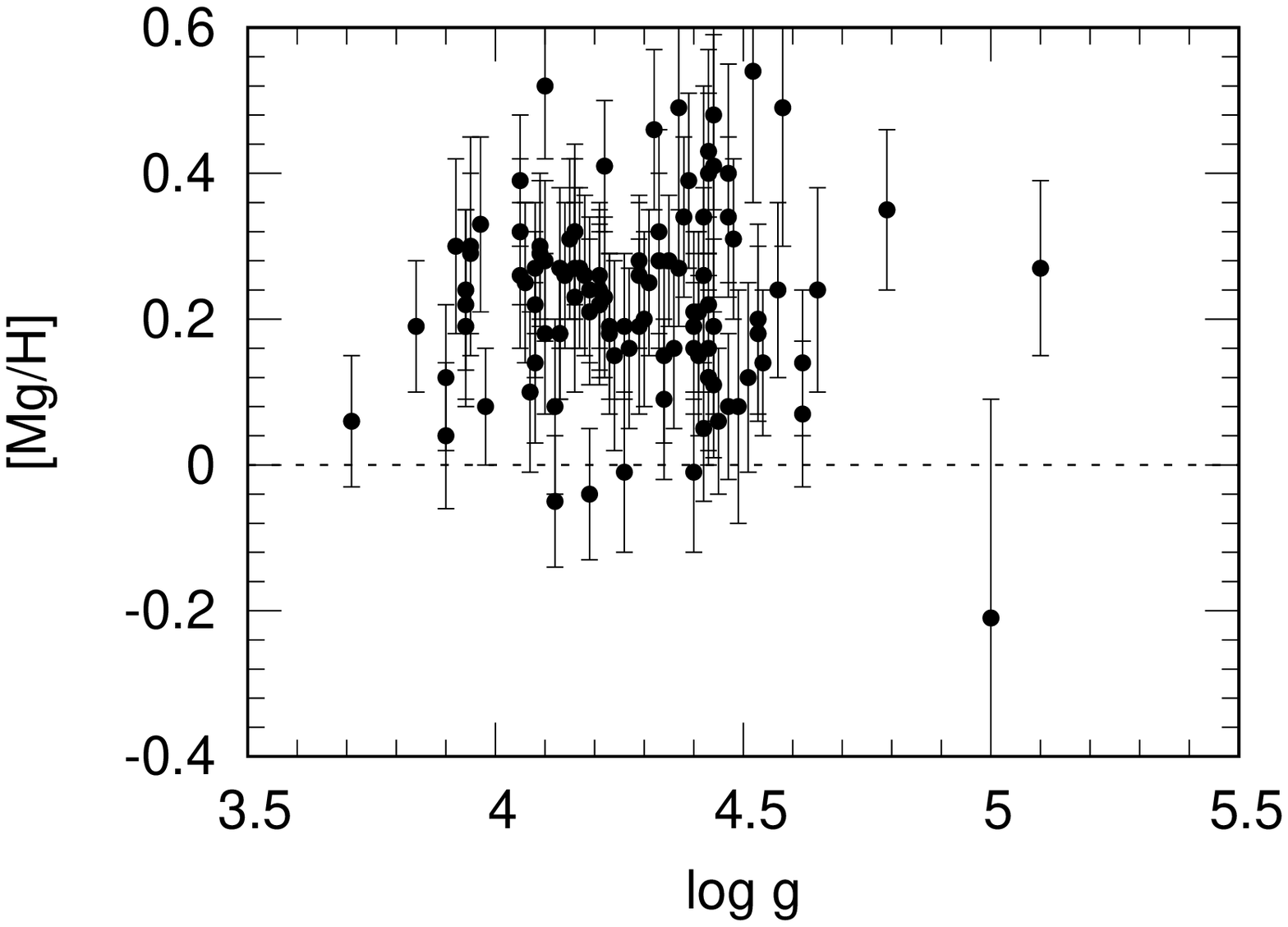} 
\includegraphics[width=55mm]{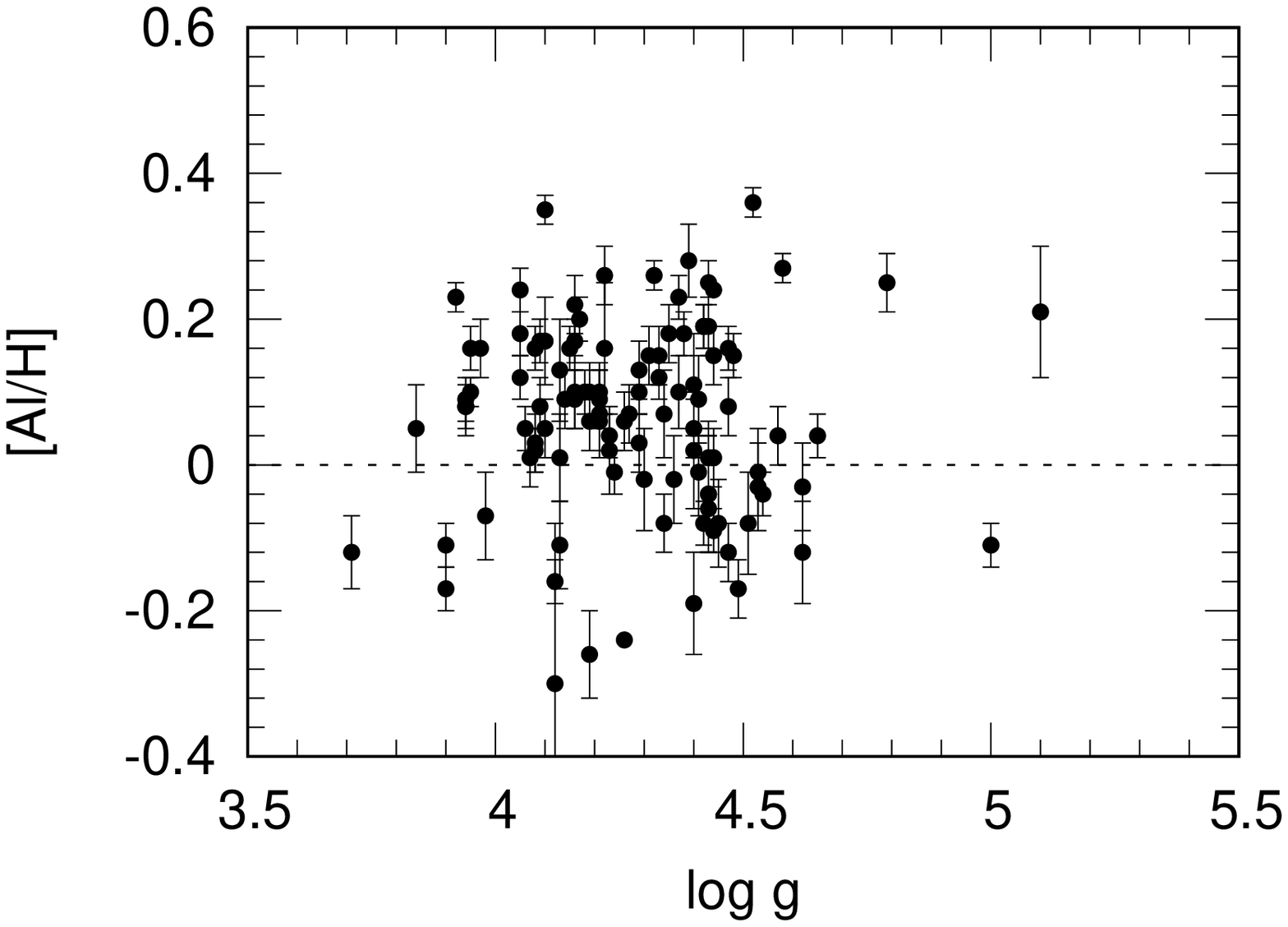} 
\includegraphics[width=55mm]{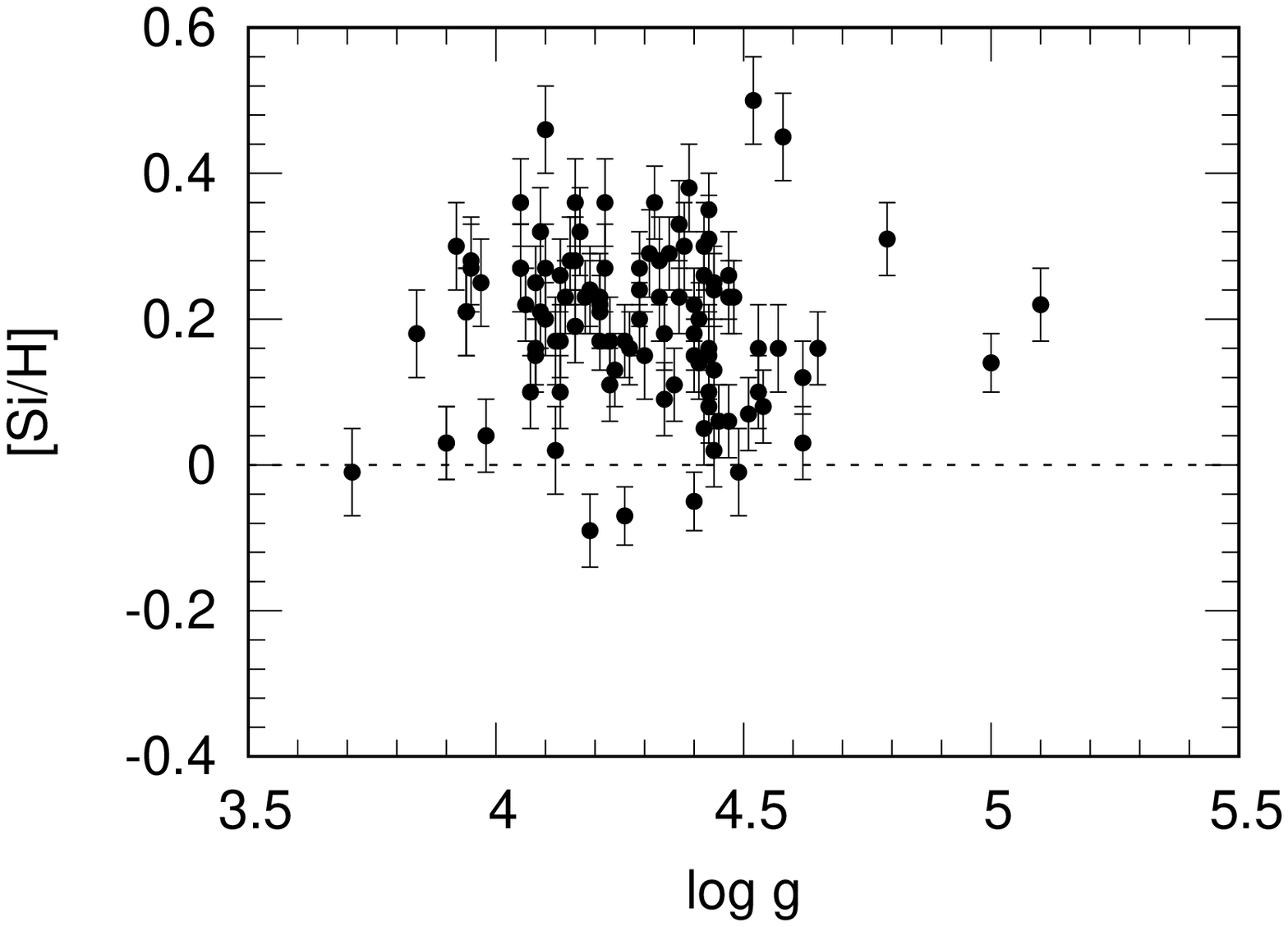} 
\includegraphics[width=55mm]{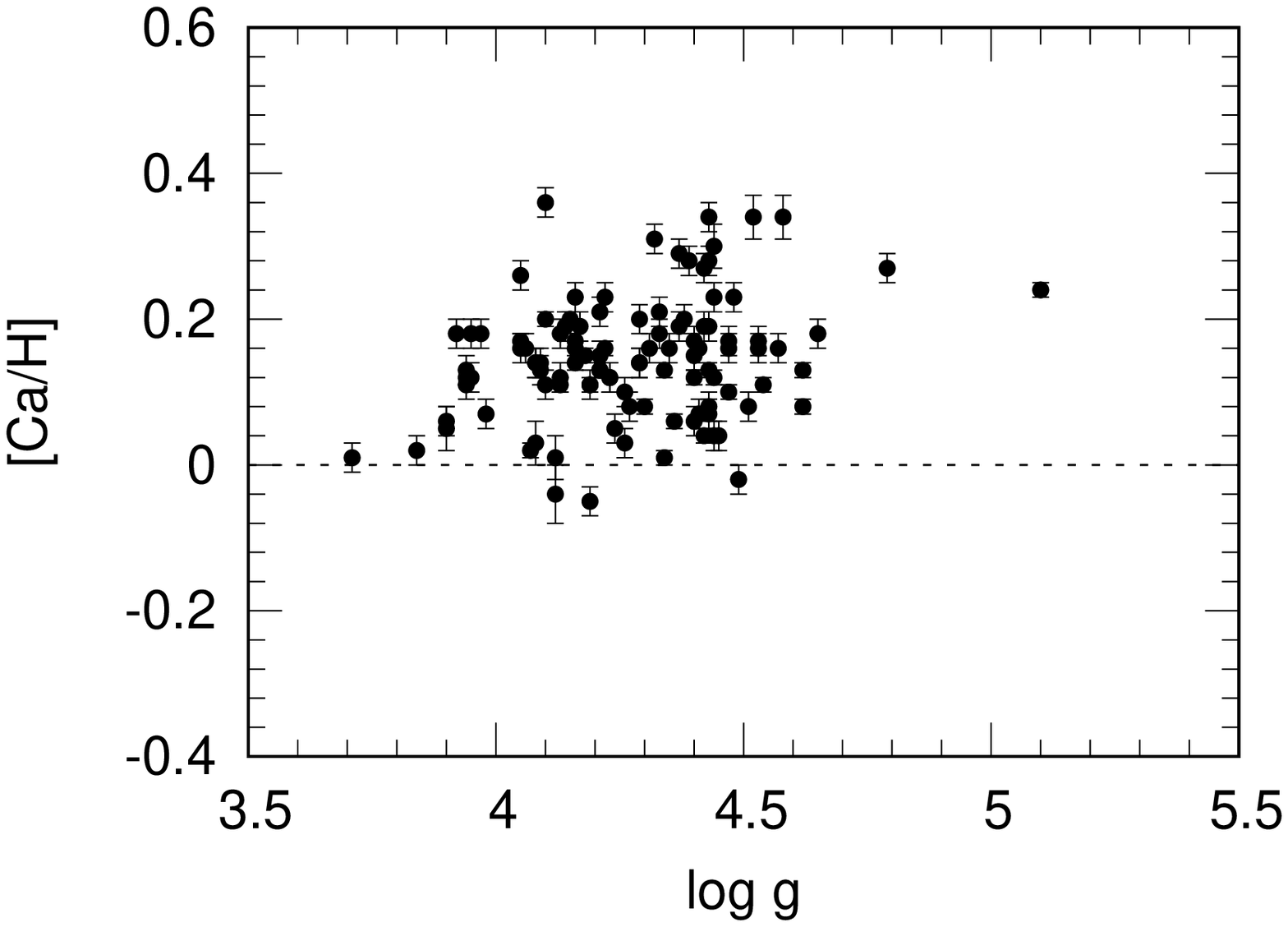} 
\includegraphics[width=55mm]{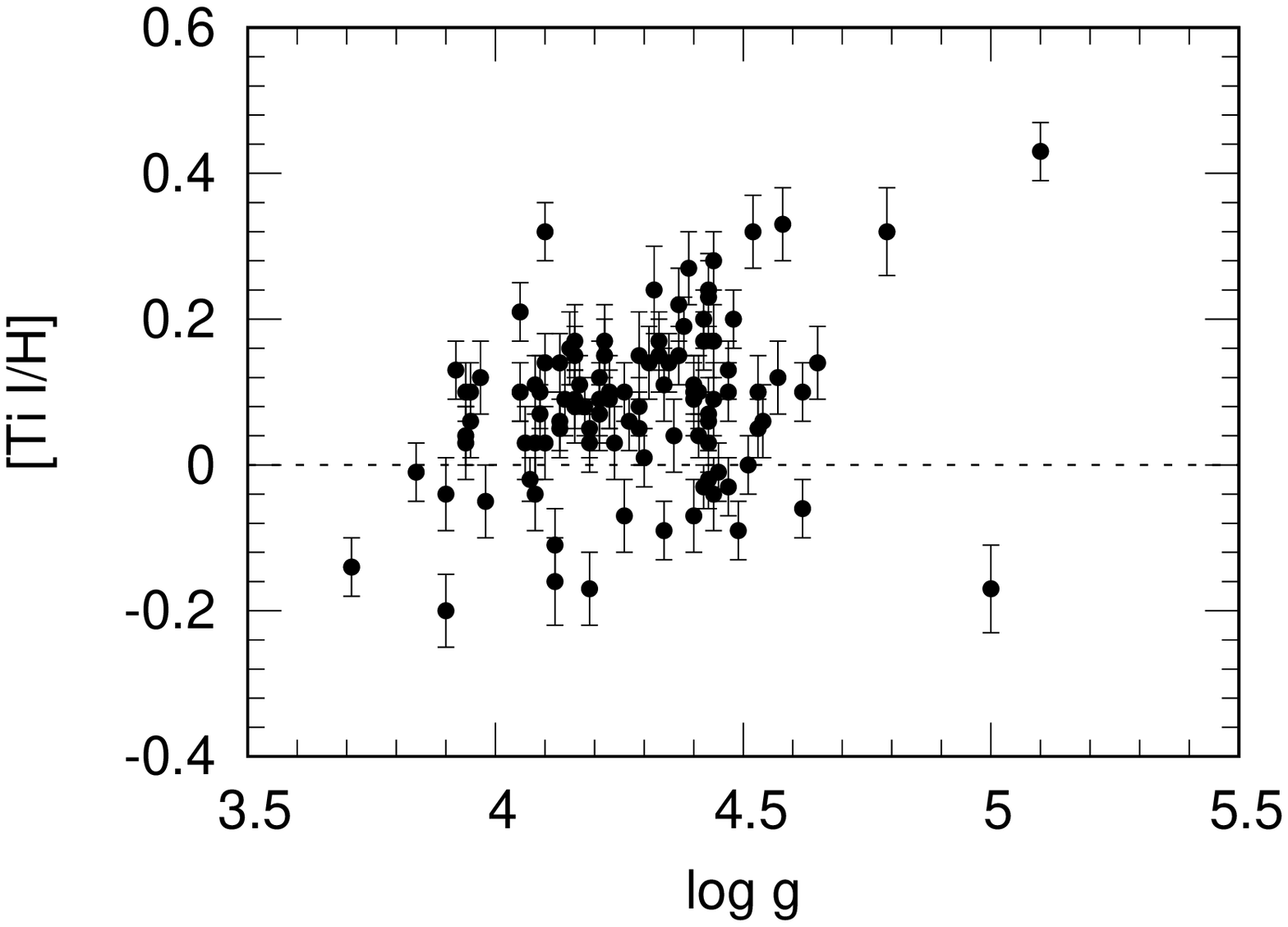} 
\includegraphics[width=55mm]{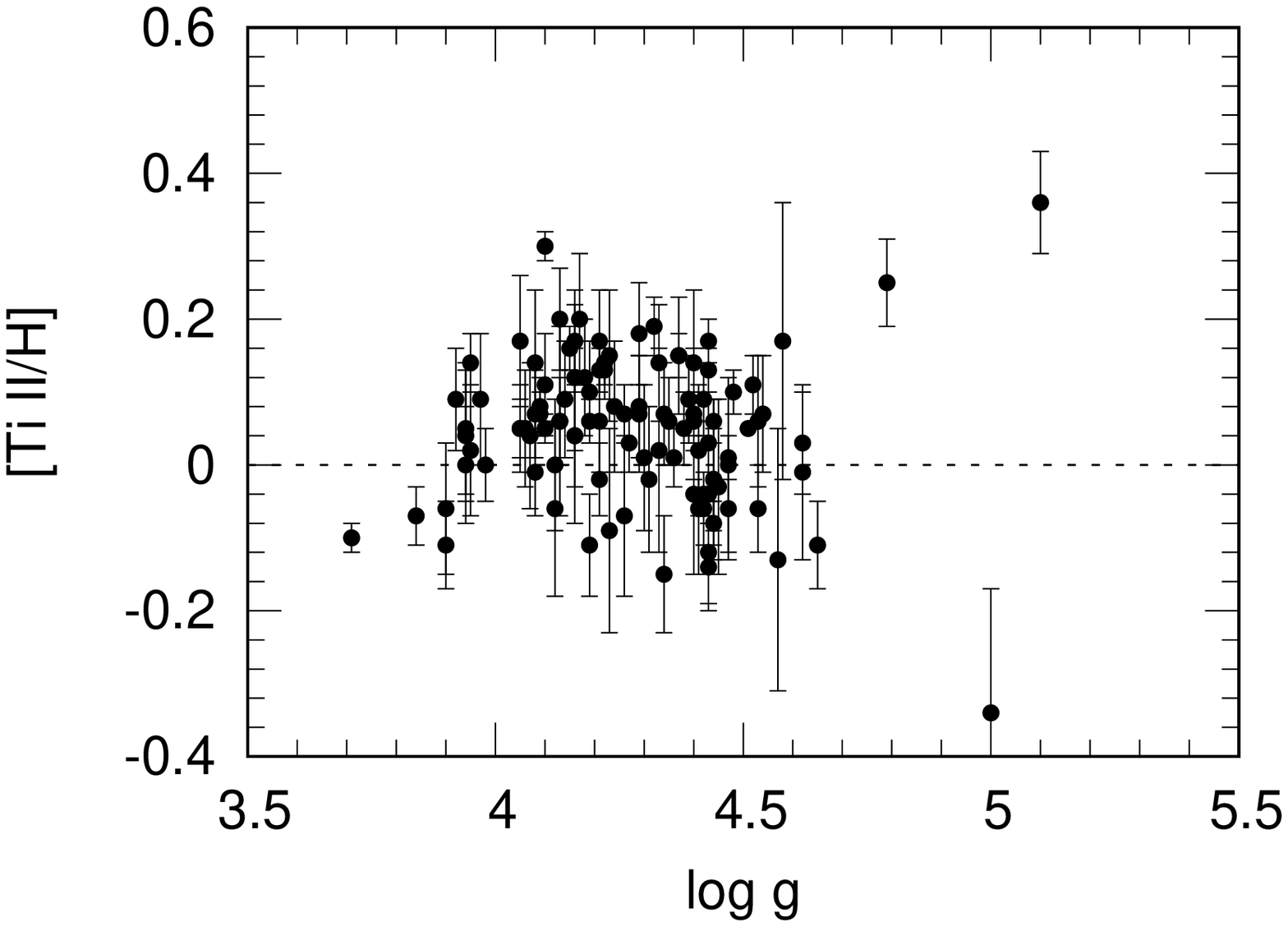} 
\includegraphics[width=55mm]{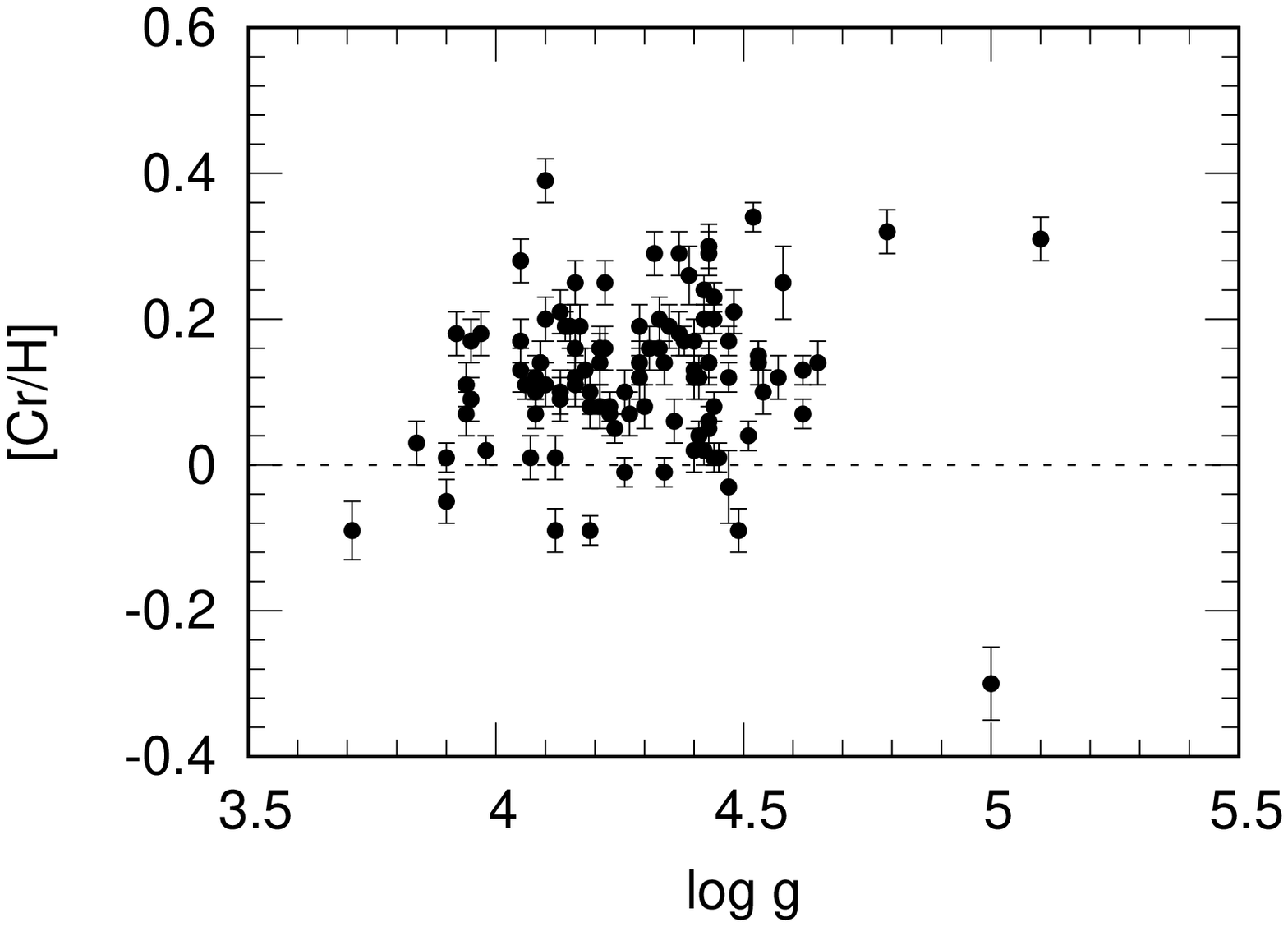} 
\includegraphics[width=55mm]{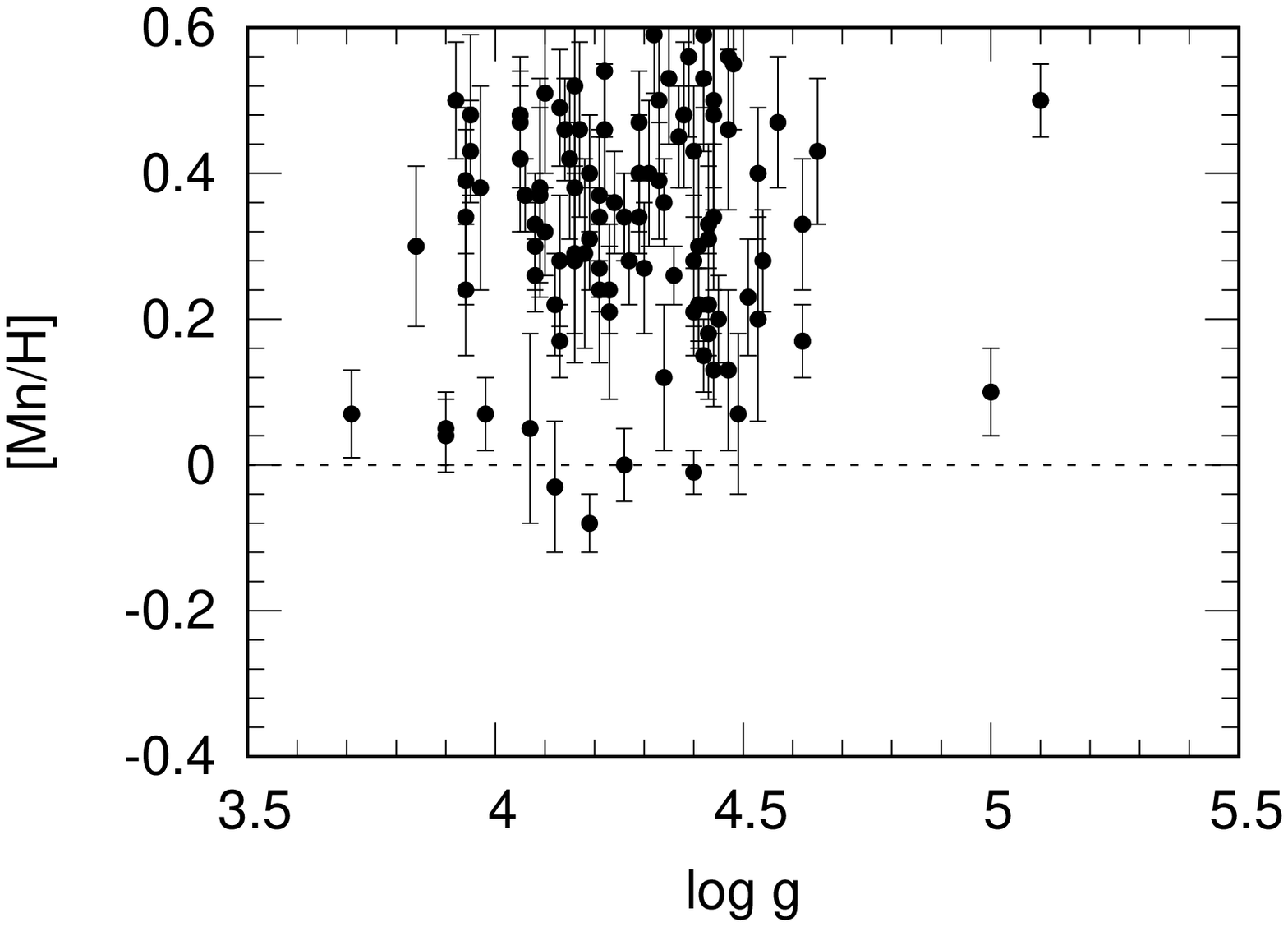} 
\includegraphics[width=55mm]{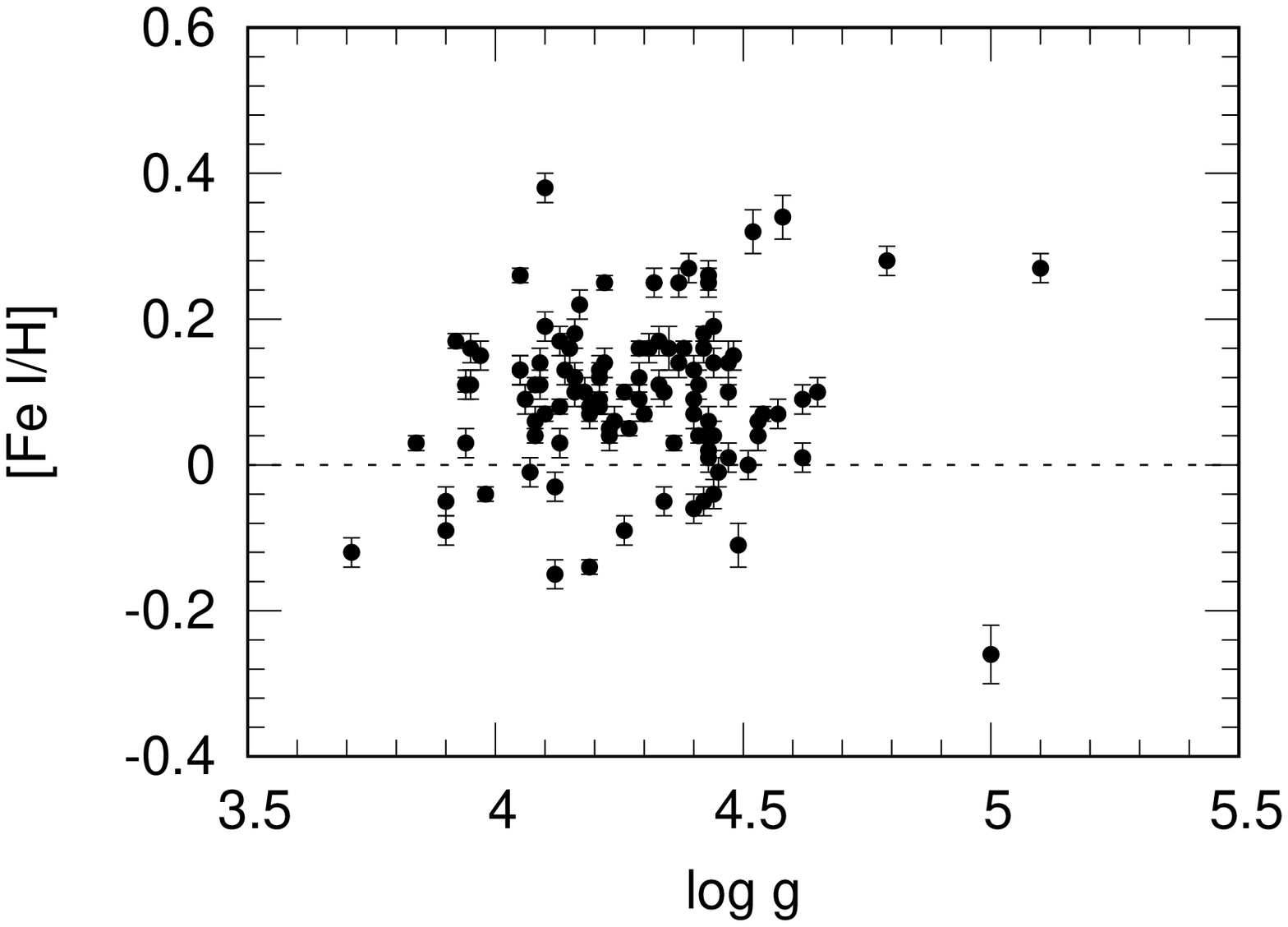} 
\includegraphics[width=55mm]{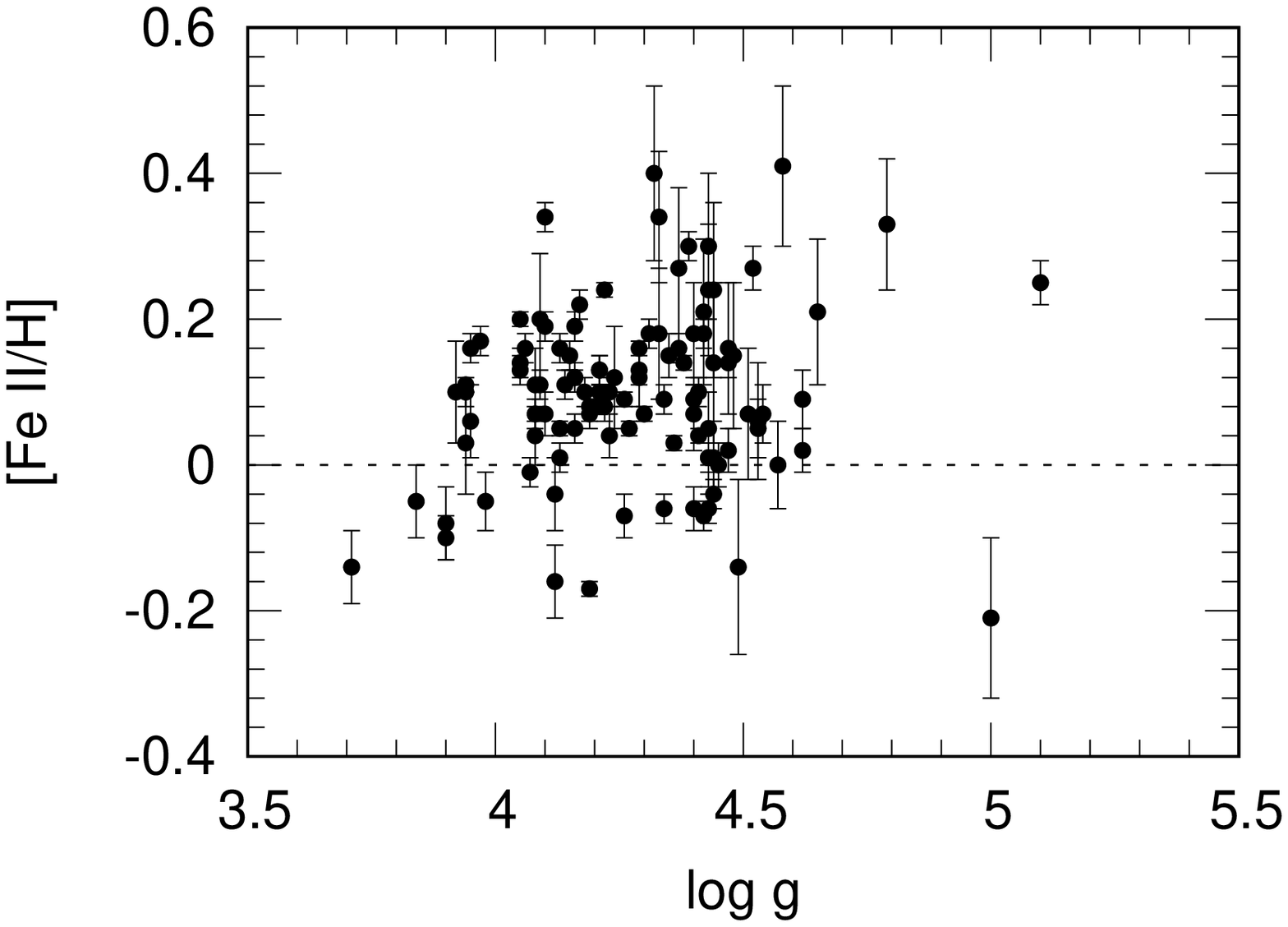} 
\includegraphics[width=55mm]{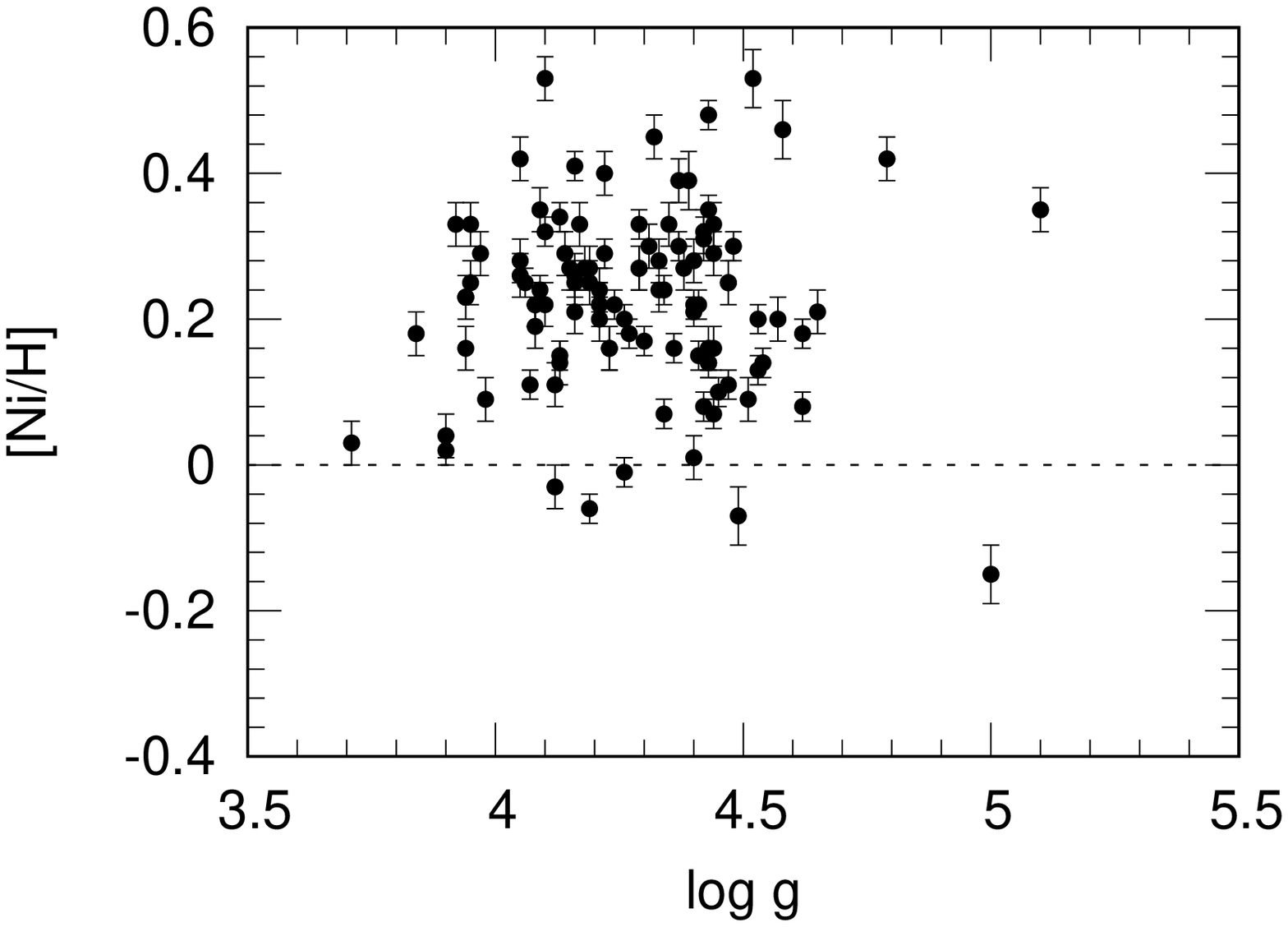} 
\includegraphics[width=55mm]{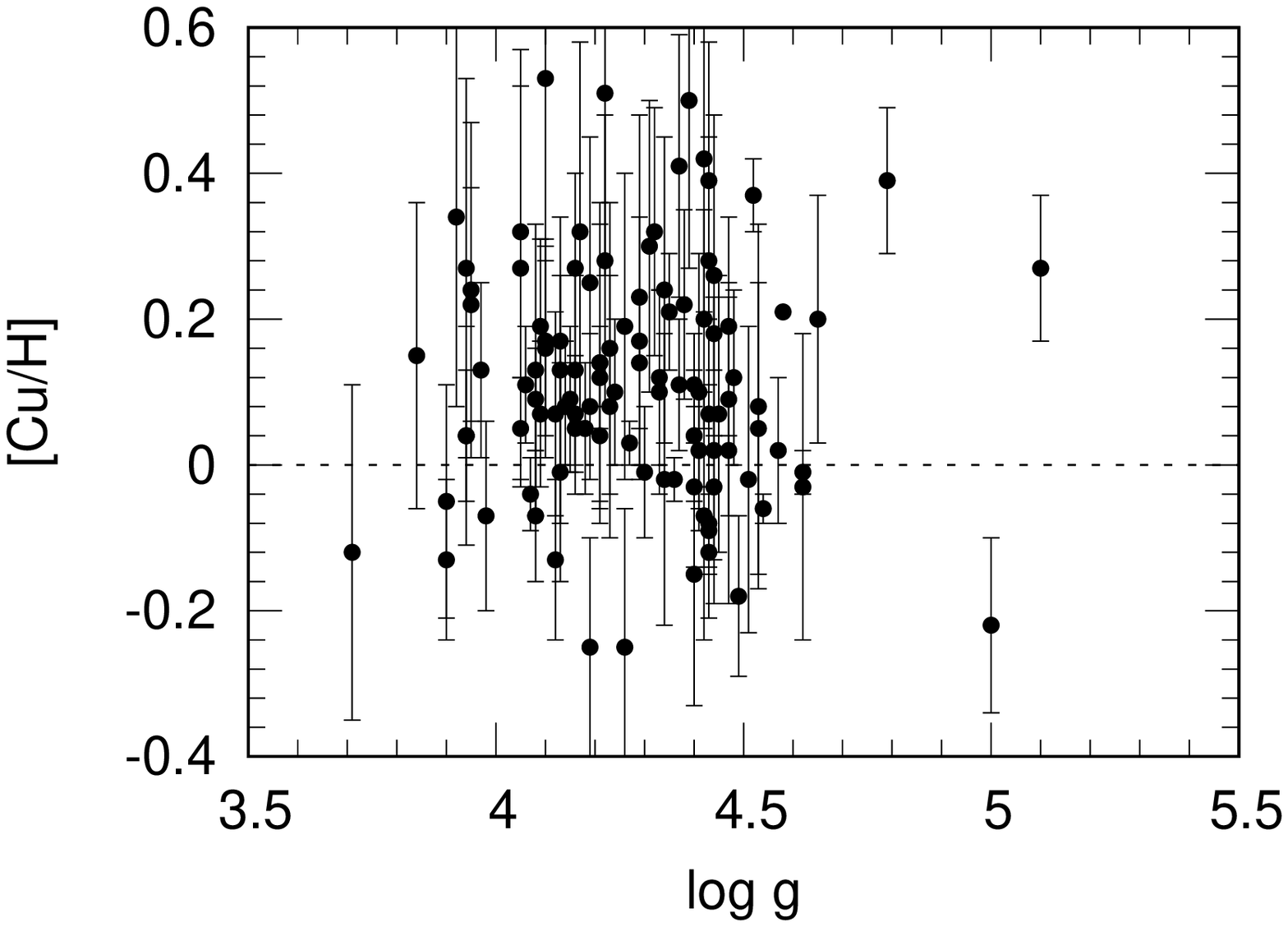} 
\includegraphics[width=55mm]{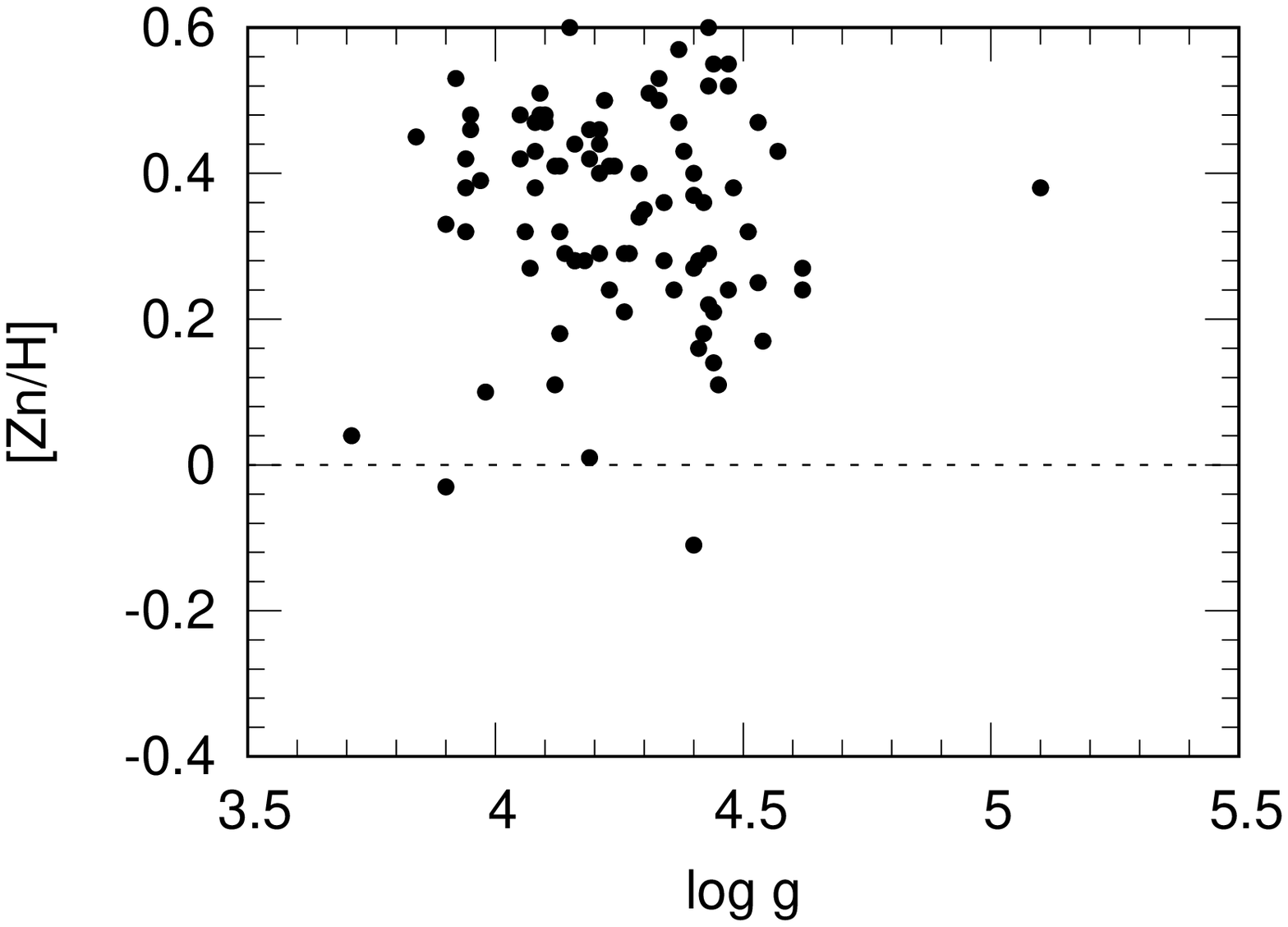} 
\caption{The dependencies of [X/H] versus \logg\, in our sample.}
\label{_figure_gravity_dependence}
\end{figure*}

\begin{figure*}
\centering
\includegraphics[width=55mm]{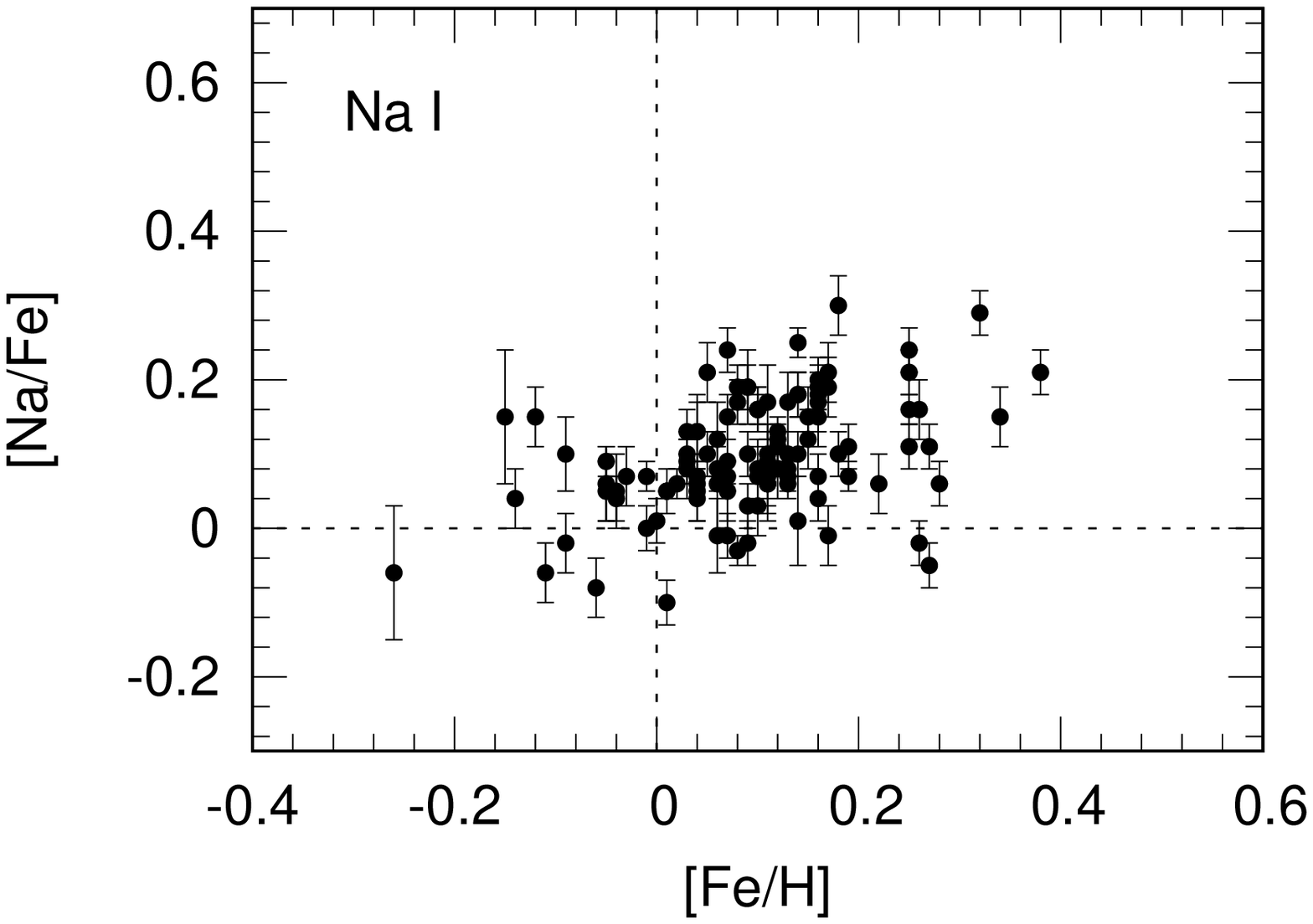} 
\includegraphics[width=55mm]{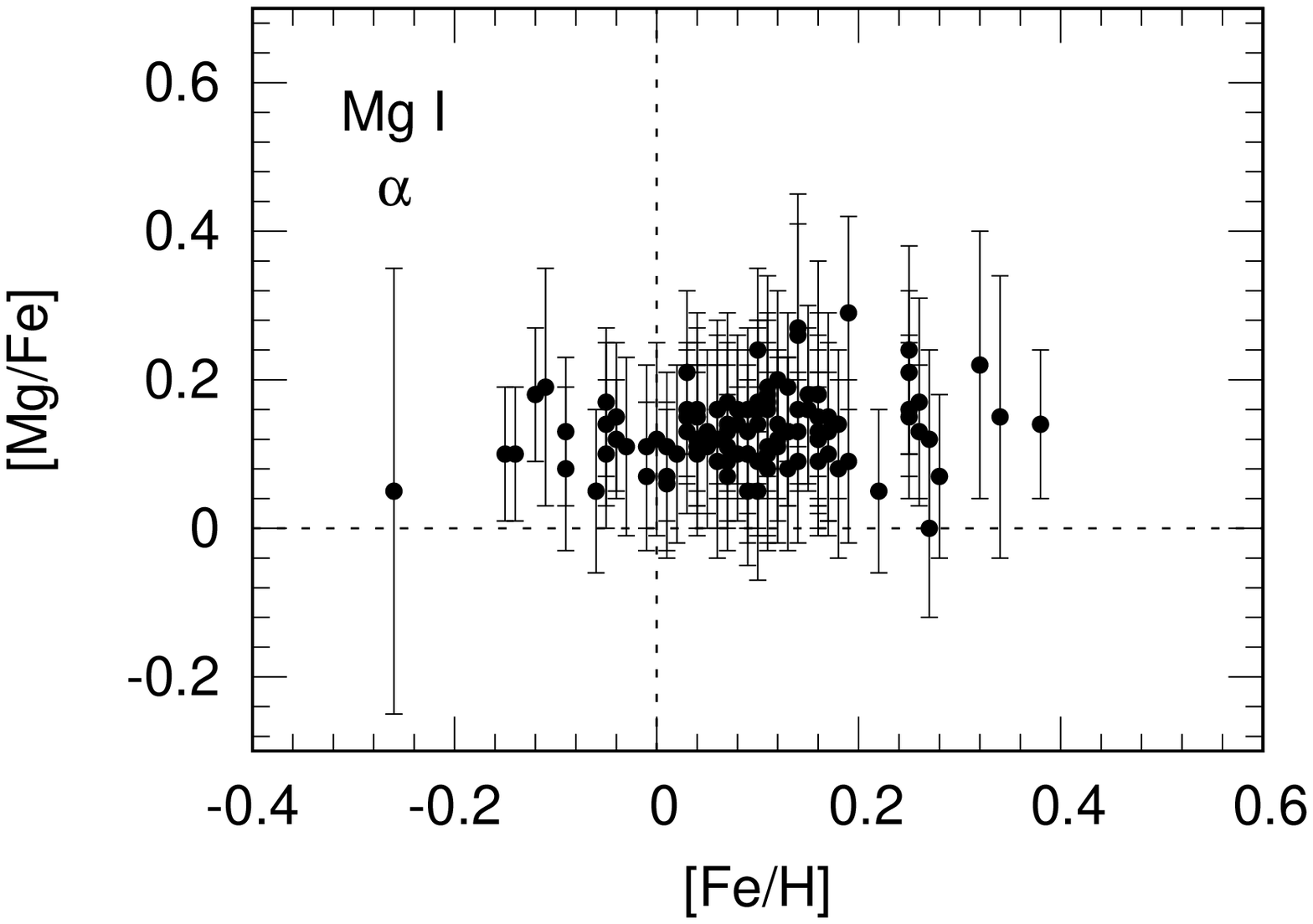} 
\includegraphics[width=55mm]{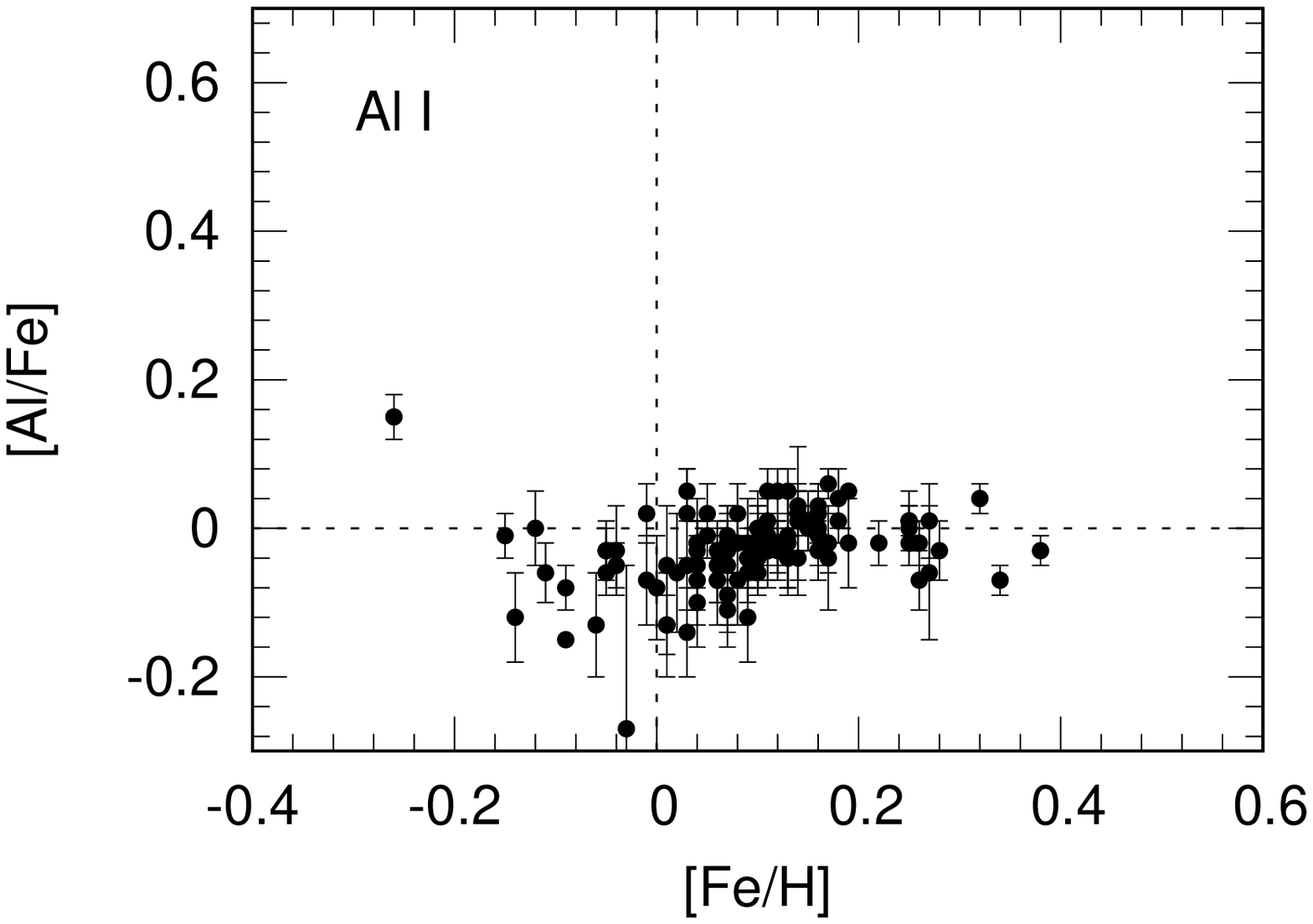} 
\includegraphics[width=55mm]{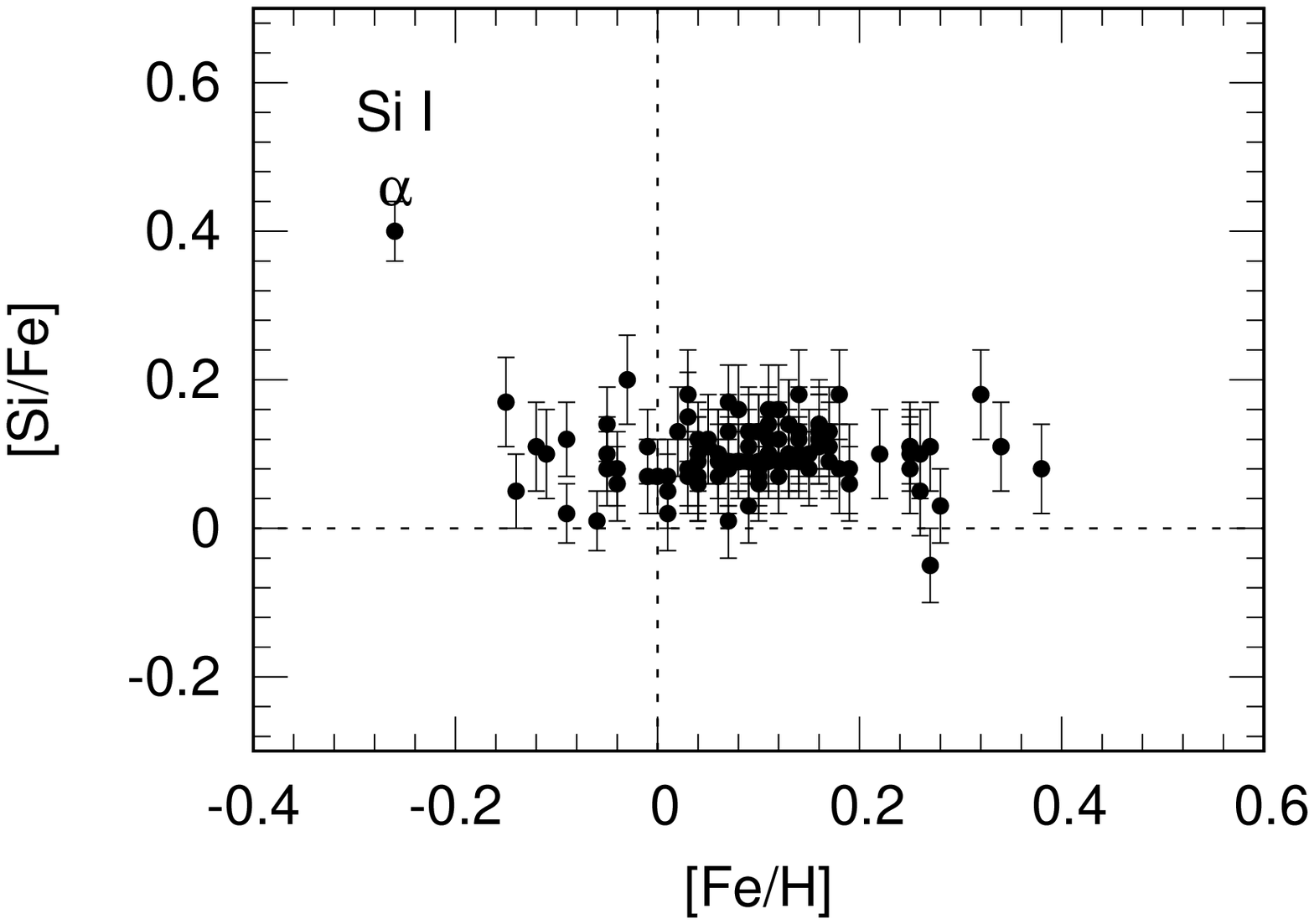} 
\includegraphics[width=55mm]{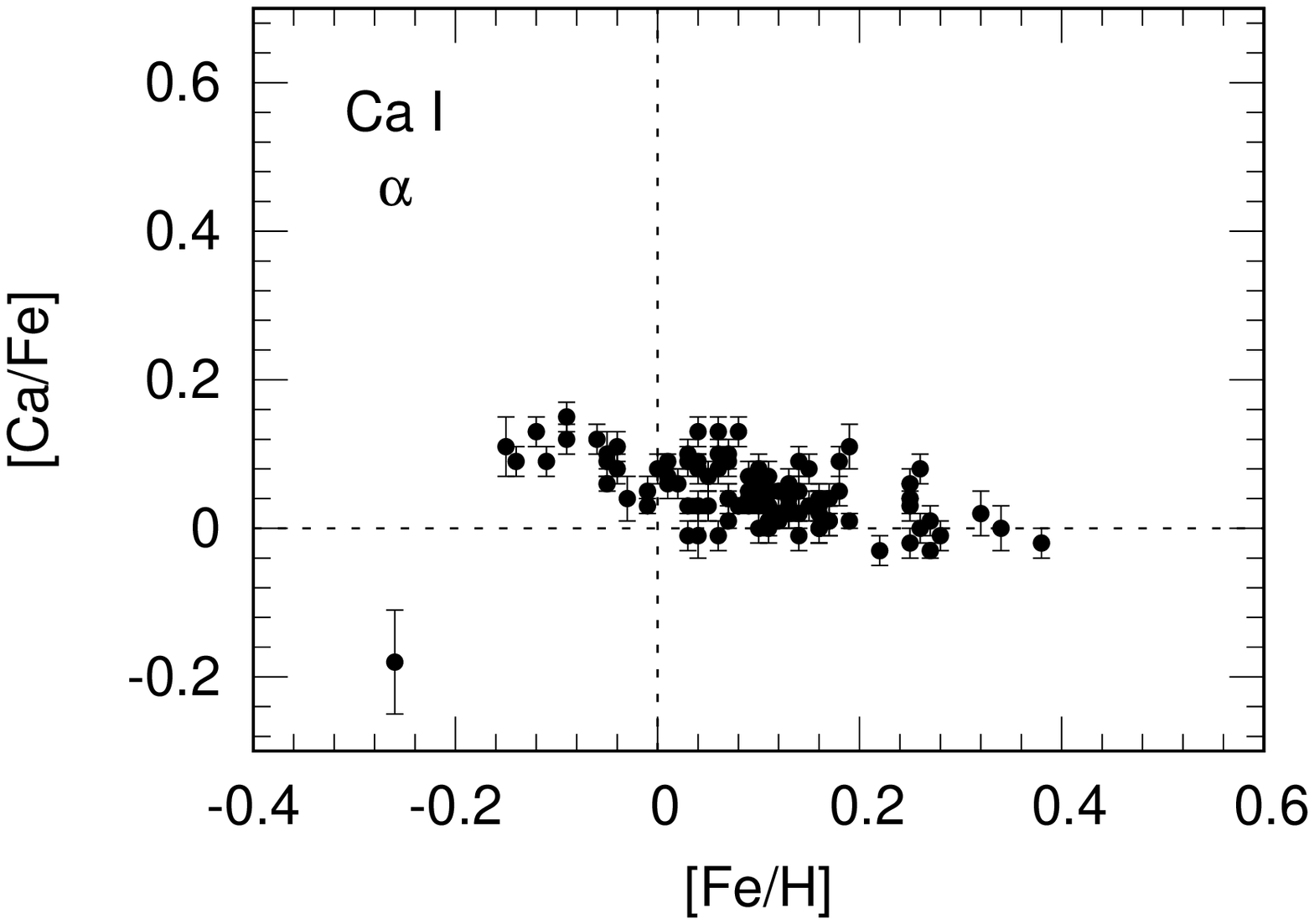} 
\includegraphics[width=55mm]{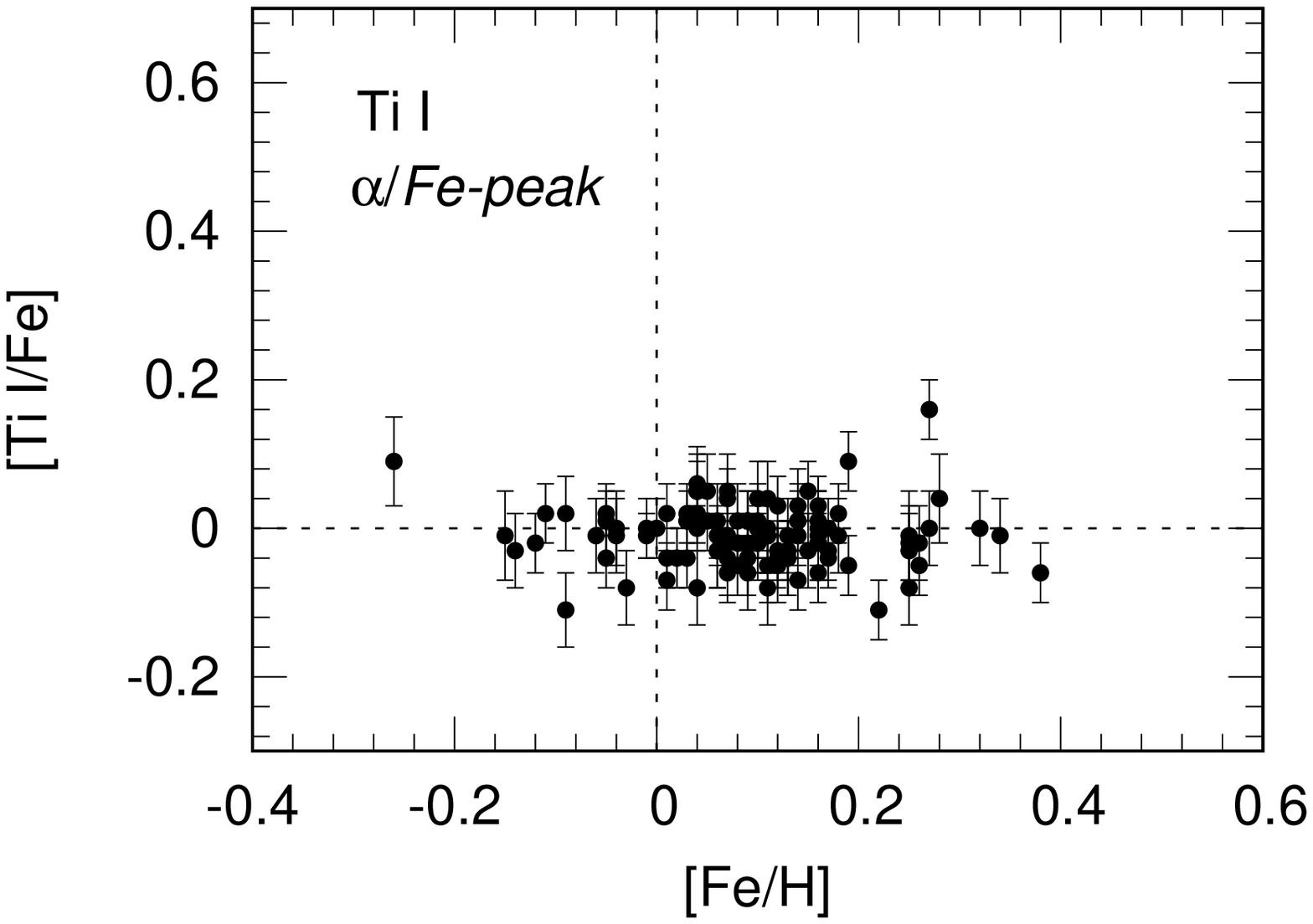} 
\includegraphics[width=55mm]{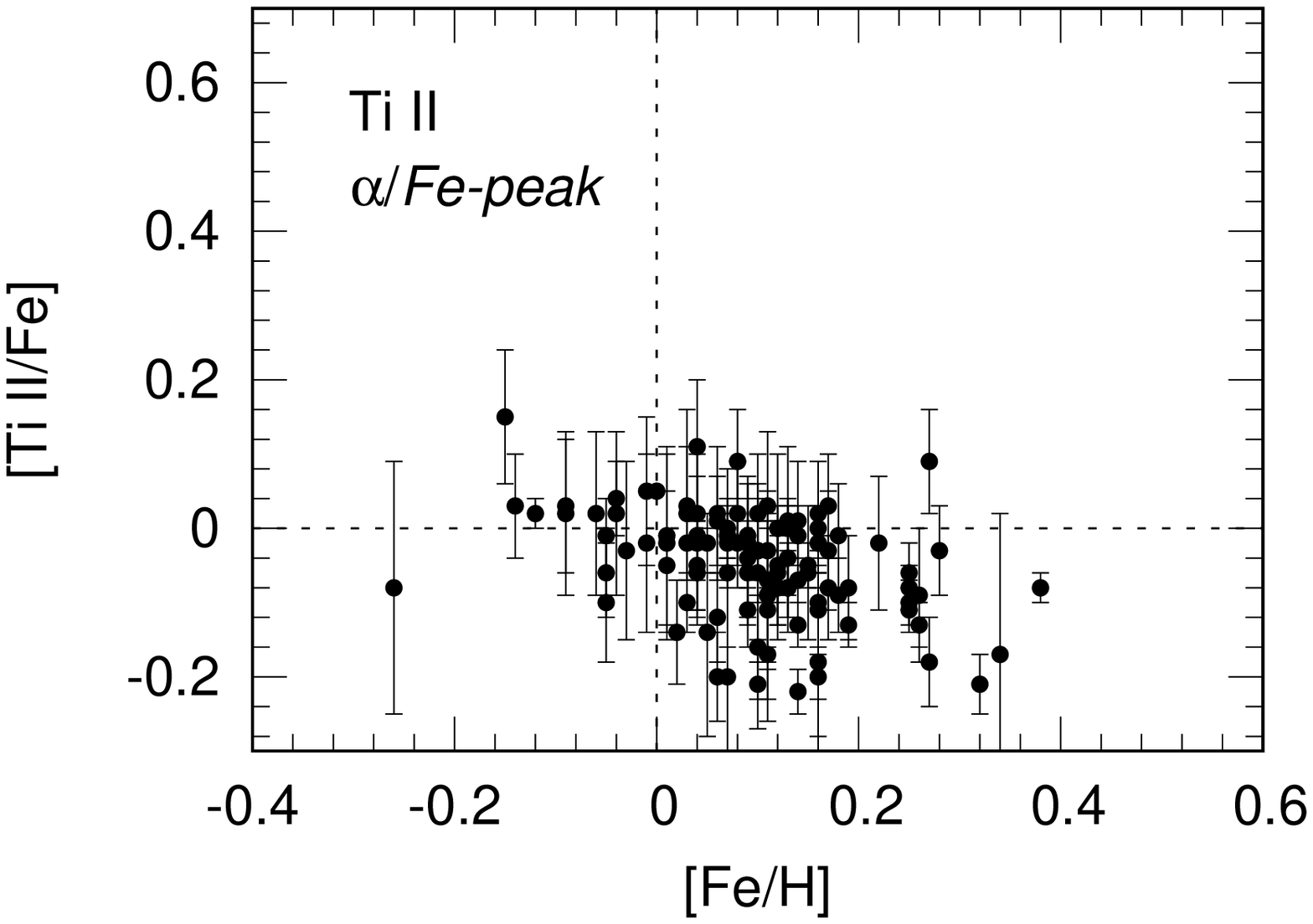} 
\includegraphics[width=55mm]{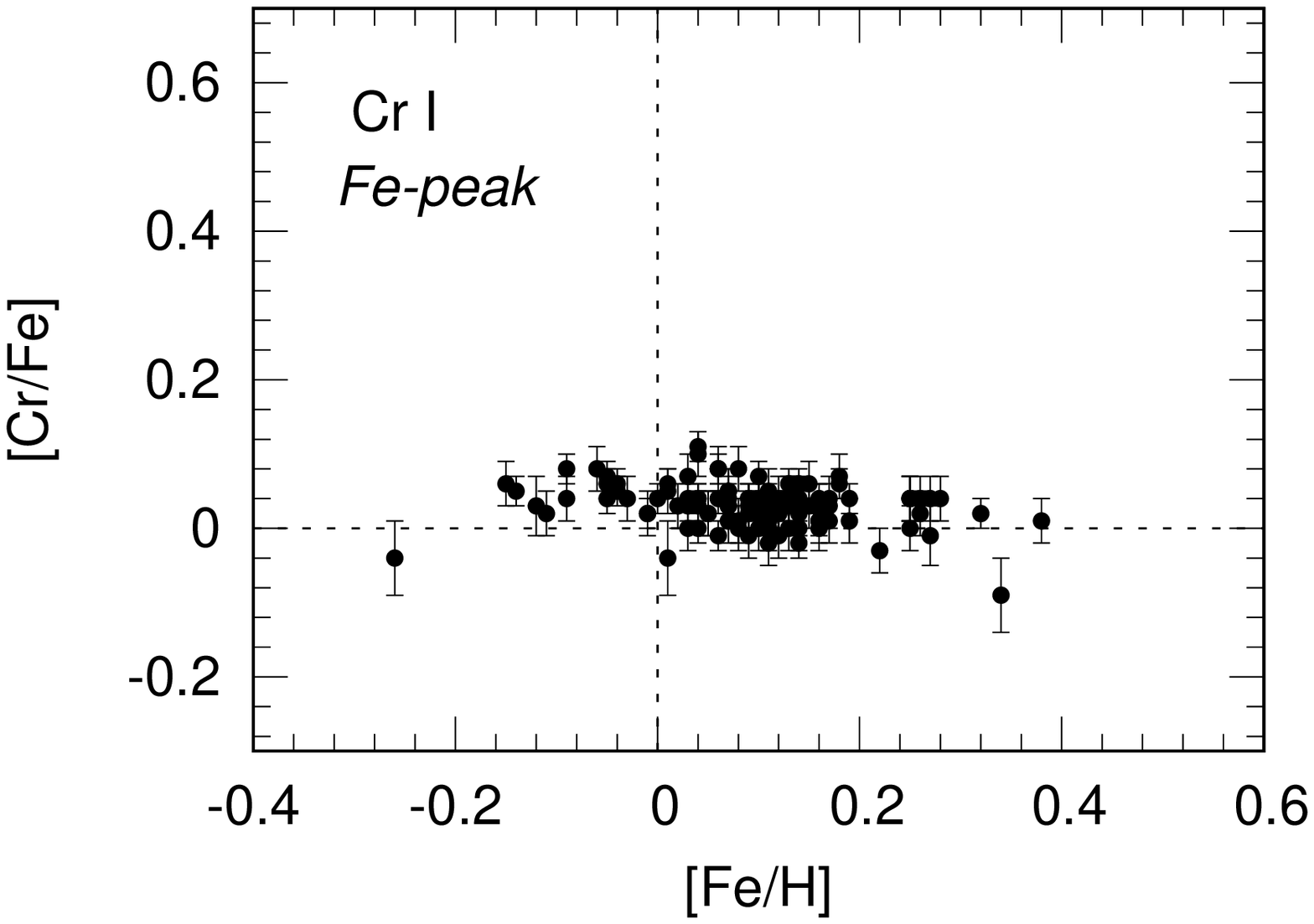} 
\includegraphics[width=55mm]{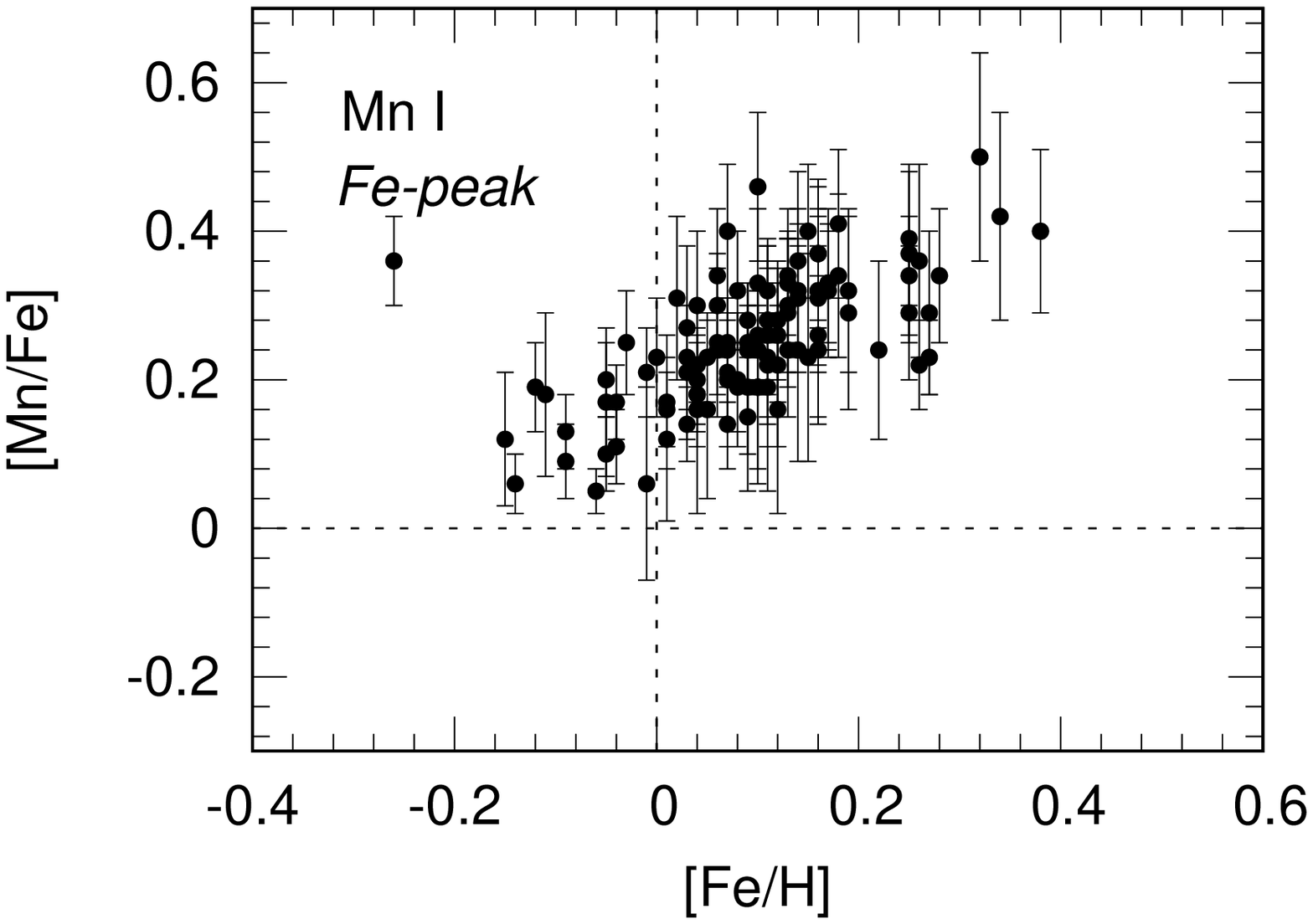} 
\includegraphics[width=55mm]{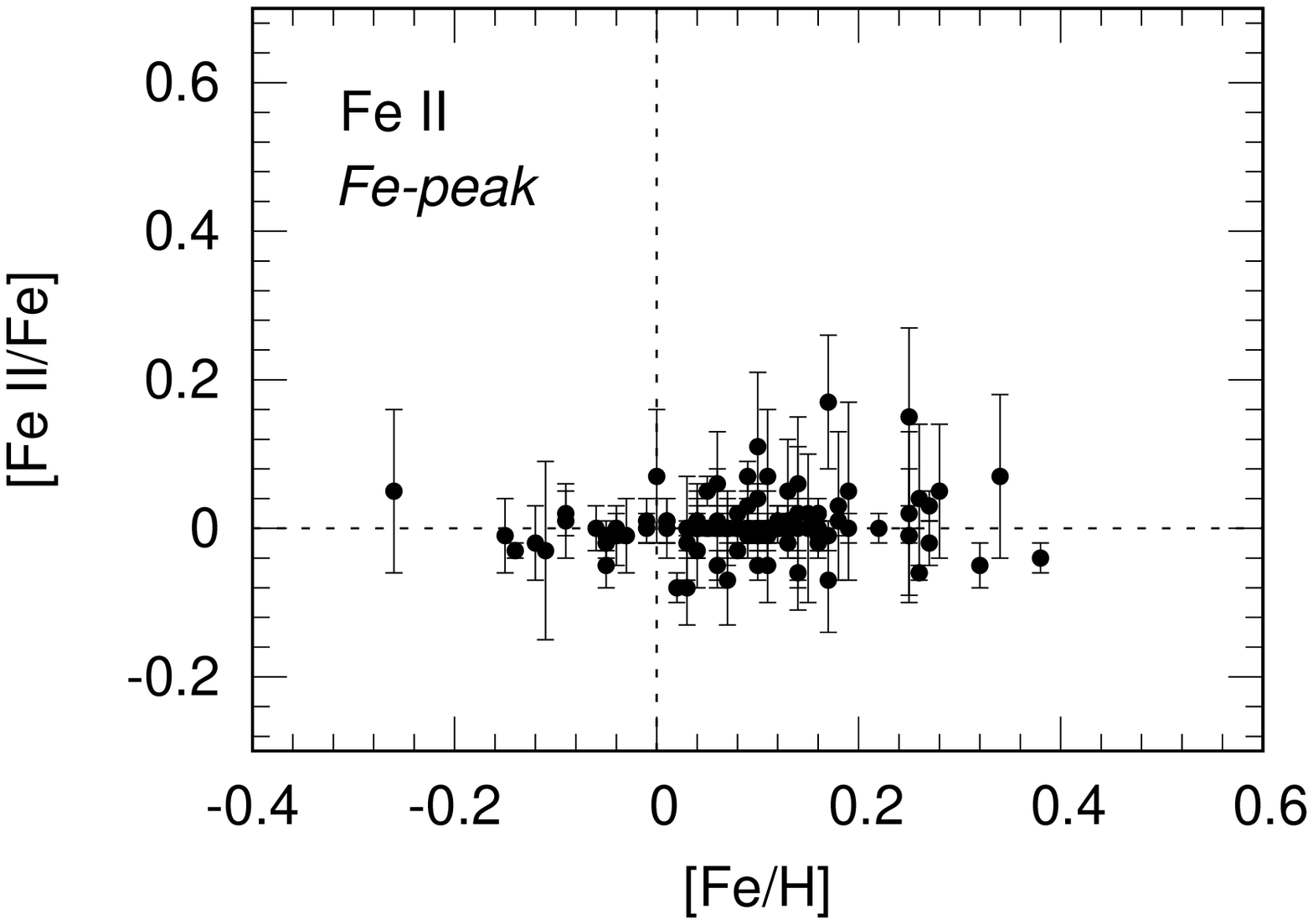} 
\includegraphics[width=55mm]{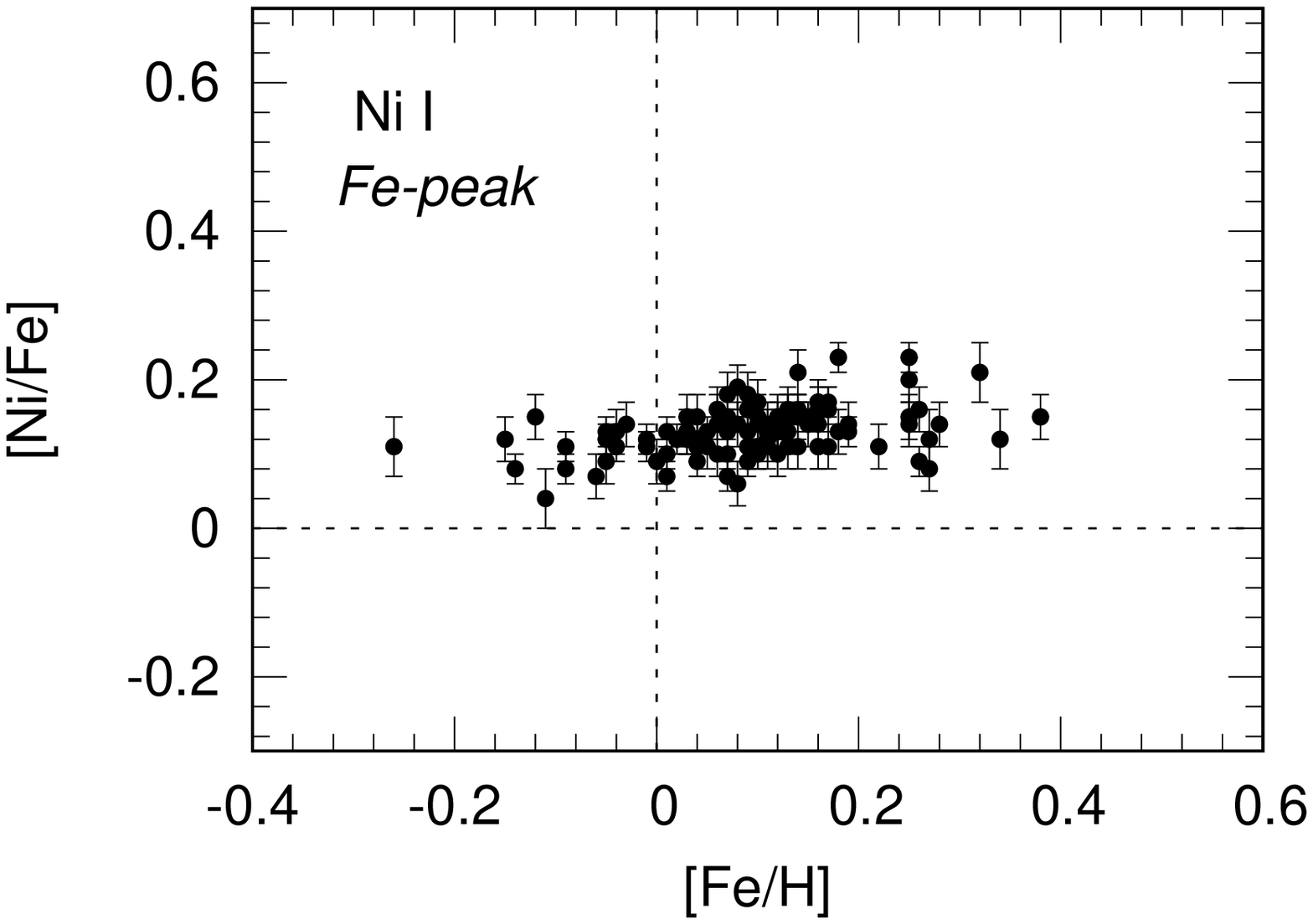} 
\includegraphics[width=55mm]{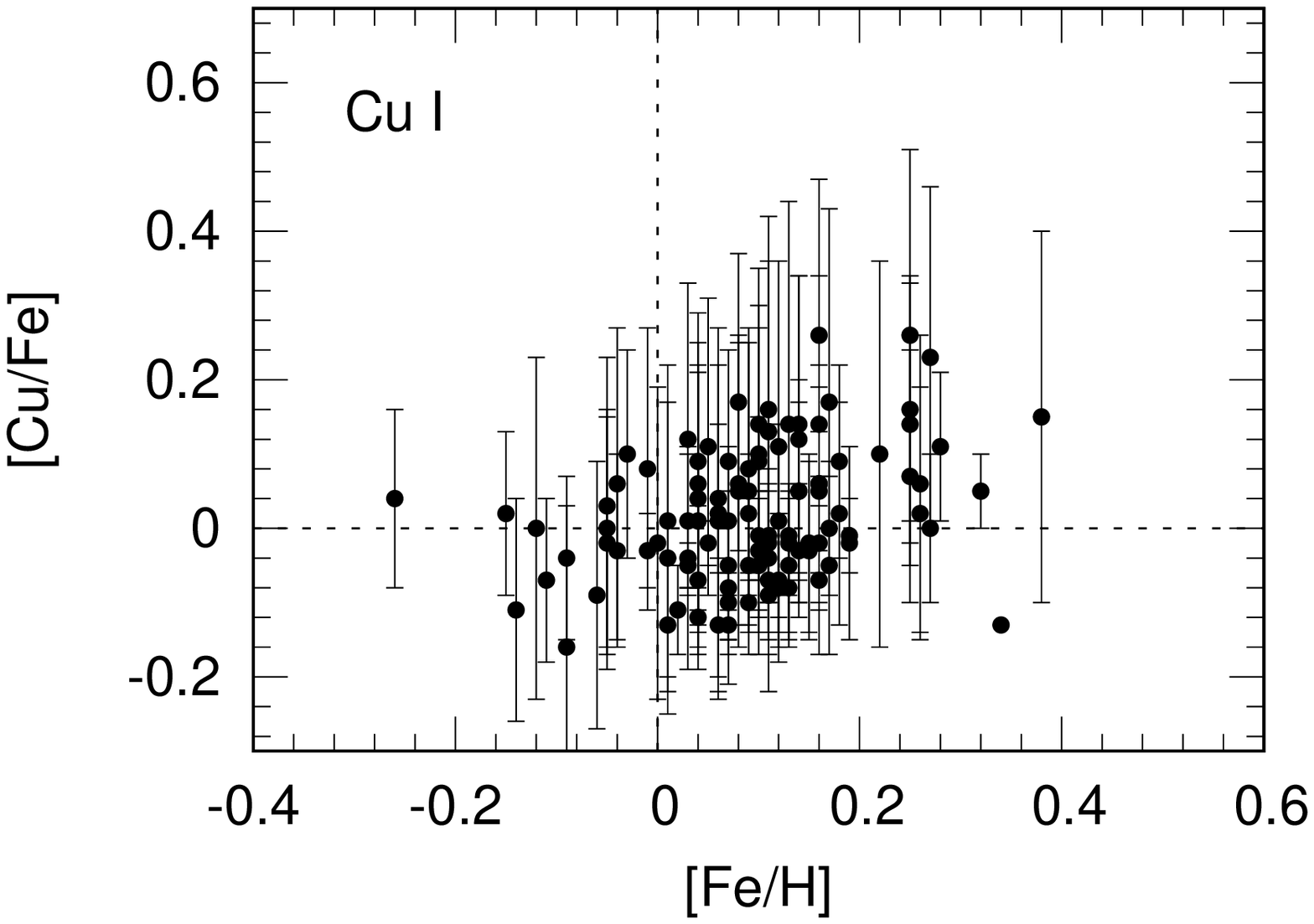} 
\includegraphics[width=55mm]{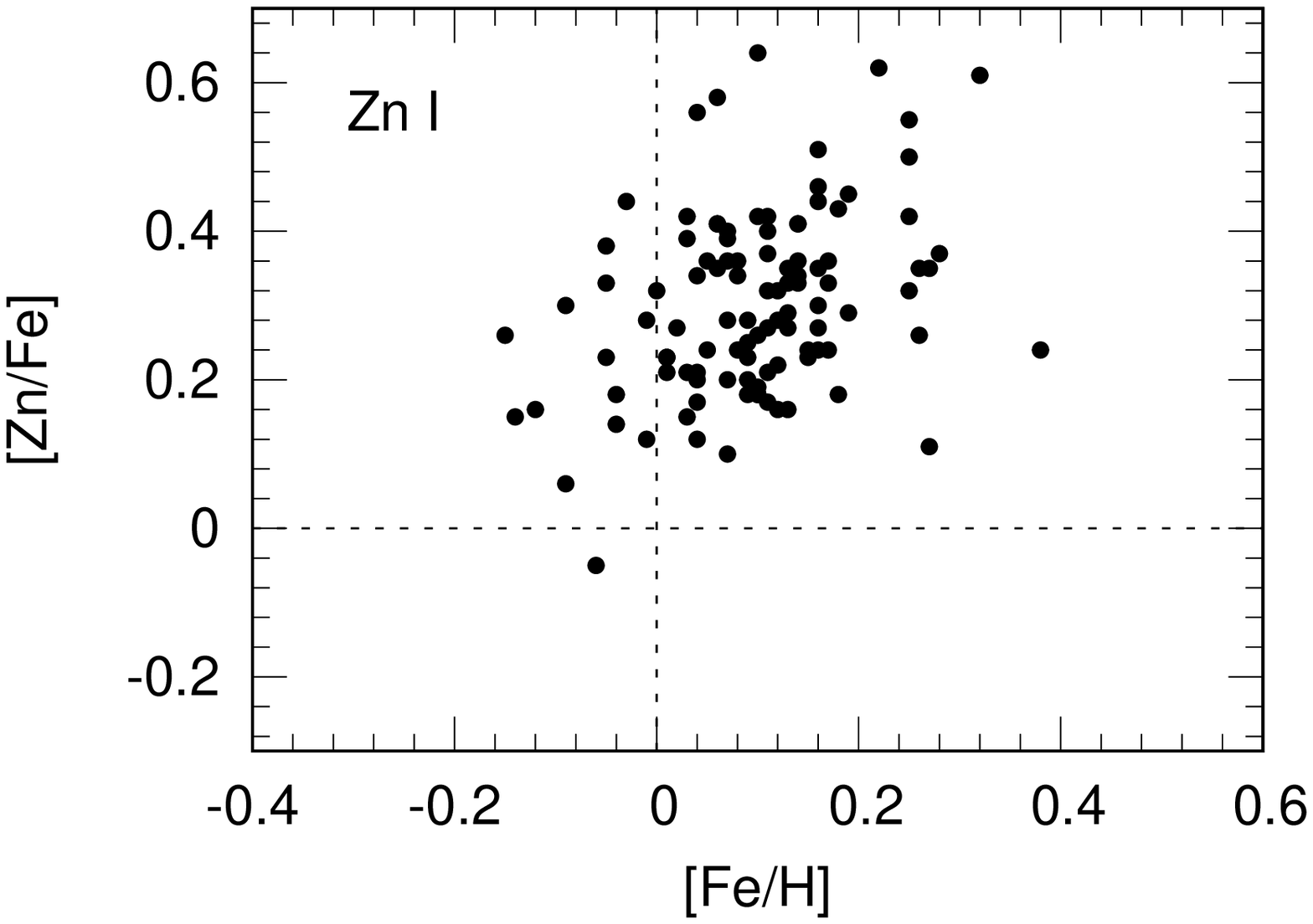} 
\caption{The dependencies of [X/Fe] versus [Fe/H] in our sample.}
\label{_figure_overiron}
\end{figure*}

\label{lastpage}
\end{document}

%% file: ms.bbl
{}